\documentclass[
	amsmath,
	amssymb,
	a4paper,
	aps,		
	prx,		
	reprint,	
	fleqn,
	nofootinbib,
	showpacs,
	floatfix
]{revtex4-1}

\usepackage[utf8]{inputenc}

\usepackage[T1]{fontenc}
\usepackage{microtype}

\usepackage[osf]{mathpazo}
\usepackage[euler-digits,euler-hat-accent]{eulervm}
\DeclareMathAlphabet{\mathsf}{OT1}{iwona}{m}{n} 

\usepackage{tabularx}
\usepackage{booktabs}
\usepackage{multirow}

\usepackage{bm}
\usepackage{amsthm}
\usepackage{braket}
\usepackage{kbordermatrix}
	\renewcommand{\kbldelim}{(}
	\renewcommand{\kbrdelim}{)}
\usepackage{graphicx}
\usepackage{dcolumn}
\usepackage[caption=false]{subfig}
\usepackage{color}

\usepackage[version=3]{mhchem}

\usepackage{xpatch}
\makeatletter
\xpatchcmd{\@ssect@ltx}{\@xsect}{\protected@edef\@currentlabelname{#8}\@xsect}{}{}
\xpatchcmd{\@sect@ltx}{\@xsect}{\protected@edef\@currentlabelname{#8}\@xsect}{}{}
\makeatother

\usepackage{hyperref}

\newcommand{\myTitle}{Conservation Laws and Work Fluctuation Relations in Chemical Reaction Networks}

\newcommand{\myName}{Riccardo Rao}
\newcommand{\myAffiliation}{Complex Systems and Statistical Mechanics, Physics and Materials Science Research Unit, University of Luxembourg, L-1511 Luxembourg, G.D.~Luxembourg}

\newcommand{\myAdvisor}{Massimiliano Esposito}

\newcommand{\derpart}[2]{\frac{\partial{#1}}{\partial{#2}}}

\newcommand{\de}{\mathrm{d}}
\newcommand{\dt}{\mathrm{d}_{t}}

\newcommand{\at}[2]{\left.{#1}\right|_{#2}}
\newcommand{\ave}[1]{\left\langle {#1} \right\rangle}

\newcommand{\transpose}{^{\mathrm{T}}}

\newcommand{\kt}{k_\mathrm{B}T}
\newcommand{\kb}{k_\mathrm{B}}

\DeclareMathOperator{\coker}{coker}

\DeclareMathOperator{\diag}{diag}

\theoremstyle{definition} 
\newtheorem{example}{Example}

\definecolor{webgreen}{rgb}{0,.5,0}
\definecolor{webbrown}{rgb}{.6,0,0}
\definecolor{grigio}{rgb}{.85,.85,.85} 
\definecolor{RoyalBlue}{rgb}{0.0, 0.14, 0.4}
\definecolor{skyblue3}{rgb}{0.13,0.29,0.53}

\hypersetup{%
    colorlinks=true, linktocpage=true, pdfstartpage=1, pdfstartview=FitV,%
    breaklinks=true, pdfpagemode=UseNone, pageanchor=true, pdfpagemode=UseOutlines,%
    plainpages=false, bookmarksnumbered, bookmarksopen=true, bookmarksopenlevel=1,%
    hypertexnames=true, pdfhighlight=/O,
    urlcolor=webbrown, linkcolor=RoyalBlue, citecolor=webgreen, 
    pdftitle={\myTitle},%
    pdfauthor={\textcopyright\ \myName},%
    pdfsubject={},%
    pdfkeywords={},%
    pdfcreator={pdfLaTeX},%
    pdfproducer={LaTeX REVTeX}%
}

\definecolor{aluminium5}{rgb}{0.33,0.34,0.32}
\newcommand{\greyt}[1]{\textcolor{aluminium5}{#1}}

\newcommand{\internal}{\mathbf{z}}
\newcommand{\internalx}{\mathbf{x}}
\newcommand{\exchanged}{\mathbf{y}}
\newcommand{\external}{\mathbf{Y}}

\newcommand{\Rf}{\mathbf{Y}_{\mathrm{p}}}
\newcommand{\Frc}{\mathbf{Y}_{\mathrm{f}}}
\newcommand{\rf}{\mathbf{y}_{\mathrm{p}}}
\newcommand{\frc}{\mathbf{y}_{\mathrm{f}}}
\newcommand{\lyp}{{\mathrm{y}_{\mathrm{p}}}}
\newcommand{\lyf}{{\mathrm{y}_{\mathrm{f}}}}
\newcommand{\iyf}{{y_{\mathrm{f}}}}

\newcommand{\concY}{[\mathbf{Y}]}

\newcommand{\cy}{\alpha}
\newcommand{\ecy}{\eta}

\newcommand{\semigp}{\mathfrak{g}}
\newcommand{\Semigp}{\mathcal{G}}

\newcommand{\ir}{{\rho_{\mathrm{i}}}}
\newcommand{\er}{{\rho_{\mathrm{e}}}}
\renewcommand{\l}{l}

\newcommand{\Y}{{\mathrm{Y}}}
\newcommand{\y}{{\mathrm{y}}}
\newcommand{\x}{{\mathrm{x}}}

\newcommand{\Yp}{{\mathrm{Y}_{\mathrm{p}}}}
\newcommand{\Yf}{{\mathrm{Y}_{\mathrm{f}}}}
\newcommand{\yrf}{{\mathrm{y}_{\mathrm{p}}}}
\newcommand{\yfrc}{{\mathrm{y}_{\mathrm{f}}}}

\newcommand{\st}[1]{\{#1\}}

\newcommand{\stoich}{\mathbf{S}}
\newcommand{\stoichM}{\mathbb{S}}
\newcommand{\rate}[2]{w_{#1}(#2)}
\newcommand{\trjdep}{[\trj]}

\newcommand{\nof}[1]{\mathsf{N}_{#1}}

\newcommand{\trj}{\mathfrak{n}_{t}}

\newcommand{\avef}[1]{\langle {#1} \rangle}

\renewcommand{\transpose}{^{\mathsf{T}}}

\begin{document}

\title{\myTitle}

\author{\myName}
\affiliation{\myAffiliation}
\author{\myAdvisor}
\affiliation{\myAffiliation}

\date{\today. Published in \emph{J.~Chem.~Phys.}, DOI:~\href{https://doi.org/10.1063/1.5042253}{10.1063/1.5042253}} 

\begin{abstract}
	We formulate a nonequilibrium thermodynamic description for open chemical reaction networks (CRN) described by a chemical master equation.
	The topological properties of the CRN and its conservation laws are shown to play a crucial role.
	They are used to decompose the entropy production into a potential change and two work contributions, the first due to time dependent changes in the externally controlled chemostats concentrations and the second due to flows maintained across the system by nonconservative forces.
	These two works jointly satisfy a Jarzynski and Crooks fluctuation theorem.
	In absence of work, the potential is minimized by the dynamics as the system relaxes to equilibrium and its equilibrium value coincides with the maximum entropy principle.
	A generalized Landauer's principle also holds:
	the minimal work needed to create a nonequilibrium state is the relative entropy of that state to its equilibrium value reached in absence of any work.
\end{abstract}


\maketitle

\section{Introduction}

Nonequilibrium thermodynamic descriptions of stochastic (bio-)chemical processes have long since been developed.
Among the first, T.L.~Hill and coworkers studied bio-catalysts as small fluctuating machines operating at steady-state.
They introduced the concept of free energy transduction and analyzed how one form of chemical work can drive another one against its spontaneous direction \cite{hill77,*hill05}.
The importance of decomposing currents into network cycles (\emph{i.e.} cyclic sets of transitions) was already emphasized.
These results were however limited to steady-state systems described by linear chemical reaction networks (CRN).
The stochastic as well as the deterministic dynamics of these CRNs is described by the same linear rate equations for, respectively, probabilities or concentrations.
They model for instance conformational changes of an enzyme or of a membrane transporter.
Inspired by these seminal works, J.~Schnakenberg formulated a steady-state thermodynamics for generic Markov jump processes and provided a systematic cycle decomposition for the entropy production (EP) rate \cite{schnakenberg76}.
He considered in particular the stochastic description in terms of the Chemical Master Equation (CME) \cite{mcquarrie67,gillespie92} of nonlinear chemical reaction networks, \emph{i.e.} CRNs described at the deterministic level by nonlinear rate equations for concentrations.
The Brussels school, J.~Ross and many others, focused on the connection between the thermodynamic description resulting from the stochastic and the deterministic dynamics \cite{luo84,mou86,zheng91,vlad94:fluctuation}.

With the advent of Stochastic Thermodynamics \cite{sekimoto10,jarzynski11,seifert12,vandenbroeck15}, the focus moved to the study of fluctuations, rather then focusing on the first two moments.
Gaspard first showed that EP fluctuations in nonlinear CRNs at steady state satisfy a fluctuation theorem (FT) \cite{gaspard04}.
This result was later expressed in terms of currents along Schnakenberg cycles \cite{andrieux04,andrieux07:schnakenberg}.
Fluctuations in complex chemical dynamics such as bistability was analyzed, amongst others, by Qian and coworkers \cite{ge09:maxwell,vellela09,qian10}.
A first formulation of stochastic thermodynamics for CRNs beyond steady state was done by Schmiedl and Seifert \cite{schmiedl07}.

Despite this long history none of these descriptions made use of the specific topology of the CRN encoded in its stoichiometric matrix. 
Mathematicians know however that the CRN topology plays an important role on its deterministic \cite{horn72,feinberg72} as well as stochastic dynamics \cite{anderson10,cappelletti16}.
But the question of how it affects the thermodynamic description was only studied recently:
for deterministic dynamics in Refs.~\cite{polettini14,rao16:crnThermo}, and for stochastic dynamics at steady state in \cite{polettini15}.
In this paper we address this question in full generality for CRNs whose dynamics is stochastic.
We will do so by presenting a formulation of stochastic thermodynamics for CRNs which systematically makes use of the conservation laws.
Doing so leads to a significantly more informative thermodynamic description.
In particular, we decompose the EP into three fundamental dissipative contributions:
a newly defined potential change, a driving work contribution due to time dependent changes in the externally controlled chemostats concentrations, and a nonconservative work contribution due to a minimal set of flows maintained across the system by nonconservative forces.
In contrast to the traditional chemical work given by minus the free energy change in the chemostats, these two new work contributions are shown to jointly satisfy a finite-time detailed and integral FT, when the CRN is initially prepared in an equilibrium state.
In turn, the importance of the potential lies in the fact that it is minimized by the relaxation dynamics towards equilibrium in absence of the first two work contributions, \emph{i.e.} when the system is detailed-balanced.
It can be seen as a Legendre transform with respect to those conservation laws that are broken by the chemostats.
At equilibrium, it coincides with the potential obtained from maximizing entropy with broken conservation laws as constrains.
We also discuss the connection of our findings to absolute irreversibility \cite{murashita14}, to free energy transduction in nonlinear CRNs, and to cycle decompositions of the entropy production.
Finally, we derive a nonequilibrium Landauer's principle for the driving and nonconservative work which generalizes the previous ones to nondetailed-balanced dynamics \cite{esposito11,parrondo15}.

\paragraph*{Outline}
The paper is organized as follows.
In \S~\ref{sec:cn} (\nameref{sec:cn}) we review the stochastic description of closed and open CRNs and introduce conservation laws and stoichiometric cycles. 
In \S~\ref{sec:ST} (\nameref{sec:ST}) the connection with thermodynamics is made. 
The stochastic reaction rates are expressed in terms of Gibbs potentials via the equilibrium distribution of the closed CRN.
Enthalpy and entropy balance are defined along stochastic trajectories and Jarzynski-like FTs for the chemical work are discussed.
In \S~\ref{sec:NSST} (\nameref{sec:NSST}) the EP is partitioned into its three contributions.
In \S~\ref{sec:dbn} (\nameref{sec:dbn}) we analyze open detailed balanced CRNs, more specifically their relaxation to equilibrium as chemostats are successively introduced.
In \S~\ref{sec:fr} (\nameref{sec:fr}), finite-time detailed FTs for the driving and nonconservative work are derived.
In \S~\ref{sec:ead} (\nameref{sec:ead}) the ensemble average description is presented and the nonequilibrium Landauer's principle is derived.
Finally in \S~\ref{sec:examples}, our results are applied on a simple model to show the importance of our formulation for free energy transduction.
Throughout the paper, our formalism is illustrated using a simple enzymatic scheme, whereas some technical derivations are given in appendices.
We also provide a table which lists the symbols used throughout the paper, Tab.~\ref{tab:glossary}.

\section{Stochastic Dynamics and CRN Topology}
\label{sec:cn}

\subsection{Chemical Reaction Networks}
\label{sec:ocn}

\begin{figure}[t]
	\centering
	\includegraphics[width=.45\textwidth]{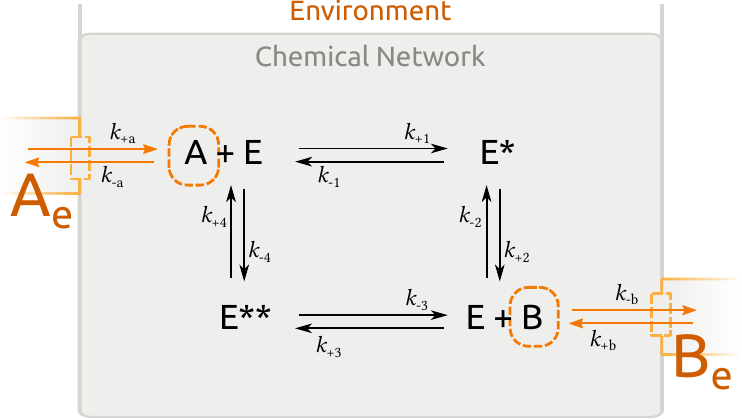}
	\caption{
		Pictorial representation of an open CRN modeling an enzymatic scheme discused in Ex.~\ref{ex:preliminary}
	}
	\label{fig:openMM}
\end{figure}

We consider a homogeneous, isobaric, and isothermal \emph{ideal dilute solution} made of $\nof{\mathrm{z}}$ chemical species, encoded in a vector $\internal$.
Their integer-valued \emph{population} $\bm n$ changes due to internal reactions which we label by $\set{\ir}$ for $\ir = \pm 1,\dots,\pm \nof{\mathrm{i}}$,
\begin{equation}
		\bm \nu_{\ir} \cdot \internal
		\ce{<=>[k_{\ir}][k_{-\ir}]}
		\bm \nu_{- \ir} \cdot \internal \, . 
	\label{ce:CRN}
\end{equation}
In \emph{open} CRNs, the population of a subset of species, named \emph{exchanged} species and denoted by $\exchanged$ where $\internal \equiv (\internalx, \exchanged)$, varies also due to exchanges with external \emph{chemostats} denoted by $\external$.
Their effect is modeled by exchange reactions, $\st{\er}$ for $\er = \pm 1,\dots,\pm \nof{\mathrm{y}}$, see Fig.~\ref{fig:openMM},
\begin{equation}
	\bm \nu^{\y}_{\er} \cdot \exchanged \ce{<=>[k_{\er}][k_{-\er}]} \bm \nu^{\Y}_{-\er} \cdot \external \, .
	\label{ce:CRNexch}
\end{equation}
The non-negative integer-valued vectors $\st{\bm \nu_{\rho} \equiv (\bm \nu^{\x}_{\rho}, \bm \nu^{\y}_{\rho})}$ for $\rho \in \st{\ir} \cup \st{\er}$, encode the \emph{stoichiometric coefficients} of each reaction.
Note that each entry of ${\bm \nu^{\y}_{\er}}$ and ${\bm \nu^\Y_{\er}}$ is nonzero and equal to one only if it corresponds to the species exchanged by $\er$.
Note also that all reactions are assumed \emph{elementary} and \emph{reversible}. 
For any reaction $\rho$, $-\rho$ denotes its backward counterpart, and the sums over $\rho$ includes both $+$ and $-$.
The different types of species are summarized in Tab.~\ref{tab:species}.

\begin{table}
	\centering
	\begin{tabular}{rcccl}
		\toprule
		\textbf{species} & \textbf{symbol} & \textbf{number} & \textbf{abundance} & \\
		\midrule
		\multirow{2}*{internal $\Big\{\quad\quad\quad\quad\quad$}		& $\internalx$ & $\nof{\mathrm{x}}$ & $\bm n_{\mathrm{x}}$ & \multirow{2}*{$\Big\}\bm n$} \\
		exchanged 		& $\exchanged$ & $\nof{\mathrm{y}}$ & $\bm n_{\mathrm{y}}$ & \\
		chemostatted	& $\external$ & $\nof{\mathrm{y}}$ & $[\external]$ & \\
		\bottomrule
	\end{tabular}
	\caption{
		\label{tab:species}
		In the second column the symbols used for the various species are listed.
		The corresponding total number of entries and symbols used to denote their abundance are given in the third and fourth column, respectively.
		The first column summarizes the name used to refer to these species, while the last one lists the symbol used to collect the abundances of the internal species.
		Internal species, $\internalx$ and $\exchanged$, are characterized by low populations, $\bm n$.
		The population of $\internalx$ can change only because of reactions, whereas that of $\exchanged$ are also exchanged with chemostats, which are identified by $\external$, Eq.~\eqref{ce:CRN}.
	}
\end{table}

The topology of the CRN is encoded in its \emph{stoichiometric vectors},
\begin{equation}
	\begin{aligned}
		\stoich_{\rho} &:= \bm \nu_{-\rho} - \bm \nu_{\rho} \, , \; \text{and} &
		\stoich^{\Y}_{\rho} &:= \bm \nu^{\Y}_{-\rho} - \bm \nu^{\Y}_{\rho} \, .
	\end{aligned}
	\label{eq:stoichiometricVector}
\end{equation}
The former quantifies the change of population induced by a given reaction $\rho$, whereas the latter the corresponding amount of chemostatted species that is exchanged.
By definition, $\stoich_{\rho} = - \stoich_{-\rho}$ and $\stoich^{\Y}_{\rho} = - \stoich^{\Y}_{-\rho}$.
Collecting the column vectors $\stoich_{\rho}$ (resp. $\stoich^{\Y}_{\rho}$) corresponding to arbitrarily-chosen forward reactions defines the internal (resp. external) \emph{stoichiometric matrix} denoted by $\stoichM$ (resp. $\stoichM^{\Y}$).
It is not difficult to see that these can be decomposed as
\begin{equation}
	\stoichM
	\equiv
	\begin{pmatrix}
		\stoichM_{\mathrm{i}} & \stoichM_{\mathrm{e}}
	\end{pmatrix}
	\equiv
	\begin{pmatrix}
		\stoichM^{\x}_{\mathrm{i}} & \mathbb{O} \\
		\stoichM^\y_{\mathrm{i}} & \stoichM^\y_{\mathrm{e}}
	\end{pmatrix} \, ,
	\label{eq:SMopen}
\end{equation}
and
\begin{equation}
	\stoichM^\Y \equiv
	\begin{pmatrix}
		\stoichM^\Y_{\mathrm{i}} & \stoichM^\Y_{\mathrm{e}}
	\end{pmatrix} \equiv
	\begin{pmatrix}
		\mathbb{O} & - \stoichM^\y_{\mathrm{e}}
	\end{pmatrix} \, .
	\label{eq:SMopenY}
\end{equation}
In closed CRNs all exchange reactions disappear and the stoichiometric matrix reduces to $\stoichM_{\mathrm{i}}$.

\paragraph*{Remark}
Previous works on thermodynamics of CRNs, \emph{e.g.} Refs.~\cite{qian05,schmiedl07,polettini14,rao16:crnThermo}, describe open CRNs by assuming that the exchanged species $\exchanged$ are so abundant that they can be regarded as particle reservoirs \emph{within} the system. 
As a result the exchange reactions are disregarded, $\exchanged$ are treated as \emph{chemostatted}, and the stoichiometric matrices read
\begin{equation}
	\begin{aligned}
		\stoichM_{\mathrm{alt}} & = \stoichM^{\x}_{\mathrm{i}} \, , \; \text{and} & 
		\stoichM^\Y_{\mathrm{alt}} & = \stoichM^\y_{\mathrm{i}} \, .
	\end{aligned}
	\label{eq:SMdirect}
\end{equation}
In the closed CRNs, the stoichiometric matrix becomes $(\stoichM_{\mathrm{alt}},\stoichM^{\Y}_{\mathrm{alt}})\transpose$.
As we will see, the two approach are formally very similar, but the former has the advantage of preserving the number of internal species when the CRN is chemostatted.    
This makes it more suitable for a stochastic description.

\begin{example}
	\label{ex:preliminary}
	For the open CRN in Fig.~\ref{fig:openMM}, 
	\begin{equation}
		\hspace{-1em}
		\begin{aligned}
			\internalx & = (\ce{E}, \ce{E^{\ast}}, \ce{E^{\ast\ast}}) \, , &
			\exchanged & = (\ce{A}, \ce{B}) \, , &
			\external & = (\ce{A_{\mathrm{e}}}, \ce{B_{\mathrm{e}}}) \, ,
		\end{aligned}
		\label{}
	\end{equation}
        and
	\begin{equation}
		\bm n = (n_{\ce{E}}, n_{\ce{E^{\ast}}}, n_{\ce{E^{\ast\ast}}}, n_{\ce{A}}, n_{\ce{B}}) \, .
		\label{}
	\end{equation}
	Internal reactions, $\ir = \pm 1, \dots, \pm4$, are distinguished from the exchange ones, $\er = \pm a, \pm b$, and the stoichiometric matrices read
	\begin{equation}
		\stoichM = \kbordermatrix{
			& \greyt{+1} & \greyt{+2} & \greyt{+3} & \greyt{+4} & & \greyt{+\mathrm{a}} & \greyt{+\mathrm{b}} \\
			\greyt{\ce{E}}	& -1 & 1 & -1 & 1 & 	\omit\vrule & 0 & 0 \\
			\greyt{\ce{E}^{\ast}} &	1 & -1 & 0 & 0 & 	\omit\vrule & 0 & 0 \\
			\greyt{\ce{E}^{\ast\ast}} &	0 & 0 & 1 & -1 & 	\omit\vrule & 0 & 0 \\
			\greyt{\ce{A}}	& -1 & 0 & 0 & 1 & 	\omit\vrule & 1 & 0	\\
			\greyt{\ce{B}}	& 0	& 1 & -1 & 0 & 	\omit\vrule & 0 & 1	
		} \, ,
		\label{}
	\end{equation}
	and
	\begin{equation}
		\stoichM^{\Y} = \kbordermatrix{
			& \greyt{+1} & \greyt{+2} & \greyt{+3} & \greyt{+4} & & \greyt{+\mathrm{a}} & \greyt{+\mathrm{b}} \\
			\greyt{\ce{A_{\mathrm{e}}}}	& 0 & 0 & 0 & 0 & 	\omit\vrule & -1 & 0	\\
			\greyt{\ce{B_{\mathrm{e}}}}	& 0	& 0 & 0 & 0 & 	\omit\vrule & 0 & -1	
		} \, ,
		\label{}
	\end{equation}
	for our arbitrary choice of forward reactions.
	\qed
\end{example}

\paragraph*{Notation}
Henceforth, we will use the following notation
\begin{equation*}
	\hspace{-1.0em}
	\begin{aligned}
		\bm {a}! &= {\textstyle\prod_{i}} a_{i}! \, ,
		& \bm a^{\cdot \mathbf{b}} &= {\textstyle\prod_{i}} a_{i}^{b_{i}} \, , \text{ and}
		& c^{\cdot \mathbf{b}} &= c^{{\sum_{i}} b_{i}} \, ,
	\end{aligned}
	\label{}
\end{equation*}
for generic vectors $\bm a$ and $\bm b$, and for a generic constant $c$.
``$\ln \bm a$'' must be read as a vector whose entries are the logarithm of the entries of $\bm a$.
$\bm 1$ denotes a vector whose entries are all equal to $1$.
Total and partial time derivatives are written as $\dt$ and $\partial_{t}$, and the overdot ``~$\dot{}$~'' denotes rates of change of observables which are not state functions.

\subsection{Chemical Master Equation}
\label{sec:cme}

In our stochastic description, $\bm n$ is treated as a fluctuating variable and all reactions are regarded as stochastic events.
The probability of finding the CRN in the state $\bm n$ at time $t$ is denoted by $p_{\bm n} \equiv p_{\bm n}(t)$ and its evolution is ruled by the CME \cite{mcquarrie67,nicolis77,gillespie92}
\begin{equation}
	\hspace{-1em}
	\begin{split}
		\dt p_{\bm n} & = {\textstyle\sum_{\rho}} \big\{ 
			\rate{-\rho}{\bm n + \stoich_{\rho}} \, p_{\bm n + \stoich_{\rho}} - \rate{\rho}{\bm n} p_{\bm n}
		\big\} \\
		& = {\textstyle\sum_{\bm m}} \mathcal{W}_{\bm n \bm m} p_{\bm m} \, ,
	\end{split}
	\label{eq:CME}
\end{equation}
where the stochastic generator reads
\begin{equation}
	\mathcal{W}_{\bm {n}\bm {m}} = {\textstyle\sum_{\rho}} \rate{\rho}{\bm m} \left\{ \delta_{\bm{n}, \bm{m}+\stoich_{\rho}} - \delta_{\bm{n},\bm{m}} \right\} \, .
	\label{}
\end{equation}
Since all reactions are assumed elementary, we consider \emph{mass-action} stochastic reaction rates
\begin{equation}
	\rate{\rho}{\bm n} := 
		k_{\rho} \, \frac{V}{V^{\cdot \bm \nu_{\rho}}} \, \concY^{\cdot \bm \nu^\Y_{\rho}} \frac{\bm n!}{(\bm n - \bm \nu_{\rho})!} \, .
	\label{eq:srr}
\end{equation}
where $\st{k_{\rho}}$ denote the \emph{rate constants}.
The dependence on the volume $V$ ensures the correct scaling when taking the large particle limit and guarantees that $\st{k_{\rho}}$ are the same as in deterministic descriptions \cite{kurtz72}.
The chemostats concentrations $[\external]$ only appear in exchange reactions $\er$ and quantify the concentration of the exchanged species in the chemostats.
Hence, they are real-valued, nonfluctuating, and unaffected by the occurrence of exchange reactions.
We assume that $\concY$ can change over time and their value at each time $t$ is encoded in the driving \emph{protocol} $\pi_{t}$.
This may describe for instance, the controlled injection of certain molecules across a cell membrane.
In such situations the CRN is said to be subjected to a ``driving''. In absence of driving the CRNs is instead said to be \emph{autonomous}.

\emph{Equilibrium probability distributions} are of crucial importance for our discussion.
They satisfy the \emph{detailed balance property}
\begin{equation}
	\hspace{-1em}
	\rate{\rho}{\bm n} p^{\mathrm{eq}}_{\bm n} = 
	\rate{-\rho}{\bm n + \stoich_{\rho}} p^{\mathrm{eq}}_{\bm n+\stoich_{\rho}} \, , \quad \text{for all} \, \rho \, , \bm {n} \, .
	\label{eq:dbp}
\end{equation}
This means that the probability current of any reaction $\rho$ occurring from any state $\bm n$ vanishes.
Stochastic CRNs which admit a steady-state probability distribution satisfying Eq.~\eqref{eq:dbp} are referred to as \emph{detailed balanced}.
Their stochastic thermodynamics will be analyzed in \S~\ref{sec:dbn}.

\begin{example}
	For the CRN in Fig.~\ref{fig:openMM}, the transition rates are
	\begin{equation}
		\begin{aligned}
			w_{+1} & =  k_{+1} n_{\ce{A}} n_{\ce{E}} \, , & w_{-1} & =  k_{-1} n_{\ce{E^{\ast}}} \, , \\
			w_{+2} & =  k_{+2} n_{\ce{E^{\ast}}} \, , & w_{-2} & =  k_{-2} n_{\ce{E}} n_{\ce{B}} \, , \\
			w_{+3} & =  k_{+3} n_{\ce{E}} n_{\ce{B}} \, , & w_{-3} & =  k_{-3} n_{\ce{E^{\ast\ast}}} \, , \\
			w_{+4} & =  k_{+4} n_{\ce{E^{\ast\ast}}} \, , & w_{-4} & =  k_{-4} n_{\ce{E}} n_{\ce{A}} \, , \\
			w_{+\mathrm{a}} & =  k_{+\mathrm{a}} [\ce{A}_{\mathrm{e}}] \, , & w_{-\mathrm{a}} & =  k_{-\mathrm{a}} n_{\ce{A}} \, , \\
			w_{+\mathrm{b}} & =  k_{+\mathrm{b}} [\ce{B}_{\mathrm{e}}] \, , & w_{-\mathrm{b}} & =  k_{-\mathrm{b}} n_{\ce{B}} \, .
		\end{aligned}
		\label{}
	\end{equation}
	\qed
\end{example}

\subsection{Stochastic Trajectories}
\label{sec:st}

A stochastic trajectory of duration $t$, $\trj$, is defined as a set of reactions $\{\rho_{\l}\}$ sequentially occurring at times $\{t_{\l}\}$ starting from $\bm n_{0}$ at time $t_{0}$.
Such trajectories can be generated by a \emph{Stochastic Simulation Algorithm} \cite{gillespie07}.
Given the initial state, a trajectory is completely characterized by
\begin{equation}
	j_{\rho}(\bm n,\tau) := {\textstyle\sum_{\l}} \delta_{\rho \rho_{\l}} \delta_{\bm n \bm n_{t_{\l}}} \delta(\tau - t_{\l}) \, ,
	\label{eq:instTrnsRate}
\end{equation}
which encodes the reactions that occur ($\st{\rho_{\l}}$), the states from which these occur ($\st{\bm n_{t_{\l}}}$), and the reaction times ($\st{t_{\l}}$).
The transition index $\l$ runs from $\l=1$ to the last transition prior to time $t$, $\nof{t}$.
The \emph{instantaneous reaction fluxes}
\begin{equation}
	J_{\rho}(\tau) := {\textstyle\sum_{\bm n}} j_{\rho}(\bm n,\tau) = {\textstyle\sum_{\l}} \delta_{\rho \rho_{\l}} \delta(\tau - t_{\l}) \, .
	\label{eq:instTrnsCrnt}
\end{equation}
quantify the instantaneous rate of occurrence of each reaction irrespectively of the state from which it occurs.
Additionally, we denote the population of the CRN at time $\tau \in [t_{0}=0, t]$ by $\bm n_{\tau}$.

The path probability of a trajectory reads
\begin{multline}
	\mathcal{P}\trjdep =
	\prod_{\l=0}^{\nof{t}} \exp\left\{ - \int_{t_{\l}}^{t_{\l+1}} \de \tau \, {\textstyle\sum_{\rho}} \rate{\rho}{\bm n_{\tau},\tau} \right\} \\
	\times \prod_{\l=1}^{\nof{t}} \rate{\rho_{\l}}{\bm n_{t_{\l}},t_{\l}} \, ,
	\label{eq:probTrajectory}
\end{multline}
where $t_{\nof{t} + 1} := t$ is the final time of the trajectory.
The first term accounts for the probability that the system spends $\st{t_{\l+1} - t_{\l}}$ time in the state $\st{\bm {n}_{t_{\l}}}$, while the second accounts for the probability of transitioning.
When averaging Eq.~\eqref{eq:instTrnsRate} over all stochastic trajectories, we obtain the transition rates, Eq.~\eqref{eq:srr},
\begin{equation}
	\avef{j_{\rho}(\bm n,\tau)} = \rate{\rho}{\bm n,\tau} p_{\bm n}(\tau) \, .
	\label{eq:aveCurrentGivenState}
\end{equation}

\begin{figure}[t]
	\centering
	\includegraphics[width=.45\textwidth]{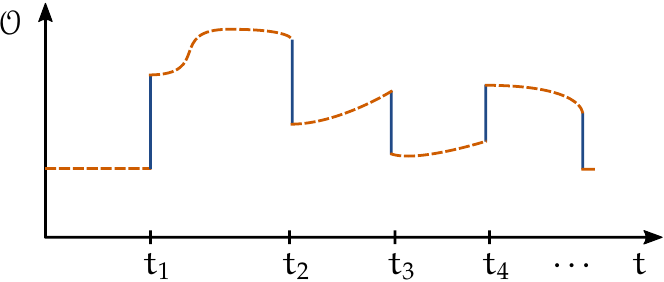}
	\caption{
		Pictorial representation of the change of a state variable observable $\mathcal{O}$ along a trajectory.
		The orange dashed curves represent the changes due to the protocol---first term in Eq.~\eqref{eq:DeltaO}---while vertical blue lines changes due to reactions---second term in Eq.~\eqref{eq:DeltaO}.
	}
	\label{fig:trj}
\end{figure}

Changes of generic observables along trajectories are written as
\begin{equation}
	\hspace{-2.2em}
	\delta\mathcal{X}\trjdep
	= \int_{0}^{t} \de \tau \Big\{ \dot{\mathcal{X}}(\bm n_{\tau},\tau)
	+ \sum_{\bm n, \rho} \delta \mathcal{X}_{\rho} (\bm n,\tau) \, j_{\rho}(\bm n, \tau) \Big\} \, .
	\label{eq:deltaX}
\end{equation}
where $\dot{\mathcal{X}}(\bm n,\pi_{\tau})$ denotes its change in time while the CRN dwells in the state $\bm n$ (it need not be an exact time derivative), and $\delta \mathcal{X}_{\rho} (\bm n, \pi_{\tau})$ denotes its finite change along the reaction $\rho$ occurring while in $\bm n$.
By contrast, the changes of state observables $\mathcal{O}(\bm n,t)$ can be written as
\begin{multline}
	\Delta\mathcal{O}\trjdep = \mathcal{O}(\bm n_{t},t) - \mathcal{O}(\bm n_{0},0) \\
	\hspace{-2.9em}
	= \int_{0}^{t} \de \tau \Big\{ \at{\left[ \partial_{\tau} {\mathcal{O}}(\bm n,\tau) \right]}{\bm n_{\tau}}
	+ \sum_{\bm n, \rho} \Delta_{\rho} \mathcal{O}(\bm n,\tau) \, j_{\rho}(\bm n,\tau) \Big\} \, .
	\label{eq:DeltaO}
\end{multline}
where $\partial_{\tau} {\mathcal{O}}(\bm n,\tau)$ is the time derivative of ${\mathcal{O}}(\bm n,\tau)$, and
\begin{equation}
	\Delta_{\rho} \mathcal{O}(\bm n,\tau) := \mathcal{O}(\bm n + \stoich_{\rho},\tau) - \mathcal{O}(\bm n,\tau) \, ,
	\label{eq:DeltaRhoO}
\end{equation}
is the difference of ${\mathcal{O}}(\bm n,\tau)$ along reactions, see Fig.~\ref{fig:trj}.

\subsection{Conservation Laws}
\label{sec:cl}

The topological properties of CRNs are encoded in the matrices $\stoichM$ and $\stoichM^{\Y}$ and can be identified via their cokernels and kernels.
\emph{Conservation laws} $\bm{\ell}$ are defined as vectors in $\coker \stoichM$,
\begin{equation}
	\bm \ell \cdot \stoich_{\rho} = 0 \, , \quad \text{for all} \, \rho \, .
	\label{eq:cl}
\end{equation}
They identify conserved quantities, called \emph{components} \cite{alberty03}
\begin{equation}
	L_{\bm n} := \bm \ell \cdot \bm {n} \, .
	\label{eq:component}
\end{equation}
Despite the fact that $L_{\bm n}$ depends on the stochastic variable $\bm n$, the probability of observing any specific value $L$,
\begin{equation}
	P(L) := {\textstyle\sum_{\bm {n}}^{}} p_{\bm {n}} \, \delta[L_{\bm n},L] \, ,
	\label{eq:probabilityComponent}
\end{equation}
is constant over time, \emph{i.e.} $\dt P(L) = 0$.
$\delta$ is a Kronecker delta.
More generally, any observable of type $\mathcal{O}(L_{\bm n})$ \emph{does not fluctuate},
\begin{equation}
	\dt {\textstyle\sum_{\bm {n}}} p_{\bm n} \, \mathcal{O}(L_{\bm n}) = 0 \, ,
	\label{eq:invariance}
\end{equation}
as a direct consequence of the fact that $\Delta_{\rho} \mathcal{O}(L_{\bm n}) = 0$.
Clearly, $P(L)$ can be deduced from the initial conditions $p_{\bm n}(0)$ and only those states for which $P(L_{\bm n},0)$ is nonvanishing have a finite probability of being observed during the subsequent stochastic dynamics.

In closed CRNs, conservation laws \eqref{eq:cl} follow from
\begin{equation}
	\bm \ell^{\x} \cdot \stoich^{\x}_{\ir} + \bm \ell^\y \cdot \stoich^\y_{\ir} = 0 \, ,
	\quad \text{for all} \, \ir \, .
	\label{eq:clClosed}
\end{equation}
We denote a set of linearly independent conservation laws of the closed CRN by $\left\{ \bm{\ell}_{\lambda} \right\}$, and the corresponding components by $\left\{ L^{\lambda}_{\bm n} := \bm \ell_{\lambda} \cdot \bm {n} \right\}$, for $\lambda = 1, \dots, \nof{\lambda} := \dim \coker \mathbb{S}_{\mathrm{i}}$.
The choice of this set is not unique, and different choices have different physical meanings.
This set is never empty since the total mass is always conserved.
The latter corresponds to a $\bm \ell$ whose entries are the masses of each species.
Physically, the conservation laws of closed CRNs can always be chosen so as to correspond to \emph{moieties}, which are parts of molecules exchanged between species along reactions or subject to isomerization \cite{haraldsdottir16}.

For open CRNs, the condition identifying conservation laws, Eq.~\eqref{eq:cl}, becomes
\begin{subequations}
		\begin{align}
			\bm \ell^{\x} \cdot \stoich^{\x}_{\ir} + \bm \ell^\y \cdot \stoich^\y_{\ir} & = 0 \, , \quad \text{for all} \, \ir \, , \label{eq:clOpenI} \\
			\bm \ell^\y \cdot \stoich^\y_{\er} & = 0 \, , \quad \text{for all} \, \er \label{eq:clOpenE} \, .
		\end{align}
		\label{eq:clOpen}
\end{subequations}
We now recall that for all $\er$ there is one and only one exchanged species for which the corresponding entry of $\stoich^\y_{\er}$ is different from zero.
Hence, Eq.~\eqref{eq:clOpenE} demands that $\bm \ell^\y = \bm 0$ and Eq.~\eqref{eq:clOpen} become $\bm \ell^{\x} \cdot \stoich^{\x}_{\ir} = 0$ for all $\ir$.

Crucially, any set of independent conservation laws of the open CRN, Eq.~\eqref{eq:clOpen}, denoted by $\st{\bm \ell_{\lambda_{\mathrm{u}}}}$, for $\lambda_{\mathrm{u}} = 1, \dots, \nof{\lambda_{\mathrm{u}}} := \dim \coker \mathbb{S} < \nof{\lambda}$, can be regarded as a subset of the conservation laws of the closed CRN, $\st{\bm \ell_{\lambda}} \equiv \st{\bm \ell_{\lambda_{\mathrm{u}}}} \cup \st{\bm \ell_{\lambda_{\mathrm{b}}}}$, since they satisfy Eq.~\eqref{eq:clClosed}, too.
In view of this, we call them \emph{unbroken conservation laws}.
The remaining independent conservation laws, labeled as $\st{\bm \ell_{\lambda_{\mathrm{b}}}}$ and referred to as \emph{broken}, satisfy Eq.~\eqref{eq:clClosed} while not Eq.~\eqref{eq:clOpen}. 
They involve exchanged species, $\bm \ell_{\lambda_{\mathrm{b}}}^\y \neq \bm 0$, hence $\bm \ell^\y_{\lambda_{\mathrm{b}}} \cdot \stoich^\y_{\er} \neq \bm 0$ and the probability distribution of any set $\{L^{\lambda_{\mathrm{b}}}_{\bm n} \equiv \bm \ell_{\lambda_{\mathrm{b}}} \cdot \bm {n}\}$,
\begin{equation}
	P(\st{L_{\lambda_{\mathrm{b}}}}) := {\textstyle\sum_{\bm {n}}} p_{\bm n} {\textstyle\prod_{\lambda_{\mathrm{b}}}} \delta\big[L^{\lambda_{\mathrm{b}}}_{\bm n},L_{\lambda_{\mathrm{b}}}\big] \, ,
	\label{eq:componentBroken}
\end{equation}
changes in time.

Summarizing, in open CRNs, the chemostatting breaks a subset of the conservation laws of the corresponding closed CRN, $\{\bm\ell_{\lambda_{\mathrm{b}}}\}$.
Only the probability distribution of the \emph{unbroken components} $\{L^{\lambda_{\mathrm{u}}}_{\bm n} \equiv \bm \ell_{\lambda_{\mathrm{u}}} \cdot \bm {n}\}$,
\begin{equation}
	P(\st{L_{\lambda_{\mathrm{u}}}}) := {\textstyle\sum_{\bm {n}}} p_{\bm n} {\textstyle\prod_{\lambda_{\mathrm{u}}}} \delta\big[L^{\lambda_{\mathrm{u}}}_{\bm n},L_{\lambda_{\mathrm{u}}}\big] \, ,
	\label{}
\end{equation}
is invariant and completely determined by the initial probability distribution $p_{\bm {n}}(0)$.
The state space identified by one particular set of values for $\st{L_{\lambda_{\mathrm{u}}}}$ is called \emph{stoichiometric compatibility class}.

\begin{example}
	\label{ex:cl}
	The CRN in Fig.~\ref{fig:openMM} has two conservation laws,
	\begin{subequations}
		\begin{align}
			\bm \ell_{\ce{E}} & =
			\kbordermatrix{
				& \greyt{\ce{E}} & \greyt{\ce{E^{\ast}}} & \greyt{\ce{E^{\ast\ast}}} & \greyt{\ce{A}} & \greyt{\ce{B}} \\
				& 1 & 1 & 1 & 0 & 0
			} \, , \label{eq:clMMu} \\
			\bm \ell_{\mathrm{b}} & =
			\kbordermatrix{
				& \greyt{\ce{E}} & \greyt{\ce{E^{\ast}}} & \greyt{\ce{E^{\ast\ast}}} & \greyt{\ce{A}} & \greyt{\ce{B}} \\
				& 0 & 1 & 1 & 1 & 1
			} \, , \label{eq:clMMb}
		\end{align}
		\label{eq:clMM}
	\end{subequations}
	among which the second is broken.
	The unbroken conservation law identifies the enzyme moiety and corresponds to the total number of enzyme molecules populating the CRN, $L^{\ce{E}}_{\bm {n}} = n_{\ce{E}} + n_{\ce{E^{\ast}}} + n_{\ce{E^{\ast\ast}}}$.
	Instead, the broken one identifies the moiety $\ce{A}$---or equivalently $\ce{B}$---, $L^{\mathrm{b}}_{\bm n} =  n_{\ce{E^{\ast}}} + n_{\ce{E^{\ast\ast}}} + n_{\ce{A}} + n_{\ce{B}}$.
	\qed
\end{example}

\subsection{Stoichiometric Cycles}
\label{sec:cycles}

We can now set the stage for the thermodynamic description based on a stoichiometric cycle decomposition.
This section, as well as the other ones discussing cycles, may be omitted at a first reading.

Additional information about the CRN topology is provided by the \emph{stoichiometric cycles} $\bm c = \st{c_{\rho}}$ as they are vectors in $\ker \stoichM$.
Equivalently, these satisfy
\begin{equation}
	{\textstyle\sum_{\rho}} \stoich_{\rho} c_{\rho} = \bm 0 \, ,
	\label{eq:cycle}
\end{equation}
and at most one entry for each forward--backward transition pair is nonzero.
Since $\stoichM$ is integer-valued, any $\mathbf{{c}}$ can always be chosen non-negative-integer-valued.
In this way, its entries denote the number of times each transition occurs along a transformation which overall leaves the state $\bm n$ unchanged.
Alternatively, a stoichiometric cycle can be seen as a set of reactions $\st{\rho_{\mathbf{c}1}, \rho_{\mathbf{c}2}, \dots, \rho_{\mathbf{c}\nof{\mathrm{c}}}}$ identifying a closed loop in the state space
\begin{equation}
	\bm {n} \rightarrow \bm {n} + \stoich_{\rho_{\mathbf{c}1}} \rightarrow \dots \rightarrow \bm {n} + \sum_{i=1}^{\nof{\mathbf{c}}} \stoich_{\rho_{\mathbf{c}i}} = \bm {n} \, ,
	\label{eq:cycleArrow}
\end{equation}
where $\sum_{i=1}^{\nof{\mathbf{c}}} \stoich_{\rho_{\mathbf{c}i}} = \sum_{\rho} \stoich_{\rho} c_{\rho} = 0$.

We now relate cycles of the closed and open CRNs as previously done for conservation laws.
In the closed CRN, the stoichiometric cycles are given by
\begin{subequations}
	\begin{align}
		\textstyle \sum_{\ir} \stoich^{\x}_{\ir} c_{\ir} & = \bm 0 \\
		\textstyle \sum_{\ir} \stoich^\y_{\ir} c_{\ir} & = \bm 0 \, .
	\end{align}
	\label{eq:cycleClosed}
\end{subequations}
The entries corresponding to the exchange reactions are taken equal to $0$: $c_{\er} = 0$, for all $\er$.
Let us denote by $\st{\bm {c}^{\cy}}$, for $\cy = 1, \dots, \nof{\cy} := \dim \ker \mathbb{S}_{\mathrm{i}}$, a set of independent stoichiometric cycles of the closed CRN.

In the open CRN, the condition identifying cycles, Eq.~\eqref{eq:cycle}, reads
\begin{subequations}
	\begin{align}
		\textstyle \sum_{\ir} \stoich^{\x}_{\ir} c_{\ir} & = \bm 0 \\
		\textstyle \sum_{\ir} \stoich^\y_{\ir} c_{\ir} + \sum_{\er} \stoich^\y_{\er} c_{\er} & = \bm 0 \, .
	\end{align}
	\label{eq:cycleOpen}
\end{subequations}
Since the cycles of the closed CRN satisfy Eq.~\eqref{eq:cycleOpen}, they can be regarded as a subset of an independent set of cycles for the open CRN, $\st{\bm c^{\cy}, \bm c^{\ecy}}$.
We refer to the additional cycles $\st{\bm c^{\ecy}}$, for $\ecy = 1, \dots, \nof{\ecy} := \dim \ker \mathbb{S} - \dim \ker \mathbb{S}_{\mathrm{i}}$, as \emph{emergent}.
They are characterized by at least one nonzero entry for $\st{\er}$, and the vectors
\begin{equation}
	\bm C^\Y_{\ecy} := {\textstyle\sum_{\rho}} \big( - \stoich^\Y_{\rho} \big) c^{\ecy}_{\rho} = {\textstyle\sum_{\er}} \stoich^\y_{\er} c^{\ecy}_{\er} \neq \bm 0
	\label{eq:C}
\end{equation}
quantify the amount of exchanged species flowing in the system from the corresponding chemostats upon completion of $\bm c^{\ecy}$.
As the concentrations of the chemostats are unaffected by the exchange of particles with the system, the emergent stoichiometric cycles can be thought of as pathways transferring chemicals across chemostats while leaving the internal state of the CRN unchanged.

As first proved in Ref.~\cite{polettini14}, by applying the rank--nullity theorem to the stoichiometric matrices of the open and closed CRNs, one can show that
\begin{equation}
	\nof{\mathrm{y}} = \nof{\lambda_{\mathrm{b}}} + \nof{\ecy} \, .
	\label{eq:rankNullity}
\end{equation}
In words, for any exchanged species either a conservation law is broken, or an emergent cycle is created.

\begin{example}
	\label{ex:cycles}
	The CRN in Fig.~\ref{fig:openMM} has one cycle
	\begin{equation}
		\bm c_{\mathrm{int}} =
		\kbordermatrix{
			& \greyt{+1} & \greyt{+2} & \greyt{+3} & \greyt{+4} & \greyt{+a} & \greyt{+b} \\
			& 1 & 1 & 1 & 1 & 0 & 0
		} \, ,
		\label{eq:cycleMMint}
	\end{equation}
	and one emergent cycle
	\begin{equation}
		\bm c_{\mathrm{ext}} =
		\kbordermatrix{
			& \greyt{+1} & \greyt{+2} & \greyt{+3} & \greyt{+4} & \greyt{+a} & \greyt{+b} \\
			& 1 & 1 & 0 & 0 & 1 & -1
		} \, .
		\label{eq:cycleMM}
	\end{equation}
	Negative entries must be interpreted as reactions occurring in the backward direction.
	The latter cycle corresponds to the injection of one molecule of $\ce{A}$, its conversion into one of $\ce{B}$ passing via $\ce{E^{\ast}}$, and its ejection,
	\begin{equation}
		\bm C_{\mathrm{ext}} = \kbordermatrix{
			& \greyt{\ce{A}} & \greyt{\ce{B}} \\
			& 1 & -1
		} \, .
		\label{eq:CMM}
	\end{equation}
	
	We can also check the validity of Eq.~\eqref{eq:rankNullity}, as the number of chemostats, $2$, equals the number of broken conservation laws, $1$, see Ex.~\ref{ex:cl}, plus the number of emergent cycles, $1$, Eq.~\eqref{eq:cycleMM}.
	\qed
\end{example}

\paragraph*{Remark}
Stoichiometric cycles must be distinguished from graph-theoretic cycles, also called \emph{loops} see \emph{e.g.} Ref.~\cite{schnakenberg76}.
To elucidate this point, we note that the network of transitions of a CRN can be regarded as a semi-infinite graph whose vertices are the accessible states $\bm n$, and whose directed edges are given by the reactions---which are encoded in the stoichiometric matrix, $\stoichM$.
Hence, one can see that loops are recursive appearance of stoichiometric cycles, as in Eq.~\eqref{eq:cycleArrow}.
However, they may not be complete at the boundaries of the graph (low $\bm n$) due to peculiar topological properties of the CRN, see \emph{e.g.} Ref.~\cite{polettini15}.
These observations will be used later to relate different approaches for cycle decomposition of thermodynamic quantities.

\section{Stochastic Thermodynamics}
\label{sec:ST}

We now build a nonequilibrium thermodynamic description on top of the stochastic dynamics.
We assume that the solvent acts as a thermal reservoir by keeping the temperature, $T$, and the pressure constant everywhere.
Since particle numbers are low, we can assume that that the time scale in which molecules spatially homogenize is much faster than that of reactions.
Therefore, if all reactions could be instantaneously shut down, we would observe an equilibrium mixture of inert species at all times.
However, due to reactions, the populations of species and their probability distribution can be far from equilibrium.
These hypotheses can be regarded as a special case of \emph{local equilibrium} \cite{*[][{, \S~15.1.}] {kondepudi14},*prigogine49}, since temperature, pressure, and density are not only locally well defined, but also constant.

\subsection{Equilibrium of Closed CRNs}
\label{sec:CCRN}

Equilibrium statistical mechanics requires that the equilibrium distribution of a closed CRN with given values of $\st{L_{\lambda}}$ reads
\begin{equation}
	p^{\mathrm{eq}}(\bm {n} | \st{L_{\lambda}}) = \frac{\exp\left\{ -\beta g_{\bm n} \right\}}{{Z}(\st{L_{\lambda}})}
	{\textstyle\prod_{\lambda}} \delta\big[ L^{\lambda}_{\bm n},L_{\lambda} \big] \, ,
	\label{eq:peq(n|L)}
\end{equation}
where
\begin{equation}
	g_{\bm n} = \left( \bm \mu^\circ - \bm 1 \kt \ln n_{\mathrm{s}} \right) \cdot \bm {n} + \kt \ln \bm n!
	\label{eq:g}
\end{equation}
is the \emph{Gibbs free energy} of the state $\bm n$ derived in App.~\ref{sec:potentials}.
The first term quantifies the energetic contribution of each single molecule:
$\bm \mu^\circ \equiv \bm \mu^\circ (T)$ is the vector of \emph{standard-state chemical potentials}, whereas $- \bm 1 \kt \ln n_{\mathrm{s}}$ is an entropic contribution---constant for all species---since $n_{\mathrm{s}}$ is the population of the solvent.
The last term is purely entropic and accounts for the indistinguishability of molecules of the same species.
In Eq.~\eqref{eq:peq(n|L)},
\begin{equation}
	Z\left( \st{L_{\lambda}} \right)
	= {\textstyle \sum_{\bm {m}}} \exp\left\{ -\beta g_{\bm m} \right\} \,
	{\textstyle\prod_{\lambda}} \delta\big[ L^{\lambda}_{\bm m},L_{\lambda} \big]
	\label{eq:Zclosed}
\end{equation}
is the partition function, while $\beta = 1/(\kt)$.
When taking into account an ensemble of components, $P(\st{L_{\lambda}})$, Eq.~\eqref{eq:peq(n|L)} allows us to write
\begin{equation}
	\begin{split}
		p^{\mathrm{eq}}_{\bm {n}} & =
		{\textstyle\sum_{\st{L_{\lambda}}}} \, p^{\mathrm{eq}}(\bm {n} | \st{L_{\lambda}}) \, P(\st{L_{\lambda}}) \\
		& = p^{\mathrm{eq}}(\bm {n} | \st{L^{\lambda}_{\bm n}}) \, P(\st{L^{\lambda}_{\bm n}}) \, ,
	\end{split}
	\label{eq:peq(n)}
\end{equation}
which can be regarded as a \emph{constrained} equilibrium distribution.
Hence, $p^{\mathrm{eq}}(\bm {n} | \st{L^{\lambda}_{\bm n}})$ is the conditional probability of observing $\bm n$ given the stoichiometric compatibility class it identifies.

Equation~\eqref{eq:peq(n)} can also be written as
\begin{equation}
	p^{\mathrm{eq}}_{\bm {n}} =
	\exp\left\{ - \beta \big[ g_{\bm n} - G_{\mathrm{eq}}(\st{L^{\lambda}_{\bm n}}) \big] \right\} \, ,
	\label{}
\end{equation}
in terms of the \emph{equilibrium Gibbs potential} of the CRN
\begin{equation}
	G_{\mathrm{eq}}(\st{L_{\lambda}}) = \kt \ln {P(\st{L_{\lambda}})} - \kt \ln Z(\st{L_{\lambda}}) \, .
	\label{eq:Geq}
\end{equation}
It is worth emphasizing that $G_{\mathrm{eq}}(\st{L_{\lambda}})$ is function solely of the set of components, and that $G_{\mathrm{eq}}(\st{L^{\lambda}_{\bm n}})$ needs to be understood as $G_{\mathrm{eq}}$ evaluated in $\st{L^{\lambda}_{\bm n}}$.
Invoking the hypothesis of local equilibrium, we extend $G_{\mathrm{eq}}$ to arbitrary probability distributions $p_{\bm {n}}$,
\begin{equation}
	G(\bm {n}) := \kt \ln p_{\bm n} + g_{\bm n} \, ,
	\label{eq:G}
\end{equation}
and we call it \emph{stochastic Gibbs potential}, as it is the far-from-equilibrium fluctuating expression of $G_{\mathrm{eq}}$.
In addition to the Gibbs free energy of the state $\bm n$, $g_{\bm n}$, it accounts for the entropic contribution due to the uncertainty of $p_{\bm n}$:
$\kt \ln p_{\bm n}$ can indeed be written as $- T (- \kb \ln p_{\bm n})$, where the term in parentheses is the \emph{self-information} measured in $\kb$ units \cite{cover06}.
For closed CRNs at equilibrium, using Eq.~\eqref{eq:peq(n)}, $G(\bm n)$ reduces to $G_{\mathrm{eq}}$ in Eq.~\eqref{eq:Geq}.
Also, its average value, the \emph{nonequilibrium Gibbs potential}
\begin{equation}
	\avef{G} = {\textstyle\sum_{\bm n}} p_{\bm n} \left[ \kt \ln p_{\bm {n}} + g_{\bm n} \right] \, ,
	\label{eq:aveG}
\end{equation}
takes its minimum value at equilibrium
\begin{equation}
	\begin{split}
		\ave{G} - \avef{G_{\mathrm{eq}}}_{\mathrm{L}} & = \ave{G - G_{\mathrm{eq}}} \\
		& = \kt {\textstyle\sum_{\bm {n}}} p_{\bm {n}} \ln \frac{p_{\bm {n}}}{p^{\mathrm{eq}}_{\bm {n}}} \\
		& \equiv \kt \, \mathcal{D}(p\|p^{\mathrm{eq}}) \ge 0 \, .
	\end{split}
	\label{eq:delta<G>}
\end{equation}
In the first equality, we used the fact that the equilibrium Gibbs potential only depends on the components,
\begin{equation}
	\begin{split}
		& \avef{G_{\mathrm{eq}}}_{\mathrm{L}} \equiv {\textstyle\sum_{\st{L_{\lambda}}}} P(\st{L_{\lambda}}) G_{\mathrm{eq}}(\st{L_{\lambda}}) \\ 
		& = {\textstyle\sum_{\st{L_{\lambda}}}} \left[ {\textstyle\sum_{\bm n}} p_{\bm n} {\textstyle\prod_{\lambda}} \delta\big[ L^{\lambda}_{\bm n}, L_{\lambda} \big] \right] G_{\mathrm{eq}}(\st{L_{\lambda}}) \\
		& = {\textstyle\sum_{\bm {n}}} p_{\bm {n}} G_{\mathrm{eq}}(\st{L^{\lambda}_{\bm n}}) \, .
	\end{split}
	\label{}
\end{equation}
In the last equality of Eq.~\eqref{eq:delta<G>}, $\mathcal{D}(p\|p^{\mathrm{eq}})$ is the relative entropy of the transient probability distribution $p_{\bm n}$ with respect to the equilibrium one $p^{\mathrm{eq}}_{\bm n}$.
It is always positive and vanishes only when $p_{\bm n} = p^{\mathrm{eq}}_{\bm {n}}$.
We will see later (\S~\ref{sec:ead}) that Eq.~\eqref{eq:delta<G>} quantifies exactly the average dissipation of the relaxation to equilibrium.

\subsection{Local Detailed Balance}
\label{sec:ldb}

The \emph{zero-th law of thermodynamics for CRNs} requires that closed CRNs relax to equilibrium.
To ensure this, the dynamical requirement for detailed balance, Eq.~\eqref{eq:dbp}, is combined with the equilibrium distribution, Eq.~\eqref{eq:peq(n)}.
As a result, the \emph{local detailed balance} ensues
\begin{equation}
	\ln \frac{w_{\ir} (\bm n)}{w_{-\ir} (\bm n + \stoich_{\ir})} = - \beta \Delta_{\ir} g_{\bm n} \, ,
	\label{eq:ldbI}
\end{equation}
where $\Delta_{\ir} \cdot$ is defined as in Eq.~\eqref{eq:DeltaRhoO}.
In agreement with deterministic descriptions, see \emph{e.g.} Ref.~\cite{rao16:crnThermo}, we recover the relation between the rate constants and the standard-state chemical potentials
\begin{equation}
	\ln \frac{k_{\ir}}{k_{-\ir}} = - \beta \, \left( \bm \mu^\circ - \kt \bm 1 \ln [\mathrm{s}] \right) \cdot \stoich_{\ir} \, ,
	\label{eq:lnk}
\end{equation}
in which $[\mathrm{s}]:=n_{\mathrm{s}}/V$ denotes the concentration of solvent.
The local detailed balance \eqref{eq:ldbI} should be regarded as a fundamental property of the stochastic reaction rates of elementary reactions valid beyond closed CRNs.
This central concept is well known in stochastic thermodynamics because it provides the connection between stochastic dynamics and nonequilibrium thermodynamics.

In open CRNs,
\begin{equation}
	\ln \frac{\rate{\rho}{\bm n}}{\rate{-\rho}{\bm n + \stoich_{\rho}}}
	= - \beta \big( \Delta_{\rho} g_{\bm n} + \bm \mu_\Y \cdot \stoich^\Y_{\rho} \big)
	\label{eq:ldb}
\end{equation}
generalizes Eq.~\eqref{eq:ldbI}, where
\begin{equation}
	\bm \mu_{\Y} = \bm \mu_{\Y}^\circ + \kt \ln \left\{ \concY / [\mathrm{s}] \right\}
	\label{eq:muYe}
\end{equation}
are the chemical potentials of the chemostats.
The first contribution accounts for the Gibbs free energy change of the internal species, while the second one for the Gibbs free energy exchanged with the chemostats.

We introduce the \emph{transition affinities} which quantify the force acting along each transition
\begin{equation}
	A_{\rho}(\bm {n}) =
	\kt \ln \frac{\rate{\rho}{\bm n} p_{\bm {n}}}{\rate{-\rho}{\bm n + \stoich_{\rho}} p_{\bm n + \stoich_{\rho}}} \, .
	\label{eq:affinityDynamic}
\end{equation}
They measure the distance from detailed balance \eqref{eq:dbp}, where they all vanish.
Using Eq.~\eqref{eq:ldb}, they can be rewritten in terms of differences of stochastic Gibbs potential \eqref{eq:G},
\begin{equation}
	A_{\rho}(\bm n)
	= - \Delta_{\rho} G(\bm {n}) + \bm \mu_\Y \cdot \big(- \stoich^\Y_{\rho} \big) \, .
	\label{eq:affinity}
\end{equation}
This fundamental relation reveals the thermodynamic nature of the dynamical forces acting along reaction.
Its early formulation for deterministic chemical kinetics is due to de~Donder \cite{donder27}.

We will prove in \S~\ref{sec:ead} that our theoretical framework based on Eq.~\eqref{eq:ldb} guarantees that closed CRNs described by a CME \eqref{eq:CME} relax to equilibrium, Eq.~\eqref{eq:peq(n)}:
the average potential $\avef{G}$ is minimized by the dynamics during the relaxation and hence plays the role of a Lyapunov function.

\subsection{Enthalpy and Entropy Balance}
\label{sec:balances}

Starting from the stochastic Gibbs potential \eqref{eq:G} and the local detailed balance \eqref{eq:ldb}, we now formulate the energy and entropy balance along stochastic trajectories.

The \emph{stochastic entropy} of the CRNs follows from the derivative of the stochastic Gibbs potential \eqref{eq:G} with respect to the temperature,
\begin{equation}
	S(\bm {n}) = - \left( \derpart{G}{T} \right)_{\bm {n}} = - \kb \ln p_{\bm n} + s_{\bm n} \, .
	\label{eq:S}
\end{equation}
Similar to $G(\bm n)$, $S(\bm {n})$ is the far-from-equilibrium fluctuating expression of the entropy \cite{seifert05}.
The first term on the rhs is the self-information, while the second is the entropy of the state $\bm n$,
\begin{equation}
	\hspace{-1em}
	s_{\bm n} = - \derpart{g_{\bm{n}}}{T} = \left( \bm {s}^\circ + \kb \ln n_{\mathrm{s}} \right) \cdot \bm {n} - \kb \ln {\bm {n}!} \, .
	\label{eq:s}
\end{equation}
It accounts for both the entropic contribution carried by each species, \emph{i.e.} the \emph{standard entropies of formation}
\begin{equation}
	\bm {s}^\circ = - \derpart{\bm \mu^\circ}{T} \, ,
	\label{eq:s0}
\end{equation}
as well as the entropic contribution due to the multiplicity of indistinguishable states.
When averaged, we recover the \emph{Gibbs--Shannon entropy} plus an internal entropy contribution,
\begin{equation}
	\avef{S} = {\textstyle\sum_{\bm n}} p_{\bm n} \left[ - \kb \ln p_{\bm {n}} + s_{\bm n} \right] \, .
	\label{eq:gibbsShannon}
\end{equation}
The enthalpy follows from
\begin{equation}
	H(\bm n) = G(\bm n) + TS(\bm n) = g_{\bm n} + T s_{\bm n} = \bm h \cdot \bm n \, ,
	\label{eq:H}
\end{equation}
where
\begin{equation}
	\bm h = \bm \mu^\circ + T \bm s^\circ = \bm h^\circ
	\label{eq:enthalpyOfFormation}
\end{equation}
denotes the vector of \emph{standard enthalpies of formation}, in agreement with traditional thermodynamics of ideal dilute solutions \cite{alberty03}.
Likewise, the chemical potentials of the chemostats, Eq.~\eqref{eq:muYe}, will be decomposed in terms of enthalpic and entropic contributions,
\begin{equation}
	\bm \mu_{\mathrm{Y}} = \bm {h}_{\mathrm{Y}} - T \bm {s}_{\mathrm{Y}} \, ,
	\label{eq:decMuY}
\end{equation}
where $\bm h_\Y = \bm h^\circ_\Y$ and $\bm {s}_{\mathrm{Y}} = \bm {s}_{\mathrm{Y}}^\circ - \kb \ln \left\{ \concY / [\mathrm{s}] \right\}$.

To recover the enthalpy balance along stochastic trajectories, we write the change of enthalpy as the sum of its changes due to reactions,
\begin{equation}
	\begin{split}
		\Delta H [\trj] & = H(\bm {n}_{t}) - H(\bm {n}_{0}) \\
		& = \int_{0}^{t} \de \tau \, \sum_{\bm n, \rho} \Delta_{\rho} H(\bm {n}) \, j_{\rho}(\bm n, \tau) \, ,
	\end{split}
	\label{eq:deltaHpreliminary}
\end{equation}
where
\begin{widetext}
	\begin{equation}
		\Delta_{\rho} H(\bm {n}) = \bm h \cdot \stoich_{\rho}
		= \underbrace{\underbrace{\bm h \cdot \stoich_{\rho} + \bm h_\Y \cdot \stoich^\Y_{\rho}}_{\textstyle =: Q^{\mathrm{thr}}_{\rho}} + \underbrace{T \bm {s}_\Y \cdot \big( - \stoich^\Y_{\rho} \big)}_{\textstyle =: Q^{\mathrm{chm}}_{\rho}} }_{\textstyle =: Q_{\rho} } + \underbrace{\bm \mu_\Y \cdot \big( - \stoich^\Y_{\rho} \big)}_{\textstyle =: W^{\mathrm{c}}_{\rho}} \, , \quad \text{for all} \, \bm n \, .
		\label{eq:deltaRhoH}
	\end{equation}
\end{widetext}
We used Eqs.~\eqref{eq:DeltaO}, \eqref{eq:enthalpyOfFormation} and \eqref{eq:decMuY}.
The first two contributions, $Q^{\mathrm{thr}}_{\rho}$, account for the \emph{heat of reaction}, \emph{i.e.} the heat flowing from the thermal reservoir (the solvent).
The third term characterizes the heat flowing from the chemostats, $Q^{\mathrm{chm}}_{\rho}$.
The first three terms, $Q_{\rho}$, integrated along the trajectory quantify the total \emph{heat flow} 
\begin{equation}
	\hspace{-1.6em}
	Q\trjdep = \int_{0}^{t} \de \tau \left\{{\textstyle\sum_{\rho}} Q^{\mathrm{thr}}_{\rho} J_{\rho}(\tau) + T \bm s_\Y (\tau) \cdot \bm I^\Y(\tau) \right\} \, ,
	\label{eq:Q}
\end{equation}
where the \emph{instantaneous external currents}
\begin{equation}
	\bm I^\Y(\tau) := {\textstyle\sum_{\rho}} \big( - \stoich^\Y_{\rho} \big) J_{\rho}(\tau)
	\label{eq:instChemoCurr}
\end{equation}
give the amount of exchanged species injected in the CRN at each time, see Eq.~\eqref{eq:instTrnsCrnt}.

The last term in Eq.~\eqref{eq:deltaRhoH}, $W^{\mathrm{c}}_{\rho}$, quantifies the Gibbs free energy exchanged with the chemostats.
Once integrated, it gives the \emph{chemical work}
\begin{equation}
	W_{\mathrm{c}}\trjdep
	= \int_{0}^{t} \de \tau \, \bm \mu_\Y(\tau) \cdot \bm I^\Y(\tau) \, .
	\label{eq:cw}
\end{equation}
From Eqs.~\eqref{eq:deltaHpreliminary}--\eqref{eq:cw}, the \emph{enthalpy balance} along a trajectory follows
\begin{equation}
	\Delta H[\trj] = Q\trjdep + W_{\mathrm{c}}\trjdep \, .
	\label{eq:firstLaw}
\end{equation}
This is the expression of the first law of thermodynamics for stochastic CRNs at the trajectory level, \emph{cf.} \cite[Eq.~2.10]{callen85}.

\begin{widetext}
To recover the entropy balance along stochastic trajectories, we notice that since the entropy is a state function, its change along a trajectory reads
\begin{equation}
	\Delta S[\trj]
	= \int_{0}^{t} \de \tau \Big\{ \at{\left[ - \partial_{\tau} \kb \ln p_{\bm n}(\tau) \right]}{\bm n_{\tau}}
	+ \sum_{\bm n, \rho} \Delta_{\rho} S (\bm n) \, j_{\rho}(\bm n, \tau) \Big\} \, ,
	\label{}
\end{equation}
as seen in Eq.~\eqref{eq:DeltaO}.
The changes along transitions can be recast into
	\begin{equation}
		\begin{split}
			T \Delta_{\rho} S (\bm n)
			& = T \Delta_{\rho} s_{\bm n} - \kt \ln \frac{p_{\bm n + \stoich_{\rho}}}{p_{\bm n}} \\
			& = \underbrace{\bm h \cdot \stoich_{\rho} + \bm h_\Y \cdot \stoich^\Y_{\rho} + T \bm s_\Y \cdot \big( - \stoich^\Y_{\rho} \big)}_{\textstyle = Q_{\rho}}
			\underbrace{- \underbrace{\left[ \Delta_{\rho} g_{\bm n} + \kt \ln \frac{p_{\bm n + \stoich_{\rho}}}{p_{\bm n}} \right]}_{\textstyle = \Delta_{\rho} G(\bm n)} + \underbrace{\bm \mu_\Y \cdot \big( - \stoich^\Y_{\rho} \big)}_{\textstyle = W^{\mathrm{c}}_{\rho}}}_{\textstyle = A_{\rho}(\bm n)} \, ,
		\end{split}
		\label{}
	\end{equation}
	where we have used Eq.~\eqref{eq:H}.
	As highlighted with underbraces, the first three terms are the heat flow along reactions, while the last three correspond to the affinity of transition, Eq.~\eqref{eq:affinity}.
	When integrating over the whole trajectory, we recover the entropy balance
	\begin{equation}
		\Delta S[\trj] = \tfrac{1}{T} Q\trjdep + \Sigma\trjdep \, ,
		\label{eq:epDef}
	\end{equation}
	where the EP (times the temperature) reads
	\begin{subequations}
		\begin{align}
			T \Sigma\trjdep
			& = \int_{0}^{t} \de \tau \Big\{ \at{\left[ - \partial_{\tau} \kt \ln p_{\bm n}(\tau) \right]}{\bm n_{\tau}} + \sum_{\bm n, \rho} A_{\rho}(\bm n, \tau) \, j_{\rho}(\bm n, \tau) \Big\} \label{eq:epAffinityTrue} \\
			& = \kt \ln \frac{p_{\bm {n}_{0}}(0)}{p_{\bm {n}_{t}}(t)} + \int_{0}^{t} \de \tau \, j_{\rho}(\bm n, \tau) \, \kt \ln \frac{\rate{\rho}{\bm n, \tau}}{\rate{-\rho}{\bm n+\stoich_{\rho}, \tau}} \label{eq:epAffinity} \\
			& = W_{\mathrm{c}}\trjdep - \Delta G[\trj] \, . \label{eq:ep=W-deltaG}
		\end{align}
		\label{eq:eps}
	\end{subequations}
\end{widetext}
The second equality follows from the definition of affinity, Eq.~\eqref{eq:affinityDynamic}, when integrating the changes of the probability distribution.
Instead, the third one readily follows from the relationship between affinity and Gibbs potential, Eq.~\eqref{eq:affinity}.
It expresses the overall energy dissipated as the difference between the Gibbs free energy supplied by the chemostats and that changing internally.

Mindful of Eq.~\eqref{eq:probTrajectory}, the EP can be rewritten as the ratio of the probability of observing the trajectory $\trj$ under a forward dynamics driven by a protocol $\pi_{t}$, over the probability of observing the backward trajectory $\trj^{\dagger}$ under a dynamics driven by the time-reversed protocol $\pi^{\dagger}$ such that $\pi^{\dagger}_{\tau} := \pi_{t - \tau}$:
\begin{equation}
	T \Sigma\trjdep = 
	\kt \ln \frac{p_{\bm n_{0}}(0) \, \mathcal{P}[\trj; \pi]}
	{p_{\bm n_{t}}(t) \, \mathcal{P}[\trj^{\dagger}; \pi^{\dagger}]} \, .
	\label{eq:ep}
\end{equation}
This central result in stochastic thermodynamics \cite{seifert05,seifert12} was formulated for CRNs in Ref.~\cite{schmiedl07} and clearly shows that the EP measures the statistical asymmetry of a trajectory under time reversal.
It implies that the EP satisfies the following integral FT
\begin{equation}
	\ave{\exp\left\{ - \Sigma / \kb \right\}} = 1 \, ,
	\label{eq:epIFT}
\end{equation}
where the \emph{ensemble average} $\ave{\cdot}$ runs over all trajectories.
It represents a refinement of the second law of thermodynamics at the trajectory level.
Using the Jensen's inequality, the second law ensues: $\ave{\Sigma} \ge 0$.

\paragraph*{Remark}
Using Eqs.~\eqref{eq:enthalpyOfFormation} and \eqref{eq:decMuY}, the local detailed balance, Eq.~\eqref{eq:ldb}, can be rewritten as
\begin{equation}
	\hspace{-1.6em}
	\kb \ln \frac{\rate{\rho}{\bm n}}{\rate{-\rho}{\bm n + \stoich_{\rho}}} =
	- \tfrac{1}{T} Q^{\mathrm{thr}}_{\rho} + \bm s_\Y \cdot \stoich^\Y_{\rho} + \Delta_{\rho} s_{\bm n} \, .
	\label{eq:ldbInterpretation}
\end{equation}
The first term is the entropy change in the thermal bath, the second one the entropy change in the chemostats, whereas the last one the internal entropy change of the CRN.

\paragraph*{Remark}
Chemical work and Gibbs potential are defined up to a gauge, which accounts for the choice of the standard-state chemical potentials.
Indeed, let us consider the following transformation,
\begin{equation}
	\begin{aligned}
		\bm \mu^\circ &\rightarrow \bm \mu^\circ + {\textstyle\sum_{\lambda}} a_{\lambda} \bm \ell_\lambda \\
		\bm \mu^\circ_\Y &\rightarrow \bm \mu^\circ_\Y + {\textstyle\sum_{\lambda}} a_{\lambda} \bm \ell^\y_\lambda \, ,
	\end{aligned}
	\label{eq:gaugeMu}
\end{equation}
where the second term is a linear combination of conservation laws.
This transformation leaves affinities \eqref{eq:affinity} and EP \eqref{eq:ep} unchanged, while transforming both the chemical work \eqref{eq:firstLaw}, and the Gibbs potential \eqref{eq:G}.
The former changes as
\begin{equation}
	W_{\mathrm{c}}\trjdep \rightarrow W_{\mathrm{c}}\trjdep + {\textstyle\sum_{\lambda_{\mathrm{b}}}} a_{\lambda_{\mathrm{b}}} \bm \ell^\y_{\lambda_{\mathrm{b}}} \cdot \bm{\mathcal{I}}^\Y[\trj] \, ,
	\label{}
\end{equation}
where
\begin{equation}
	\bm{\mathcal{I}}^\Y[\trj] 
	= \int_{0}^{t} \de \tau \, \bm I^\Y(\tau) \, ,
	\label{eq:extCrnt}
\end{equation}
are the integrated currents of exchanged species flowing in the system.
Likewise, the Gibbs potential becomes
\begin{equation}
	G(\bm n) \rightarrow G(\bm n) + {\textstyle\sum_{\lambda}} a_{\lambda} L^{\lambda}_{\bm n} \, .
	\label{}
\end{equation}
Using the properties of conservation laws, \S~\ref{sec:cl}, it is easy to verify that
\begin{align}
	\Delta L_{\lambda_{\mathrm{u}}}[\trj] & = 0 \, , & \Delta L_{\lambda_{\mathrm{b}}}[\trj] & = \bm \ell^\y_{\lambda_{\mathrm{b}}} \cdot \bm{\mathcal{I}}^\Y[\trj] \, ,
	\label{}
\end{align}
which confirms that the gauge terms cancels in the EP, Eq.~\eqref{eq:ep=W-deltaG}.

Alternatively, one can apply the transformation \eqref{eq:gaugeMu} to either $(\bm h,\bm h_{\Y})$ or $(\bm s^\circ,\bm s^\circ_{\Y})$ and investigate how the terms in the entropy balance \eqref{eq:epDef} change.
In the former case, one can easily verify that both $Q[\trj]$ and $S(\bm n)$ are unaltered.
In the latter case, instead,
\begin{equation}
	\begin{aligned}
		S(\bm n) &\rightarrow S(\bm n) + {\textstyle\sum_{\lambda}} a_{\lambda} L^{\lambda}_{\bm n} \, , \\
		Q^{\mathrm{thr}}[\trj] &\rightarrow Q^{\mathrm{thr}}[\trj] \, , \quad \text{and} \\
		Q^{\mathrm{chm}}[\trj] &\rightarrow Q^{\mathrm{chm}}[\trj] + T {\textstyle\sum_{\lambda_{\mathrm{b}}}} a_{\lambda_{\mathrm{b}}} \bm \ell^\y_{\lambda_{\mathrm{b}}} \cdot \bm{\mathcal{I}}^\Y[\trj] \, ,
	\end{aligned}
	\label{eq:gaugeSQ}
\end{equation}
where we distinguished the thermal and chemical heat contributions.

We thus emphasize that, $W_{\mathrm{c}}$, $G(\bm n)$, $S(\bm n)$, and $Q^{\mathrm{chm}}$ are not uniquely defined, in contrast to $\Sigma$ and $Q^{\mathrm{thr}}$.
Despite that, once the gauge is fixed---\emph{i.e.} the values of the standard-state quantities are chosen---they are useful concept for characterizing the dissipation of the process.
Further discussions on the gauge arising in the work--potential connection will be given in \S~\ref{sec:dwGauge}.

\paragraph*{Remark}
Rather than defining the heat as \emph{minus the entropy change in the environment times $T$}, Eqs.~\eqref{eq:deltaRhoH} and \eqref{eq:Q}, we could have defined it as \emph{minus the entropy change in the thermal reservoir times $T$}, $Q^{\mathrm{thr}}$, thus leaving the chemical part aside.
Clearly, this does not affect the EP, but its expression would lose the typical Kelvin--Clausius form, Eq.~\eqref{eq:epDef}, as it would read $\Sigma\trjdep = \Delta S[\trj] - \tfrac{1}{T} Q^{\mathrm{thr}}\trjdep - \int_{0}^{t} \de \tau \, \bm s_\Y(\tau) \cdot \bm I^\Y(\tau)$.
These two different but equivalent approaches are not new to nonequilibrium thermodynamics and have been discussed in Ref.~\cite[Ch.~III, \S~3]{groot84}, for instance.

\subsection{FT for the Chemical Work\\and comparison with previous results}
\label{sec:frCW}

When combining the EP FT \eqref{eq:epIFT} with Eq.~\eqref{eq:ep=W-deltaG} we immediately obtain the integral FT for the chemical work
\begin{equation}
	\ave{\exp\left\{ -\beta (W_{\mathrm{c}} - \Delta G) \right\}} = 1 \, .
	\label{eq:ifrW}
\end{equation}
However, a Jarzynski-like integral FT \cite{bochkov77,bochkov79,jarzynski97,horowitz07} for the chemical work---\emph{i.e.} expressions such as $\ave{\exp\left\{ - \beta W_{\mathrm{c}} \right\}} = \exp\left\{ - \beta \Delta G_{\mathrm{eq}} \right\}$---does not ensue.
This relation would require that \emph{(i)} the process starts and finishes at equilibrium in a closed network, $\Delta G = \Delta G_{\mathrm{eq}}$---the condition on the final state can be relaxed, though---, and \emph{(ii)} $\Delta G_{\mathrm{eq}}$ is a nonfluctuating quantity along the process, so that its exponential can be moved out of the average.
However, due to broken conservation laws, $G_{\mathrm{eq}}$ fluctuates along any trajectory of open CRNs.

\begin{figure}[t]
	\centering
	\subfloat[][]
	{\includegraphics[width=.23\textwidth]{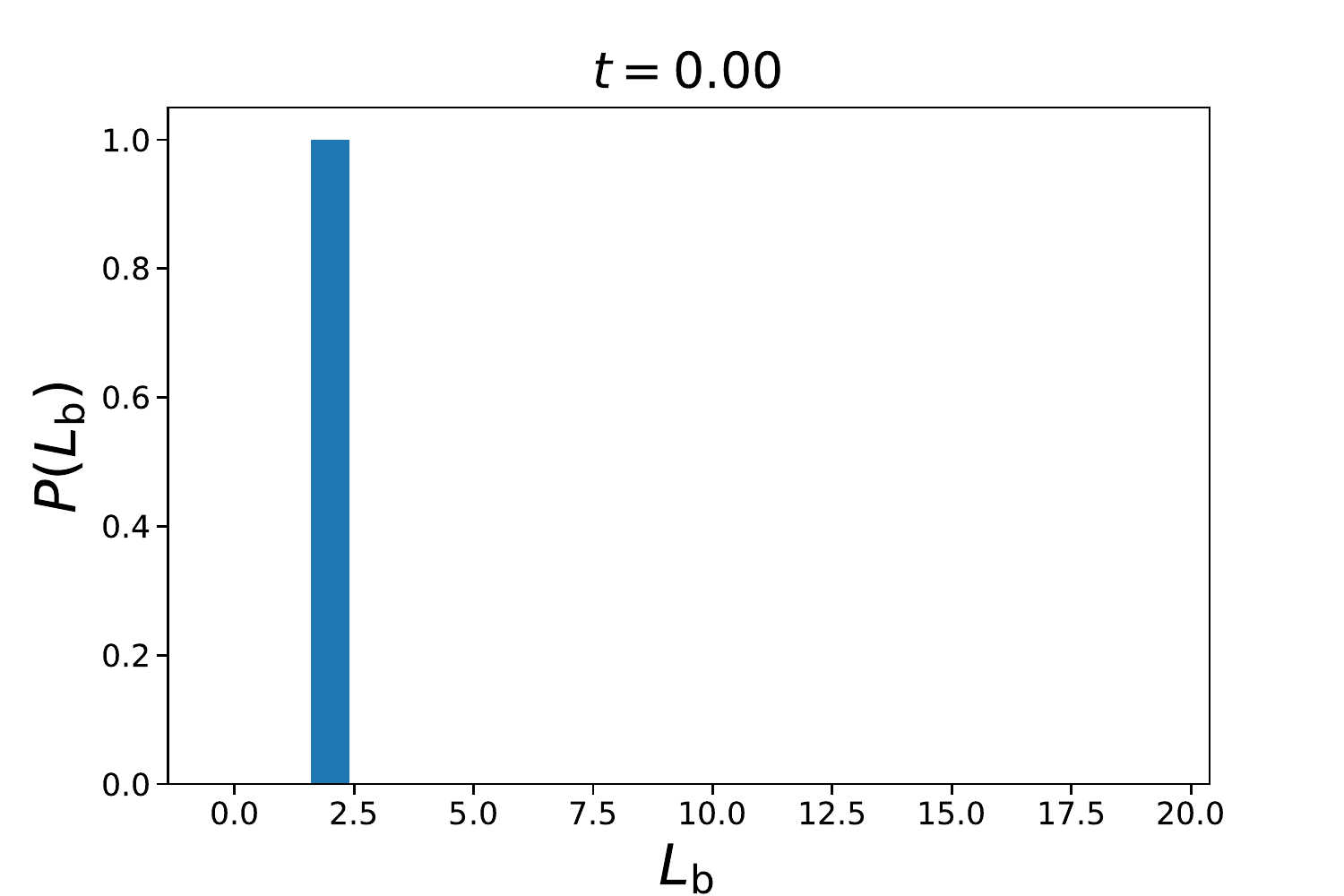} } \,
	\subfloat[][]
	{\includegraphics[width=.23\textwidth]{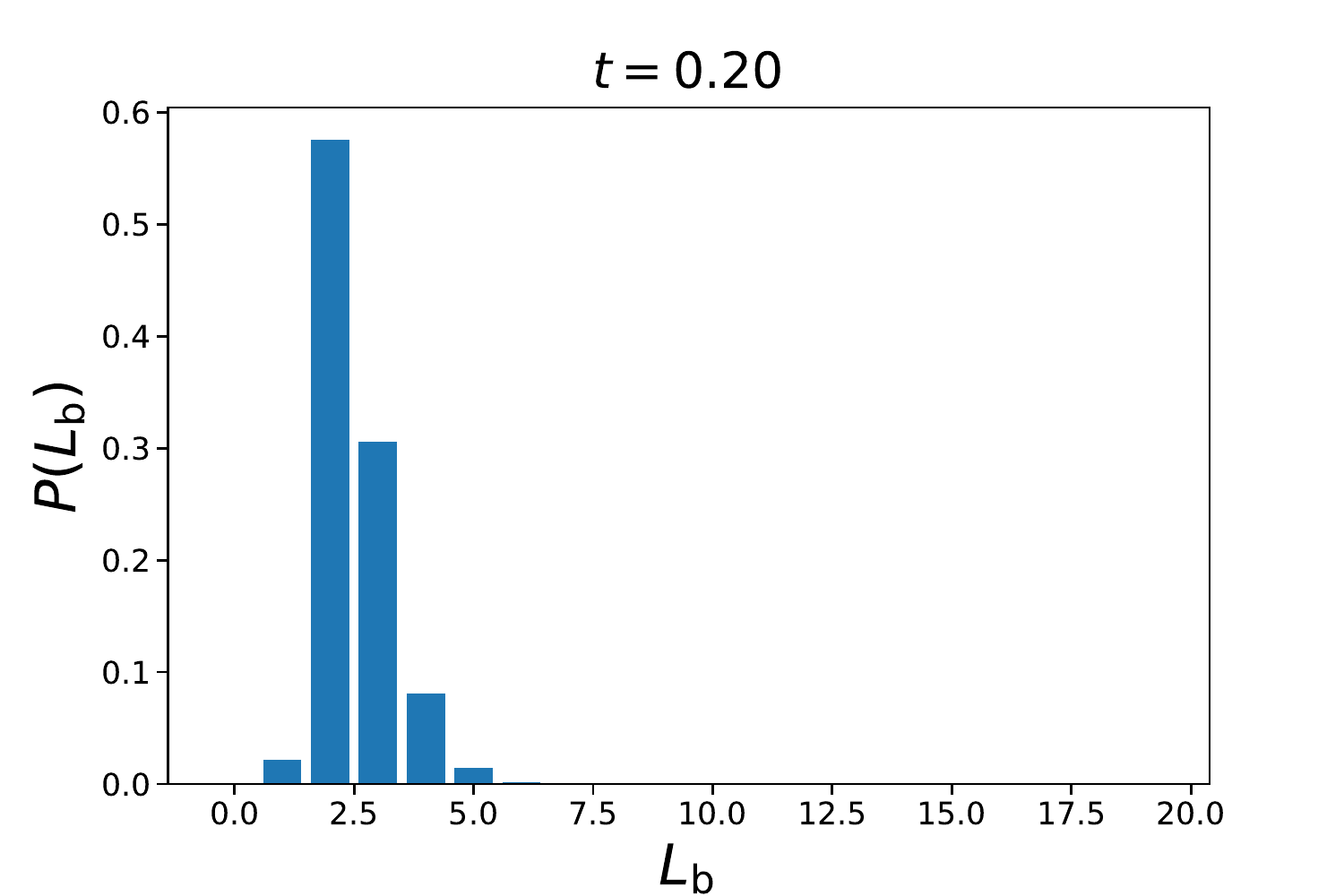} } \\
	\subfloat[][]
	{\includegraphics[width=.23\textwidth]{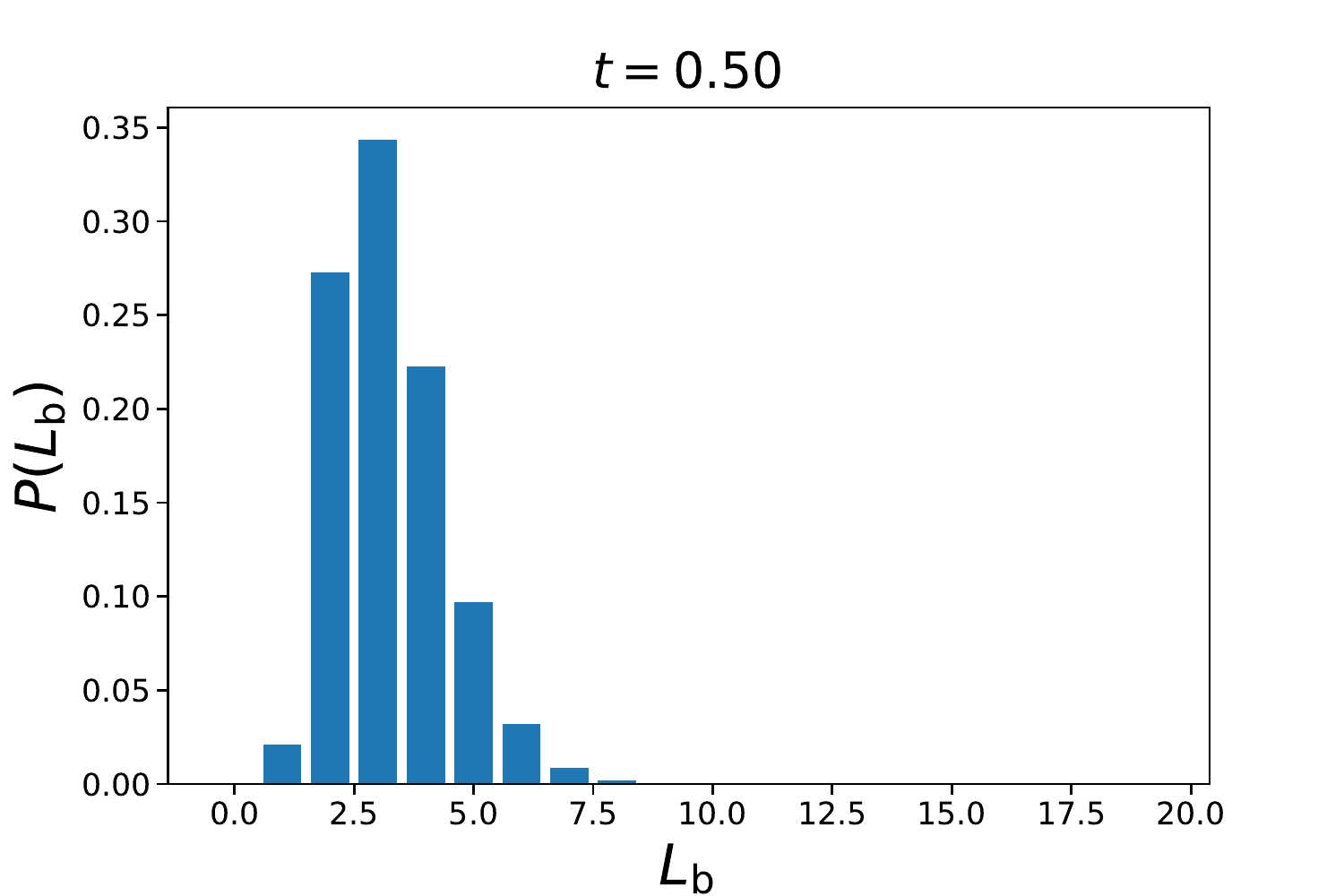} } \,
	\subfloat[][]
	{\includegraphics[width=.23\textwidth]{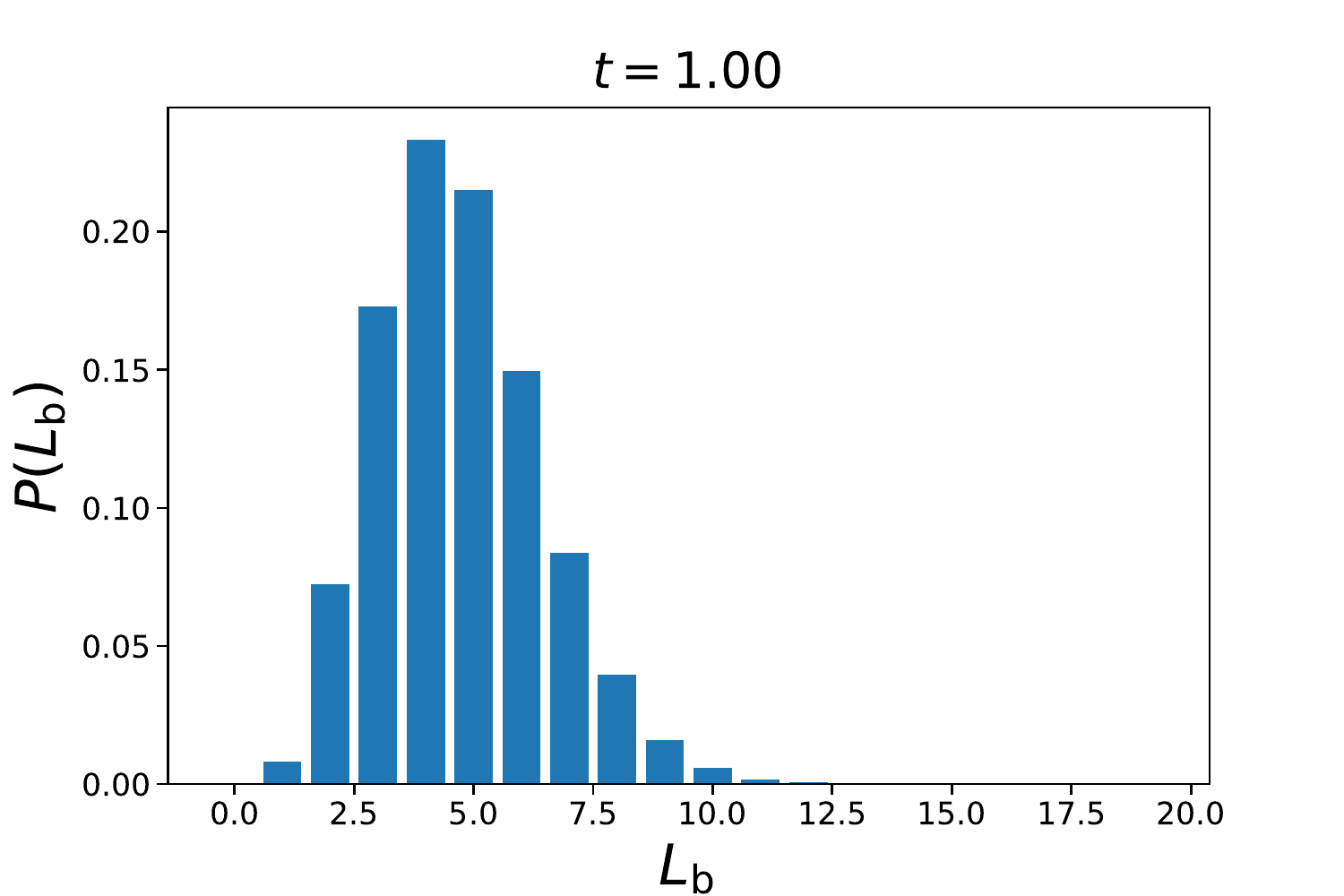} }
	\caption{
		Illustration of the evolution of the probability distribution of the broken components associated to Eq.~\eqref{eq:clMMb} in the CRNs in Fig.~\ref{fig:openMM}.
		As the CRN evolves, the state space enlarges, as the stochastic dynamics explores states corresponding to different broken components, $L_{\mathrm{b}}$.
		The four distribution are obtained by means of $10^{6}$ trajectories simulated using the stochastic simulation algorithm.
		All rate constants are equal to $1$, whereas the concentrations of the chemostatted species are $[\ce{A_{\mathrm{e}}}] = 17$ and $[\ce{B_{\mathrm{e}}}] = 10$.
		The value of the enzyme moiety is $L_{\mathrm{E}} = 5$.
	}
	\label{fig:breaking}
\end{figure}

Let us consider a generic process in which the CRNs is initially closed and at equilibrium, Eq.~\eqref{eq:peq(n)}, with a Gibbs free energy ${\textstyle\sum_{\st{L_{\lambda}}}} P(\st{L_{\lambda}}) G_{\mathrm{eq}}(\st{L_{\lambda}})$.
The CRN is then open and driven according to some time-dependent protocol, $\pi_{\tau}$ for $\tau \in [0,t]$.
At time $t$ the CRN is closed again, and let to relax to a new equilibrium distribution $p^{\mathrm{eq}_{t}}_{\bm n}$.
Since the chemostatting procedure unavoidably breaks some conservation laws, the accessible state space suddenly increases.
The final distribution of broken components, $P(\st{L_{\lambda_{\mathrm{b}}}};t)$, will thus have a support broader than that of the initial distribution, $P(\st{L_{\lambda_{\mathrm{b}}}};0)$, see \emph{e.g.} Fig.~\ref{fig:breaking}.
This process is akin to the free expansion of a gas that is initially at equilibrium in a constrained region of space.
The crucial point is that the initial state is a \emph{constrained}, or \emph{local}, equilibrium with respect to the state space where the dynamics subsequently evolves.

The stochastic thermodynamics of these processes is characterized by \emph{absolute irreversibility} \cite{murashita14}.
Namely, when the EP \eqref{eq:ep} is integrated over all trajectories to obtain the FT \eqref{eq:epIFT}, there are some backward trajectories whose corresponding forward probability is vanishing.
These are the trajectories leading to values of the broken components not in $\mathrm{supp}\left\{ P(\st{L_{\lambda_{\mathrm{b}}}};0) \right\}$.
Since the EP of these trajectories diverges negatively, see Eq.~\eqref{eq:ep}, the expression of the integral FTs \eqref{eq:epIFT}, as well as \eqref{eq:ifrW}, is invalidated, but can be replaced by $\avef{\exp\left\{ - \Sigma / \kb \right\}} = 1 - \lambda_{\mathrm{S}}$, where $0 \leq \lambda_{\mathrm{S}} \leq 1$ measures the probability of those backward trajectories whose forward one has zero probability \cite{murashita14}.

Hence, let us assume that $\mathrm{supp}\left\{ P(\st{L_{\lambda_{\mathrm{b}}}};0) \right\}$ spans all possible values of $\st{L_{\lambda_{\mathrm{b}}}}$, so that no absolute irreversibility occurs.
By conditioning the average in Eq.~\eqref{eq:ifrW} upon observation of specific initial and final components ($\ave{\cdot}_{\st{L_{\lambda}},\st{L_{\lambda}'}}$) we obtain
\begin{multline}
	{\textstyle\sum_{\st{L_{\lambda}}}} {\textstyle\sum_{\st{L'_{\lambda}}}} \,
	P(\st{L_{\lambda}};0) \, P(\st{L_{\lambda}'};t) \\
	\exp \left\{ \beta [G_{\mathrm{eq}_{t}}(\st{L_{\lambda}'}) - G_{\mathrm{eq}_{0}}(\st{L_{\lambda}})] \right\} \\
	\ave{\exp\left\{ -\beta W_{\mathrm{c}} \right\}}_{\st{L_{\lambda}},\st{L_{\lambda}'}} = 1 \, .
\end{multline}
However, this equation cannot be simplified further:
since the Gibbs potential depends on the broken components, it fluctuates during the transient dynamics and an average over all components must be taken.
As a result, no Jarzynski FT for the chemical work in the Gibbs ensemble can be derived.

In Ref.~\cite{schmiedl07} a Jarzynski relation for the chemical work is derived using the grand canonical ensemble \cite[Eq.~(61)]{schmiedl07}.
Translated into our notation, this result reads
\begin{equation}
	\hspace{-1em}
	\ave{\exp\left\{ -\beta [W_{\mathrm{c}} - \Delta(\bm \mu^{\mathrm{eq}} \cdot \bm n) ]  \right\}} = \exp\left\{ - \beta \Delta \mathfrak{G}_{\mathrm{eq}} \right\} \, ,
	\label{eq:IFTseifert}
\end{equation}
where the initial and final equilibrium states are grand canonical:
\begin{equation}
	p_{\bm n}^{\mathrm{eq}} = \exp\left\{ \beta \left[ \mathfrak{G}_{\mathrm{eq}} - g_{\bm n} + \bm \mu^{\mathrm{eq}} \cdot \bm n \right] \right\} \, .
	\label{eq:eqSeifert}
\end{equation}
The grand potential is defined as
\begin{equation}
	\mathfrak{G} := G - \bm \mu^{\mathrm{eq}} \cdot \bm n \, ,
	\label{eq:GrandSeifert}
\end{equation}
and $\bm \mu^{\mathrm{eq}}$ are implicitly defined by
\begin{equation}
	\bm \mu^{\mathrm{eq}}_\x \cdot \stoich^\x_{\ir} + \bm \mu_\y^{\mathrm{eq}} \cdot \stoich^\y_{\ir} = 0 \, , \quad \text{for all} \, \ir \, ,
	\label{eq:eqCP}
\end{equation}
\cite[Eq.~(27)]{schmiedl07}.
The absence of the exchange transition is due to a different form of chemostatting, see remark in \S~\ref{sec:ocn}.
The grand potential is naturally suited to describe CRNs in which all species are chemostatted and $\bm \mu^{\mathrm{eq}}$ are their chemical potentials.
But for most CRNs, where only a subset of species are typically chemostatted, the grand potential is not the most convenient and intuitive potential to work with.
The physical interpretation of the contribution $-\Delta(\bm \mu^{\mathrm{eq}} \cdot \bm n)$ is for instance not transparent.
In the following sections we will make use of conservation laws to identify the potential which best describes CRNs where only a subset of species are chemostatted.
New work contributions with a transparent physical interpretation will ensue.

\section{CRN-Specific Stochastic Thermodynamics}
\label{sec:NSST}

We now proceed with our main results.
Making use of the conservation laws identified in \S~\ref{sec:cl}, we decompose the EP into three fundamental contributions:
a potential difference, a contribution due to time-dependent driving, and a minimal set of contributions due to nonconservative chemical forces.
To do so, we first decompose the local detailed balance and then proceed with the EP.

\subsection{Entropy Production}
\label{sec:dissBalance}

We start our EP decomposition by partitioning the set of chemostatted species $\mathbf{Y}$ into two groups, denoted by $\Rf$ and $\Frc$.
Likewise, the corresponding exchanged species are denoted by $\rf$ and $\frc$, respectively.
The former group is composed by a minimal set of chemostatted species which---when starting from the closed CRN---break all broken conservation laws.
In other words, each entry of $\Rf$ breaks exactly one distinct conservation law.
The remaining chemostatted species form the latter group.
For a given CRN, our partitioning is not unique but the number of $\rf$ and $\frc$ is uniquely defined:
$\nof{\lyp} = \nof{\lambda_{\mathrm{b}}}$ and $\nof{\lyf} = \nof{\y} - \nof{\lambda_{\mathrm{b}}}$, respectively, see Ex.~\ref{sec:epAwesomeMM}.

We now notice that the linear independence of $\st{\bm \ell_{\lambda}}$ implies that the matrix whose rows are $\st{\bm \ell^\yrf_{\lambda_{\mathrm{b}}}}$ is nonsingular.
We will denote by $\st{\overline{\bm \ell}^\yrf_{\lambda_{\mathrm{b}}}}$ the column vectors of the inverse of the latter matrix.
By making use of this important property, we can recast the identity
\begin{equation}
	\hspace{-1.2em}
	\Delta_{\rho} L^{\lambda_{\mathrm{b}}}_{\bm n}
	\equiv \bm \ell_{\lambda_{\mathrm{b}}} \cdot \stoich_{\rho}
	\equiv \bm \ell^{\mathrm{x}}_{\lambda_{\mathrm{b}}} \cdot \stoich_{\rho}^{\mathrm{x}}
	+ \bm \ell^\yrf_{\lambda_{\mathrm{b}}} \cdot \stoich_{\rho}^\yrf
	+ \bm \ell^\yfrc_{\lambda_{\mathrm{b}}} \cdot \stoich_{\rho}^\yfrc
	\label{}
\end{equation}
into
\begin{equation}
	\hspace{-1.2em}
	\stoich_{\rho}^\yrf
	= \Delta_{\rho} \bm M_{\bm n}^\yrf
- {\textstyle\sum_{\lambda_{\mathrm{b}}}} \overline{\bm \ell}^\yrf_{\lambda_{\mathrm{b}}} \left[ \bm \ell^{\mathrm{x}}_{\lambda_{\mathrm{b}}} \cdot \stoich_{\rho}^{\mathrm{x}} + \bm \ell^\yfrc_{\lambda_{\mathrm{b}}} \cdot \stoich_{\rho}^\yfrc \right] \, ,
	\label{eq:stoichYpDec}
\end{equation}
where
\begin{equation}
	\bm M^\yrf_{\bm n} := {\textstyle\sum_{\lambda_{\mathrm{b}}}} \overline{\bm \ell}^\yrf_{\lambda_{\mathrm{b}}} L^{\lambda_{\mathrm{b}}}_{\bm n} \, .
	\label{eq:moietyV}
\end{equation}
Mindful that $\stoich^{\Y}_{\rho} = - \stoich^{\y}_{\rho}$ and $\bm \ell^{\mathrm{x}}_{\lambda_{\mathrm{b}}} \cdot \stoich_{\er}^{\mathrm{x}} = 0$ for all $\er$, one can use Eq.~\eqref{eq:stoichYpDec} to rewrite the chemical work along reactions as
\begin{equation}
	- \bm \mu_\Y \cdot \stoich^\Y_{\rho}
	= \Delta_{\rho} \left[ \bm \mu_\Yp \cdot \bm M^\yrf_{\bm n} \right] 
	- \bm{\mathcal{F}}_\Yf \cdot \stoich^\Yf_{\rho} \, ,
	\label{eq:effectiveWork}
\end{equation}
where
\begin{equation}
	\bm{\mathcal{F}}_\Yf := \bm \mu_\Yf - \bm \mu_\Yp \cdot {\textstyle\sum_{\lambda_{\mathrm{b}}}} \overline{\bm \ell}^\yrf_{\lambda_{\mathrm{b}}} \bm \ell^\yfrc_{\lambda_{\mathrm{b}}} \, .
	\label{eq:fundForces}
\end{equation}
A reformulation of the local detailed balance Eq.~\eqref{eq:ldb} readily ensues
\begin{equation}
	\ln \frac{\rate{\rho}{\bm n}}{\rate{-\rho}{\bm n + \stoich_{\rho}}}
	= - \beta \big( \Delta_{\rho} \semigp_{\bm n} + \bm{\mathcal{F}}_\Yf \cdot \stoich^\Yf_{\rho} \big) \, ,
	\label{eq:ldbAwesome} 
\end{equation}
where
\begin{equation}
	\semigp_{\bm n} := 
	g_{\bm n} - \bm \mu_\Yp \cdot \bm M^\yrf_{\bm n} \, .
	\label{eq:semigp}
\end{equation}

We now notice that the expression of the potential $\semigp_{\bm n}$ is reminiscent of a Legendre transform of $g_{\bm n}$ with respect to $\bm M^\yrf_{\bm n}$, in which $\bm \mu_{\Yp}$ are the conjugated intensive fields.
To reveal the physical meaning of $\bm M^\yrf_{\bm n}$, let us consider the case in which the broken conservation laws correspond to moieties, see \S~\ref{sec:cl}, and hence each species can be thought of as a composition of these.
Through $\rf$, some combinations of these moieties are exchanged with the environment.
The entries of $\bm M^\yrf_{\bm n}$ quantify the total abundance of these combinations in state $\bm n$, hence we refer to $\bm M^\yrf_{\bm n}$ as the \emph{moiety population vector}.
In view of this and the fact that (in general) not all moieties are exchanged, one can interpret $\semigp_{\bm n}$ as the \emph{semigrand Gibbs free energy} of the state $\bm n$ \cite{alberty03}.
Note also that, from the definition of broken conservation law, Eq.~\eqref{eq:clClosed}, it follows that $\Delta_{\ir} \bm M_{\bm n}^\yrf = 0$, for all $\ir$---\emph{viz.} internal reactions never create or destroy moieties---whereas only for $\er$ we have that $\Delta_{\er} \bm M_{\bm n}^\yrf \neq 0$---\emph{viz.} exchange reactions introduce or remove moieties.
We also mention that an alternative interpretation of $\semigp_{\bm n}$ can be given once we rewrite it as
\begin{equation}
	\semigp_{\bm n} :=
	g_{\bm n} - {\textstyle\sum_{\lambda_{\mathrm{b}}}} f_{\lambda_{\mathrm{b}}} L^{\lambda_{\mathrm{b}}}_{\bm n} \, ,
	\label{eq:semigpL}
\end{equation}
where
\begin{equation}
	f_{\lambda_{\mathrm{b}}} := \bm \mu_\Yp \cdot \overline{\bm \ell}^\yrf_{\lambda_{\mathrm{b}}} \, .
	\label{eq:f}
\end{equation}
In this form $\semigp_{\bm n}$ is reminiscent of a Legendre transform with respect to the broken components $\st{L^{\lambda_{\mathrm{b}}}_{\bm n}}$, in which $\st{f_{\lambda_{\mathrm{b}}}}$ are the conjugated intensive fields.

In the second term on the rhs of Eq.~\eqref{eq:ldbAwesome}, $\bm{\mathcal{F}}_{\Yf}$ identifies chemical potential gradients imposed by the chemostats on the CRN.
Its entries, denoted by $\set{\mathcal{F}_{\iyf}}$, for $\iyf = 1,\dots,\nof{\lyf}$, are a maximal independent set of nonconservative chemical forces:
if and only if $\bm{\mathcal{F}}_{\Yf} = \bm 0$, then the rhs of Eq.~\eqref{eq:ldbAwesome} is conservative.
In this case, the CRN is detailed-balanced since the steady-state probability distribution defined by $p_{\bm n}^{\mathrm{eq}} \propto \exp\left\{ - \beta \semigp_{\bm n} \right\}$ satisfies the detailed balance property, Eq.~\eqref{eq:dbp}.
Since $\set{\mathcal{F}_{\iyf}}$ make the CRN non-detailed balanced, we refer to them as \emph{fundamental nonconservative chemical forces}.
Equation~\eqref{eq:ldbAwesome} is our first major result.

To proceed with our EP decomposition, we combine Eqs.~\eqref{eq:epAffinity} and \eqref{eq:ldbAwesome},
\begin{multline}
	T \Sigma\trjdep
	=
	\kt \ln \frac{p_{\bm {n}_{0}}(0)}{p_{\bm {n}_{t}}(t)} \\
	\hspace{-1.6em}
	- \int_{0}^{t} \de \tau {\sum_{\rho,\bm n}} \Delta_{\rho} \semigp_{\bm n}(\tau) \, j_{\rho}(\bm n, \tau)
	+ {\textstyle\sum_{\iyf}} W^{\mathrm{nc}}_{\iyf}\trjdep \, ,
	\label{eq:epProof}
\end{multline}
where
\begin{equation}
	W^{\mathrm{nc}}_{\iyf}\trjdep
	:= \int_{0}^{t} \de \tau \, \mathcal{F}_{\iyf}(\tau) I_{\iyf}(\tau) \, .
	\label{eq:pCW}
\end{equation}
$\st{I_{\iyf}(\tau)}$, for $\iyf = 1,\dots,\nof{\lyf}$, denote the entries of the instantaneous external currents corresponding to $\Frc$, Eq.~\eqref{eq:instChemoCurr}.
We now recall that $\semigp_{\bm n}$ is a state function, hence
\begin{equation}
	\Delta \semigp\trjdep
	=
	W_{\mathrm{d}}\trjdep +
	\int_{0}^{t} \de \tau {\sum_{\rho,\bm n}} \Delta_{\rho} \semigp_{\bm n}(\tau) \, j_{\rho}(\bm n, \tau)
	\, ,
	\label{eq:deltasemigProof}
\end{equation}
where
\begin{equation}
	\begin{split}
		W_{\mathrm{d}}\trjdep
		&:= \int_{0}^{t} \de \tau \at{\left[ \partial_{\tau} \semigp_{\bm n}(\tau) \right]}{\bm n_{\tau}} \\
		&\phantom{:}= \int_{0}^{t} \de \tau \, \left[ - \partial_{\tau} \bm \mu_\Yp(\tau) \right] \cdot \bm M^\yrf_{\bm n_{\tau}} \, .
	\end{split}
	\label{eq:dw}
\end{equation}
Therefore, combining Eqs.~\eqref{eq:epProof} and \eqref{eq:deltasemigProof} we obtain
\begin{equation}
	\hspace{-1em}
	T \Sigma\trjdep = - \Delta \Semigp\trjdep + W_{\mathrm{d}}\trjdep + {\textstyle\sum_{\iyf}} W^{\mathrm{nc}}_{\iyf}\trjdep \, ,
	\label{eq:epAwesome}
\end{equation}
where the first term is the difference of \emph{stochastic semigrand Gibbs potential}
\begin{equation}
	\Semigp(\bm {n}) :=
	\kt \ln p_{\bm n} + \semigp_{\bm n} \, .
	\label{eq:Semigp}
\end{equation}

The EP decomposition in Eq.~\eqref{eq:epAwesome} is a major result of our paper.
The first term on the rhs constitutes the conservative force contribution of the EP.
It describes the dissipation due to overall changes of thermodynamic state variables:
enthalpy, $H(\bm n)$, entropy, $S(\bm n)$, and chemical energy $\set{\bm \mu_\Yp \cdot \bm M^\yrf_{\bm n}}$.
The second term, Eq.~\eqref{eq:dw}, arises in presence of time-dependent driving and accounts for the changes caused by manipulations of the chemical potentials $\bm \mu_{\Yp}$.
As it is a controlled way of changing the Gibbs free energy landscape of the CRN, we refer to it as \emph{driving chemical work}.
Finally, for each exchanged species $\Frc$, a nonconservative force contribution \eqref{eq:pCW} arises, $\st{W^{\mathrm{nc}}_{\iyf}}$.
All together, they account for the chemical energy flowing between different chemostats across the CRN, and we refer to them as \emph{nonconservative chemical work} contributions.
Equation~\eqref{eq:epAwesome} holds for an arbitrary CRN, yet it is CRN-specific, as it is derived using the topological properties of the CRN encoded in the conservation laws.
To gain more intuition, we now focus on specific classes of CRNs, whose resulting decomposition is summarized in Tab.~\ref{tab:EPprocesses}.
In \S~\ref{sec:CRNSEnergyBalance} we continue our discussion on the work contributions $W_{\mathrm{d}}$ and $\st{W^{\mathrm{nc}}_{\iyf}}$, whereas in Ex.~\ref{sec:epAwesomeMM} and in \S~\ref{sec:examples} we evaluate them for specific models.
Finally, in \S\S~\ref{sec:fr} and \ref{sec:ead} we will further explore the implications of Eq.~\eqref{eq:epAwesome}.

\paragraph*{Autonomous Detailed-Balanced CRNs:}
The CRN is autonomous and all fundamental forces vanish.
The trajectory EP becomes minus a potential difference,
\begin{equation}
	T \Sigma[\trj] = - \Delta \Semigp[\trj] \, .
	\label{eq:epDB}
\end{equation}
We will prove in \S~\ref{sec:ead} that this is the class of open CRNs which relax to equilibrium and in which the average potential $\avef{\Semigp}$ is minimized at equilibrium by the dynamics described by CME \eqref{eq:CME}.

\paragraph*{Unconditionally Detailed-Balanced CRNs:}
The set of species $\Frc$ is empty---\emph{i.e.} each exchanged species breaks a conservation law---and no fundamental force arises.
Hence, these CRNs are detailed-balanced irrespective of the values of $\bm \mu_{\Y}$, but the time-dependent driving prevents them from reaching equilibrium, and their EP reads
\begin{equation}
	T \Sigma\trjdep =
	- \Delta \Semigp\trjdep
	+ W_{\mathrm{d}}\trjdep
	\, .
	\label{eq:epAwesomeUDB}
\end{equation}

\paragraph*{Autonomous CRNs:}
The driving work vanishes and the forces are constant in time.
Hence, the EP becomes
\begin{equation}
	T \Sigma[\trj] =
	- \Delta \Semigp[\trj]
	+ {\textstyle\sum_{\iyf}} \mathcal{F}_{\iyf} \mathcal{I}_{\iyf}[\trj]
	\, .
	\label{eq:epAwesomeNoDriving}
\end{equation}
The nonconservative chemical work display a typical current--force structure.
In the long time limit, $\Delta \Semigp[\trj]$ is typically subextensive in time, and we obtain the EP typical of nonequilibrium steady states
\begin{equation}
	T \Sigma[\trj]
	\overset{t\rightarrow \infty}{=}
	{\textstyle\sum_{\iyf}} \mathcal{F}_{\iyf} \mathcal{I}_{\iyf}[\trj] \, ,
	\label{eq:epAwesomeLT}
\end{equation}
see Eq.~\eqref{eq:extCrnt}.
In other words, $T\Sigma[\trj]$ is dominated by the dissipative flows of chemicals across the CRN.

\paragraph*{Remark}
For CRN with infinite number of species and reactions---\emph{e.g.} aggregation--fragmentation and polymerization processes \cite{krapivsky10,rao15:denzymes,andrieux08:copolymerization}---the CRN may undergo steady growth regimes in which $\Delta \mathcal{G}$ is not subextensive in time and cannot be neglected in long-time limit.

\paragraph*{Remark}
Our EP decomposition is not unique and different expressions for $\semigp_{\bm n}$ and $\bm{\mathcal{F}}_\Yf$ correspond to different ways of partitioning $\external$ into $\Rf$ and $\Frc$.

\begin{table}
	\centering
	\begin{tabular}{rccc}
		\toprule
		\textbf{dynamics}	& { }$\bm{-\Delta \Semigp}${ } 	& { }$\bm{W_{\mathrm{d}}}${ } 	& { }$\bm{W^{\mathrm{nc}}}$ \\
		\midrule
		autonomous detailed-balanced & \checkmark		& 0	 	& 0 \\
		unconditionally detailed-balanced	 & \checkmark		& \checkmark	 	& 0 \\
		autonomous				 	 & \checkmark 	& 0		& \checkmark	\\
		nonequilibrium steady state  & 0	& 0 	& \checkmark	\\
		\bottomrule
	\end{tabular}
	\caption{
		Entropy production for specific processes.
		``0'' (resp. ``$\checkmark$'') denotes a vanishing (resp. a finite) contribution.
	}
	\label{tab:EPprocesses}
\end{table}

\begin{figure}[t]
	\centering
	\includegraphics[width=.45\textwidth]{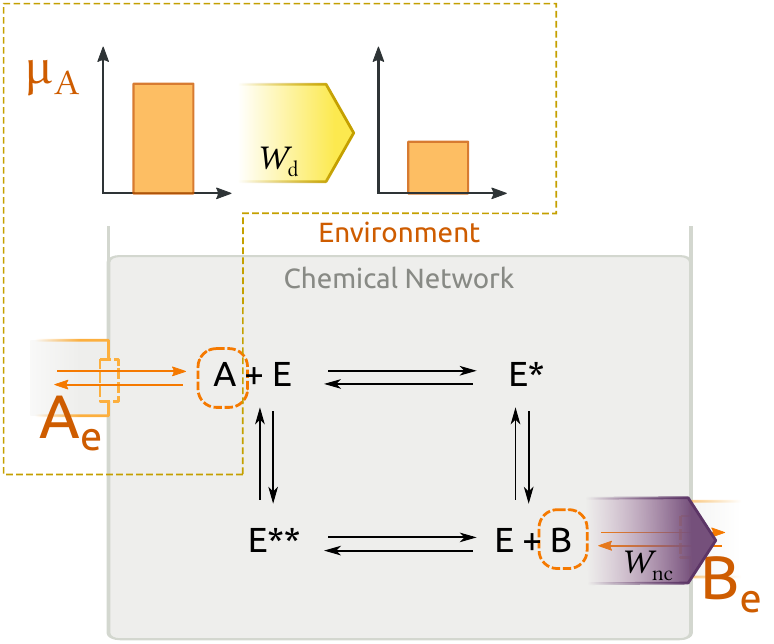}
	\caption{
		Pictorial illustration of the work contributions.
		The driving one arises when the chemical potential of the chemostat $\ce{A_{\mathrm{e}}}$ changes in time.
		The nonconservative chemical work, instead, characterizes the sustained conversion of $\ce{A}$ into $\ce{B}$.
	}
	\label{fig:epAwesomeMM}
\end{figure}

\begin{example}
	\label{sec:epAwesomeMM}
	For the open CRN in Fig.~\ref{fig:openMM}, the chemostatted species can be split into $\Rf$ and $\Frc$ in two possible---and trivial---ways:
	either $\ce{A}$ is regarded as the species breaking the conservation law \eqref{eq:clMMb}, or $\ce{B}$.
	We consider the former choice, $\rf = (\ce{A})$ and $\frc = (\ce{B})$.
	Since $\ell^{\mathrm{b}}_{\ce{A}} = 1$, the only entry of the moiety vector reads,
	\begin{equation}
		M^{\ce{A}}_{\bm n} = n_{\ce{E^{\ast}}} + n_{\ce{E^{\ast\ast}}} + n_{\ce{A}} + n_{\ce{B}} = L^{\mathrm{b}}_{\bm n} \, ,
		\label{eq:moietyMM}
	\end{equation}
	which is equal to the total abundance of the $\ce{A}$--$\ce{B}$ moiety.
	The intensive variable conjugated to the broken conservation law is equal to the chemical potential of $\ce{A_{\mathrm{e}}}$,
	\begin{equation}
		f_{\mathrm{b}} = \mu_{\ce{A_{\mathrm{e}}}} \, .
		\label{eq:f=muMM}
	\end{equation}
	The potential thus readily follows from Eq.~\eqref{eq:semigp}---or equivalently Eq.~\eqref{eq:semigpL}---,
	\begin{equation}
		\semigp_{\bm n} = g_{\bm n}
		- \mu_{\ce{A_{\mathrm{e}}}} M^{\ce{A}}_{\bm n} \, .
		\label{eq:semigpMM}
	\end{equation}
	The instantaneous driving work rate associated to any manipulation of the latter potential is
	\begin{equation}
		\dot{W}_{\mathrm{d}}(\bm n) =
		- \partial_{t} \mu_{\ce{A_{\mathrm{e}}}} M^{\ce{A}}_{\bm n} \, .
		\label{eq:MMdcw}
	\end{equation}
	Once integrated over a trajectory, it gives the driving work, Eq.~\eqref{eq:dw}.
	Since $\frc = (\ce{B})$, the conjugated fundamental chemical force reads
	\begin{equation}
		\mathcal{F}_{\ce{B_{\mathrm{e}}}}
		= \mu_{\ce{B_{\mathrm{e}}}} - \mu_{\ce{A_{\mathrm{e}}}} \, .
		\label{}
	\end{equation}
	and the instantaneous dissipative contribution due to this force is
	\begin{equation}
		\dot{W}^{\mathrm{nc}}_{\ce{B_{\mathrm{e}}}} = \mathcal{F}_{\ce{B_{\mathrm{e}}}} I_{\ce{B_{\mathrm{e}}}} \, ,
		\label{}
	\end{equation}
	where $I_{\ce{B_{\mathrm{e}}}} = J_{+\mathrm{b}} - J_{-\mathrm{b}}$.
	When integrated over a trajectory, it measures the work spent to sustain a current between $\ce{A_{\mathrm{e}}}$ and $\ce{B_{\mathrm{e}}}$ across the CRN.
	A pictorial illustration of the work contributions is given in Fig.~\ref{fig:epAwesomeMM}.
	The trajectory EP thus reads
	\begin{multline}
		T \Sigma\trjdep = \int_{0}^{t} \de \tau {\big[ - \partial_{\tau} \mu_{\ce{A_{\mathrm{e}}}}(\tau) M^{\ce{A}}_{\bm n} \big]}\big|_{\bm n_{\tau}} - \Delta \mathcal{G}\trjdep \\
		+ \int_{0}^{t} \de \tau \, \mathcal{F}_{\ce{B_{\mathrm{e}}}}(\tau) I_{\ce{B_{\mathrm{e}}}}(\tau) \, .
	\qed
	\end{multline}
\end{example}

\subsection{Energy Balance}
\label{sec:CRNSEnergyBalance}

In Eq.~\eqref{eq:epAwesome}, the driving and nonconservative chemical work, $W_{\mathrm{d}}$ and $\st{W^{\mathrm{nc}}_{\iyf}}$, emerge as dissipative contributions.
To strengthen their interpretation as work contributions, we now show that they can also be described as part of an energy balance.
For this purpose, let us introduce the \emph{semigrand enthalpy} \cite{altaner17},
\begin{equation}
	\mathcal{H}(\bm n) := H(\bm n) - \bm \mu_{\Yp} \cdot \bm M^{\yrf}_{\bm n} = \mathcal{G}(\bm n) + T S(\bm n) \, .
	\label{eq:calH}
\end{equation}
This CRN-specific potential quantifies the portion of energy which is not attributed to volume ($-pV$, where $p$ is the external pressure) and exchanged moieties, $\bm \mu_{\Yp} \cdot \bm M^{\yrf}_{\bm n}$.
It accounts for the energy stored in its internal chemical composition, \emph{i.e.} the internal species $\internalx$ and the unbroken components $\set{L_{\lambda_{\mathrm{u}}}}$.
When combining its definition with the enthalpy and entropy balances, Eqs.~\eqref{eq:firstLaw}, \eqref{eq:epDef} and \eqref{eq:epAwesome}, we obtain
\begin{equation}
	\Delta \mathcal{H}\trjdep = Q\trjdep + W_{\mathrm{d}}\trjdep + {\textstyle\sum_{\iyf}} W^{\mathrm{nc}}_{\iyf}\trjdep \, ,
	\label{eq:Hawesome}
\end{equation}
\emph{viz.} the overall change of semigrand enthalpy is equal to the sum of heat flow, driving and nonconservative chemical work.
By analogy with Eq.~\eqref{eq:firstLaw}, this can be interpreted as a CRN-specific formulation of the first law.

In \S~\ref{sec:balances}, we introduced the chemical work as the Gibbs free energy exchanged with the chemostats, Eq.~\eqref{eq:cw}.
By comparing Eqs.~\eqref{eq:H} and \eqref{eq:Hawesome}, we obtain its relationship with $W_{\mathrm{d}}$ and $\st{W^{\mathrm{nc}}_{\iyf}}$,
\begin{equation}
	\hspace{-2em}
	W_{\mathrm{c}}\trjdep - \Delta \left[ \bm \mu_{\Yp} \cdot \bm M^{\yrf}_{\bm n} \right] = W_{\mathrm{d}}\trjdep + {\textstyle\sum_{\iyf}} W^{\mathrm{nc}}_{\iyf}\trjdep \, .
	\label{}
\end{equation}
We emphasize that in contrast to the chemical work, the driving one does not account for direct exchanges of Gibbs free energy, but it captures the instantaneous changes of the chemostats Gibbs free energy.

\paragraph*{Remark}
The driving work is reminiscent of the mechanical work as defined in stochastic thermodynamics.
In this framework, $W_{\mathrm{mech}}[\trj] = \int_{0}^{t} \de \tau \at{\partial_{\tau} E_{n}(\tau)}{n_{\tau}}$, describes internal energy changes due to external time-dependent control, see \emph{e.g.} \cite{jarzynski97,crooks98}.
In CRNs, the time-dependent control is exerted via the chemostats, and $W_{\mathrm{d}}[\trj]$ indeed accounts for this fact.

\subsection{Equilibrium of open CRNs}
\label{sec:dbnEq}

We have already seen that in absence of fundamental forces, the rhs of the local detailed balance \eqref{eq:ldbAwesome} becomes a state function difference.
The steady-state probability distribution
\begin{equation}
	\hspace{-1em}
	p_{\mathrm{eq}} ( \bm {n} | \st{L_{\lambda_{\mathrm{u}}}} ) =
	\frac{\exp\left\{ -\beta \semigp_{\bm n} \right\}}{\mathcal{Z}( \st{L_{\lambda_{\mathrm{u}}}} )}
	{\textstyle\prod_{\lambda_{\mathrm{u}}}} \delta\big[L^{\lambda_{\mathrm{u}}}_{\bm n},L_{\lambda_{\mathrm{u}}}\big] \, .
	\label{eq:peq(n|L)Open}
\end{equation}
satisfies the detailed balance property \eqref{eq:ldb} and therefore characterizes the equilibrium of open CRNs.
Not accidentally, the relationship between the partition function $\mathcal{Z}( \st{L_{\lambda_{\mathrm{u}}}} )$ and that of closed CRNs, Eq.~\eqref{eq:Zclosed},
\begin{equation}
	\begin{aligned}
		\hspace{-1.6em}
		\mathcal{Z}( \st{L_{\lambda_{\mathrm{u}}}} )
		& = {\textstyle\sum_{\bm m}}
		\exp\left\{ -\beta \semigp_{\bm m} \right\} \, 
		{\textstyle\prod_{\lambda_{\mathrm{u}}}} \delta\big[L^{\lambda_{\mathrm{u}}}_{\bm m},L_{\lambda_{\mathrm{u}}}\big] \\
		& = {\textstyle\sum_{\st{L_{\lambda_{\mathrm{b}}}}}} \exp\left\{ \beta {\textstyle\sum_{\lambda_{\mathrm{b}}}} f_{\lambda_{\mathrm{b}}} L_{\lambda_{\mathrm{b}}} \right\} Z(\st{L_{\lambda}}) \, ,
	\end{aligned}
	\label{eq:Zopen} 
\end{equation}
is akin to that between canonical and grand canonical partition functions, see \emph{e.g.} \cite{*[][{, \S~3.15.}] {peliti11}}.
With an ensemble of unbroken components, $P(\st{L_{\lambda_{\mathrm{u}}}})$, the constrained equilibrium distribution reads
\begin{equation}
	\begin{split}
		p^{\mathrm{eq}}_{\bm {n}} &= {\textstyle\sum_{\st{L_{\lambda_{\mathrm{b}}}}}} \, 
		p_{\mathrm{eq}} ( \bm {n} | \st{L_{\lambda_{\mathrm{u}}}} ) \,
		P(\st{L_{\lambda_{\mathrm{u}}}}) \\
		& = p_{\mathrm{eq}}(\bm {n}|\st{L^{\lambda_{\mathrm{u}}}_{\bm n}}) \, P(\st{L^{\lambda_{\mathrm{u}}}_{\bm n}}) \, ,
	\end{split}
	\label{eq:peq(n)Open}
\end{equation}
where $p_{\mathrm{eq}}(\bm {n}|\st{L^{\lambda_{\mathrm{u}}}_{\bm n}})$ is the probability distribution of observing the state $\bm n$ given its stoichiometric compatibility class.
Equation~\eqref{eq:peq(n)Open} thus generalizes the equilibrium probability distribution \eqref{eq:peq(n)} to open CRNs.

Importantly, the average semigrand Gibbs potential \eqref{eq:Semigp} takes its minimum value at $p^{\mathrm{eq}}_{\bm {n}}$, Eq.~\eqref{eq:peq(n)Open}, where it reduces to the equilibrium semigrand Gibbs potential,
\begin{equation}
	\hspace{-1.6em}
	\Semigp_{\mathrm{eq}}(\st{L_{\lambda_{\mathrm{u}}}})
	= - \kt \ln \mathcal{Z}(\st{L_{\lambda_{\mathrm{u}}}}) + \kt \ln P(\st{L_{\lambda_{\mathrm{u}}}}) \, ,
	\label{eq:Semigpeq}
\end{equation}
averaged over $P(\st{L_{\lambda_{\mathrm{u}}}})$.
Indeed,
\begin{equation}
	\hspace{-2.0em}
	\avef{\Semigp} - \avef{\mathcal{G}_{\mathrm{eq}}}_{\mathrm{L_u}}	= \avef{\Semigp - \Semigp_{\mathrm{eq}}} = \kt \, \mathcal{D}(p\|p_{\mathrm{eq}}) \ge 0 \, ,
	\label{eq:Gt-Gteq=D}
\end{equation}
where
\begin{equation}
	\ave{\mathcal{G}_{\mathrm{eq}}}_{\mathrm{L_u}} \equiv {\textstyle\sum_{\st{L_{\lambda_{\mathrm{u}}}}}} P(\st{L_{\lambda_{\mathrm{u}}}}) \Semigp_{\mathrm{eq}}(\st{L_{\lambda_{\mathrm{u}}}}) \, .
	\label{eq:avetGeq}
\end{equation}
The first equality follows from the fact that $\Semigp_{\mathrm{eq}}$ is nonfluctuating, since it depends solely on the unbroken components.
As for the Gibbs free energy in closed CRNs, we will show later (\S~\ref{sec:ead}) that Eq.~\eqref{eq:Gt-Gteq=D} quantifies the average dissipation during the relaxation to equilibrium.

\subsection{Dissipation Balance along Stoichiometric Cycles}
\label{sec:dissBalanceSuper}

We can now formulate the EP decomposition in term of stoichiometric cycles affinities.
These are defined as the sum of the transition affinities along stoichiometric cycles $\st{\bm c \equiv \rho_{\mathrm{c}1} , \rho_{\mathrm{c}1} , \dots , \rho_{\mathrm{c}\nof{\mathrm{c}}}}$, 
\begin{equation}
	\begin{split}
		\mathcal{A} &:= 
		A_{\rho_{\mathbf{c}}1}(\bm {n}) + A_{\rho_{\mathbf{c}}2}(\bm {n}+\stoich_{\rho_{\mathrm{c}}1}) + \dots \\
		& \quad \ldots + A_{\rho_{\mathbf{c}}\nof{\mathrm{c}}}(\bm {n}+{\textstyle\sum_{j=1}^{\nof{\mathrm{c}}-1}}\stoich_{\rho_{\mathrm{c}}j}) \, .
	\end{split}
	\label{eq:cycleAffinity}
\end{equation}
Using Eq.~\eqref{eq:affinity}, and the fact that $-\Delta_{\rho}G(\bm n)$ vanishes when summed over the loop $\bm c$, we obtain
\begin{equation}
	\mathcal{A}
	= - \bm \mu_\Y \cdot \sum_{i=1}^{\nof{\mathbf{c}}} \stoich^\Y_{\rho_{\mathbf{c}}i}
	= - \bm \mu_\Y \cdot {\textstyle \sum_{\rho}} \stoich^\Y_{\rho} c_{\rho} \, .
	\label{}
\end{equation}
Since ${\textstyle \sum_{\rho}} \stoich^\Y_{\rho} c^{\cy}_{\rho} = 0$, those evaluated along the stochiometric cycles of the closed CRN, $\st{\bm c^{\cy}}$, always vanish.
In contrast, those along the emergent cycles, $\st{\bm c^{\ecy}}$, do not vanish in general,
\begin{equation}
	\mathcal{A}_{\ecy} = \bm \mu_\Y \cdot \bm C^{\Y}_{\ecy} \, ,
	\label{eq:emergentA}
\end{equation}
see Eq.~\eqref{eq:C}.
These affinities can be thus understood as the chemical potential gradient imposed by the chemostats on the cycle.

To rewrite the EP \eqref{eq:epAwesome} in terms $\st{\mathcal{A}_{\eta}}$, let us highlight their relationship with the fundamental forces,
\begin{equation}
	\mathcal{A}_{\ecy}
	= \bm{\mathcal{F}}_\Yf \cdot \bm C_{\ecy}^\Yf \, ,
	\label{eq:effecticeEmergentAffinity}
\end{equation}
which is obtained when summing the local detailed balance \eqref{eq:ldbAwesome} along $\st{\bm c^{\ecy}}$ as in Eq.~\eqref{eq:cycleAffinity}.
Since the matrix whose columns are $\st{\bm C_{\ecy}^\Yf}$ is square and nonsingular---as it can be deduced from the linear independence of the set of emergent cycles---, we can invert it and write
\begin{equation}
	\bm{\mathcal{F}}_\Yf =
	\textstyle \sum_{\ecy} \overline{\bm C}_{\ecy}^\Yf \mathcal{A}_{\ecy} \, ,
	\label{}
\end{equation}
where $\st{\overline{\bm C}_{\ecy}^\Yf}$ denote the rows of the inverse matrix.
This relation clarifies the one-to-one correspondence which lies between $\st{\mathcal{F}_{\iyf}}$ and $\st{\mathcal{A}_{\ecy}}$.
Inserting the last expression in the local detailed balance, Eq.~\eqref{eq:ldbAwesome}, we obtain
\begin{equation}
	\hspace{-1em}
	\ln \frac{\rate{\rho}{\bm n}}{\rate{-\rho}{\bm n + \stoich_{\rho}}}
	= - \beta \big( \Delta_{\rho} \semigp_{\bm n} - \textstyle \sum_{\ecy} \mathcal{A}_{\ecy} \zeta_{\ecy,\rho} \big) \, ,
	\label{eq:ldbSuperAwesome}
\end{equation}
where the coefficients
\begin{equation}
	\zeta_{\ecy,\rho} := - \overline{\bm C}_{\ecy}^\Yf \cdot \stoich^\Yf_{\rho}
	\label{eq:zeta}
\end{equation}
quantify how much each reaction contributes to the emergent cycles.
Algebraically, the row vectors $\st{\bm \zeta_{\ecy}}$ are dual to the cycles, $\st{\bm c^{\ecy}}$,
\begin{equation}
	\hspace{-1.4em}
	\bm \zeta_{\ecy} \cdot \bm c^{\ecy'} =
	- {\textstyle\sum_{\rho}} \overline{\bm C}_{\ecy}^\Yf \cdot \stoich^\Yf_{\rho} c^{\ecy'}_{\rho} =
	\overline{\bm C}_{\ecy}^\Yf \cdot {\bm C}_{\ecy'}^\Yf =
	\delta_{\ecy,\ecy'} \, .
	\label{}
\end{equation}

As previously done for Eq.~\eqref{eq:epAwesome}, when integrating the trajectory EP \eqref{eq:epAffinity} with the local detailed balance \eqref{eq:ldbSuperAwesome} we obtain
\begin{equation}
		T \Sigma\trjdep =
		- \Delta \Semigp\trjdep + W_{\mathrm{d}}\trjdep + {\textstyle\sum_{\ecy}} \Gamma_{\ecy}\trjdep \, .
	\label{eq:epSuperAwesome}
\end{equation}
The stochastic semigrand Gibbs potential and the driving work read as in Eqs.~\eqref{eq:Semigp} and \eqref{eq:dw}, respectively.
For each emergent stoichiometric cycle,
\begin{equation}
	\Gamma_{\ecy}\trjdep
	:= \int_{0}^{t} \de \tau \, \mathcal{A}_{\ecy}(\tau) {\textstyle\sum_{\rho}} \zeta_{\ecy,\rho} J_{\rho}(\tau) \, .
	\label{eq:gamma}
\end{equation}
quantifies the chemical work spent to sustain the related cyclic flow of chemicals.
For autonomous CRNs
\begin{equation}
	T \Sigma[\trj] =
	- \Delta \Semigp[\trj]
	+ {\textstyle\sum_{\ecy}} \mathcal{A}_{\ecy} \mathcal{J}_{\ecy}[\trj]
	\, ,
	\label{eq:epSuperAwesomeAuto}
\end{equation}
where
\begin{equation}
	\mathcal{J}_{\ecy}[\trj] := \int_{0}^{t} \de \tau \, {\textstyle\sum_{\rho}} \zeta_{\ecy,\rho} J_{\rho}(\tau)
	\label{eq:cycleCurrents}
\end{equation}
quantifies the integrated current along the cycle $\ecy$.
In the long-time limit, in which $\Delta \Semigp[\trj]$ is negligible, we obtain
\begin{equation}
	T \Sigma[\trj] \overset{t\rightarrow \infty}{=}  {\textstyle\sum_{\ecy}} \mathcal{A}_{\ecy} \mathcal{J}_{\ecy}[\trj] \, .
	\label{eq:epSuperAwesomeLT}
\end{equation}
When all emergent cycle affinities vanish---as well as when no emergent cycle is created---, the CRN becomes detailed-balanced, in agreement with the Kolmogorov--Wegscheider condition \cite{kolmogoroff36,kelly79,schuster89}.

We emphasize that the cycle chemical work contributions and currents, Eqs.~\eqref{eq:gamma} and \eqref{eq:cycleCurrents}, can be written as combinations of fundamental external currents, $\set{\bm{I}^{\Yf}}$ Eq.~\eqref{eq:instChemoCurr}, via Eq.~\eqref{eq:zeta}.
The added value of Eq.~\eqref{eq:epAwesome} over \eqref{eq:epSuperAwesome} lies in the fact that each force is conjugated to the external current of only one external species. 

\paragraph*{Remark}
An alternative approach that can be used for cycle EP decompositions is the graph-theoretic one based on the identification of the loops appearing in the network of transitions \cite{schnakenberg76,polettini14:cocycle}.
Once these loops are identified, they can be sorted according to the chemostats they are coupled to, as these determine their affinity, see Eq.~\eqref{eq:cycleAffinity}.
Equivalently, loops are classified according to the stoichiometric cycle they correspond to.
In Ref.~\cite{andrieux07:response}, a graph-theoretic approach based on loop affinities led to the expression analogous to Eq.~\eqref{eq:epSuperAwesomeLT}.
In contrast, our cycles EP decomposition is based on a stoichiometric approach:
emergent cycles are directly identified by the kernels of $\stoichM_{\mathrm{i}}$ and $\stoichM$.

This observation points out the redundancy which is intrinsic in bare graph-theoretic EP decompositions:
many loops may be coupled to the same set of reservoirs and thus carry the same affinity, while many others may carry a vanishing affinity---for CRN these latter are those corresponding to stoichiometric cycles of the closed network, $\set{\bm c^{\alpha}}$.
For generic networks, a systematic way of identifying these so-called \emph{symmetries} was derived in Ref.~\cite{polettini16}, whereas in Ref.~\cite{rao18:shape} they are used to formulate generic thermodynamic---rather than mere graph-theoretic---EP decompositions.

\begin{example}
	The emergent cycle affinity corresponding to the emergent stoichiometric cycle \eqref{eq:cycleMM} reads
	\begin{equation}
		\mathcal{A} = \mu_{\ce{B_{\mathrm{e}}}} - \mu_{\ce{A_{\mathrm{e}}}} = \mathcal{F}_{\ce{B_{\mathrm{e}}}} \, .
		\label{eq:affinityMM}
	\end{equation}
	The contributions to the corresponding cycle current follows from Eq.~\eqref{eq:zeta},
	\begin{equation}
		\bm \zeta = 
		\kbordermatrix{
			& \greyt{+1} & \greyt{+2} & \greyt{+3} & \greyt{+4} & \greyt{+a} & \greyt{+b} \\
			& 0 & 0 & 0 & 0 & 0 & -1
		} \, .
		\label{}
	\end{equation}
	The entries corresponding to the backward reactions are minus those of the forward.
	Notice that, since the CRN has exactly one emergent cycle, the force and cycle EP decompositions are identical, see Eq.~\eqref{eq:effecticeEmergentAffinity}.
\end{example}

\section{Semigrand Gibbs Potential}
\label{sec:dbn}

We here further elaborate on equilibrium distributions and semigrand Gibbs potentials by addressing three points:
\emph{(i)} the relationship between Eq.~\eqref{eq:peq(n)Open}, and the equilibrium distributions as expressed in chemical reaction network theory;
\emph{(ii)} the role of conservation laws for characterizing the dissipation of CRNs subject to sequential introduction of exchanged species;
\emph{(iii)} the gauge freedom intrinsic to the definition of driving work.
This section can be skipped at a first read.

\subsection{Equilibrium Distributions in\\Chemical Reaction Network Theory}
\label{sec:}

In Ref.~\cite{anderson10} (see also \cite{heuett06}) equilibrium distributions of CRNs are proven to be multi-Poissonian
\begin{equation}
	\hspace{-2em}
	p_{\mathrm{eq}} ( \bm {n} | \st{L_{\lambda_{\mathrm{u}}}} ) =
	\frac{\exp\left\{ \bm n \cdot \ln \left\{ [\mathbf{z}]_{\mathrm{eq}} V \right\} \right\}}{\bm n! \, \mathcal{Z}( \st{L_{\lambda_{\mathrm{u}}}} )} 
	{\textstyle\prod_{\lambda_{\mathrm{u}}}} \delta\big[L^{\lambda_{\mathrm{u}}}_{\bm n},L_{\lambda_{\mathrm{u}}}\big] \, ,
	\label{eq:peqACK} 
\end{equation}
where $[\mathbf{z}]_{\mathrm{eq}}$ is the equilibrium concentration distribution of the same CRN described by a set of deterministic rate equations.
$\mathcal{Z}( \st{L_{\lambda_{\mathrm{u}}}} )$ is again a normalizing factor.
To highlight the relationship between this equation and Eqs.~\eqref{eq:peq(n)Open} and \eqref{eq:eqSeifert}, we need to recall that, for deterministic CRNs, thermodynamic equilibrium is defined by the fact that chemical potential differences along all reactions vanish, see Eqs.~\eqref{eq:eqCP} and \eqref{eq:deltaCP}.
As observed in Ref.~\cite{rao16:crnThermo}, this entails that
\begin{equation}
	\bm \mu^{\mathrm{eq}} = {\textstyle\sum_{\lambda}} f_{\lambda} \bm \ell_{\lambda} \, ,
	\label{eq:mueq}
\end{equation}
where $\st{f_{\lambda}}$ are real coefficients depending on $\bm \mu_{\Y}$ and $\st{L_{\lambda_{\mathrm{u}}}}$.
Those related to the broken components, $\st{f_{\lambda_{\mathrm{b}}}}$, are indeed those appearing in Eq.~\eqref{eq:f}.
Using the expression of chemical potential valid in the thermodynamic limit, Eq.~\eqref{eq:chemPot}, we therefore have
\begin{equation}
	\hspace{-1em}
	\ln \left\{ [\mathbf{z}]_{\mathrm{eq}} V \right\} =
	- \beta \left( \bm \mu^\circ - \kt \ln n_{\mathrm{s}} - {\textstyle\sum_{\lambda}} f_{\lambda} \bm \ell_{\lambda} \right) \, ,
	\label{}
\end{equation}
from which
\begin{equation}
	\hspace{-1em}
	\begin{split}
		\bm n \cdot \ln \left\{ [\mathbf{z}]_{\mathrm{eq}} V \right\} - \ln \bm n! & = 
		- \beta \left( g_{\bm n} - \bm \mu^{\mathrm{eq}} \cdot \bm n \right) \\
		& = - \beta \big( \semigp_{\bm n} - {\textstyle\sum_{\lambda_{\mathrm{u}}}} f_{\lambda_{\mathrm{u}}} L^{\lambda_{\mathrm{u}}}_{\bm n} \big)
	\end{split}
	\label{}
\end{equation}
ensues.
At this point, Eqs.~\eqref{eq:eqSeifert}, \eqref{eq:peq(n|L)Open}, and \eqref{eq:peqACK} appear identical up to ${\textstyle\sum_{\lambda_{\mathrm{u}}}} f_{\lambda_{\mathrm{u}}} L^{\lambda_{\mathrm{u}}}_{\bm n}$.
However, since this term involves only the unbroken components it vanishes in Eq.~\eqref{eq:peqACK}.
This shows the connection between the CRN theoretical and thermodynamic expression of equilibrium distributions.

\begin{figure}[t]
	\centering
	\includegraphics[width=.45\textwidth]{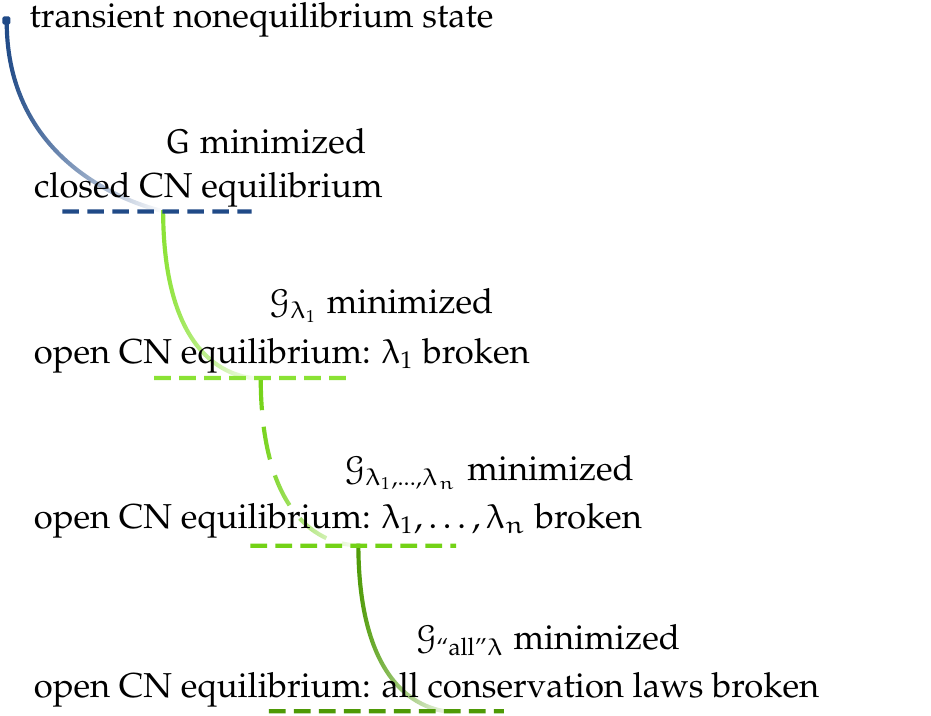}
	\caption{
		Pictorial representation of the hierarchy of equilibrium states and the semigrand Gibbs free energy drops following the relaxation to equilibrium when conservation laws are broken.}
	\label{fig:relaxation}
\end{figure}

\begin{widetext}
\subsection{Hierarchies of Equilibriums}
\label{sec:hierarchies}

We here show that when starting from a closed CRN, a sequential introduction of exchange reactions that keep the CRN detailed balanced drives it down in semigrand Gibbs potential by equilibrating previously constrained degrees of freedom: the conservation laws, see Fig.~\ref{fig:relaxation}.
Let us imagine a closed CRN whose initial probability distribution is $p_{\bm n}(0) = {\textstyle\sum_{\st{L_{\lambda}}}} \, p_{0}(\bm {n}|\st{L_{\lambda}}) \, P_{0}(\st{L_{\lambda}})$, where $P_{0}(\st{L_{\lambda}}) = \prod_{\lambda} P^{\lambda}_{0}(L_{\lambda})$, \emph{i.e.} different components are independently distributed.
As it relaxes to equilibrium, $P_{0}(\st{L_{\lambda}})$ will not change, while $p_{0}(\bm {n}|\st{L_{\lambda}})$ will relax to Eq.~\eqref{eq:peq(n|L)}.
The average dissipation is
\begin{equation}
	T\ave{\Sigma}
	= - \Delta \ave{G}
	= {\textstyle\sum_{\st{L_{\lambda}}}} \, P_0(\st{L_{\lambda}}) \, \left[ \kt \, {\textstyle\sum_{\bm {n}}} p(\bm {n}|\st{L_{\lambda}}) \ln \frac{p(\bm {n}|\st{L_{\lambda}})}{p_{\mathrm{eq}}(\bm {n}|\st{L_{\lambda}})} \right]
	\equiv {\textstyle\sum_{\st{L_{\lambda}}}} \, P_0(\st{L_{\lambda}}) \, \big[ - \Delta \ave{G(\st{L_{\lambda}})} \big] \, .
	\label{}
\end{equation}
This expression is obtained when combining the properties of the Gibbs potential, Eq.~\eqref{eq:delta<G>}, with the equilibrium distribution of closed CRNs, Eq.~\eqref{eq:peq(n)}.
It shows that the average drop of Gibbs free energy can be expressed as the weighted average of the drops of Gibbs free energy at given components, $- \Delta \ave{G(\st{L_{\lambda}})}$.

We now open the CRN by chemostatting one species.
Hence, one conservation law is broken, \emph{e.g.} the total mass $\bm \ell_{\lambda_{1}}$, and the CRN relaxes to a new equilibrium, Eq.~\eqref{eq:peq(n)Open}, whose partition function is denoted by $\mathcal{Z}_{\lambda_{1}}$, Eq.~\eqref{eq:Zopen}.
Clearly, $P^{\lambda}_{0}(L_{\lambda})$, for $\lambda \neq \lambda_{1}$, will not change during the relaxation, and we can rewrite the new equilibrium as
\begin{equation}
	\hspace{-.4em}
	p_{\mathrm{eq}}^{(\lambda_{1})}(\bm n) = 
	\frac{\exp \left\{ -\beta g_{\bm n} + \beta f_{\lambda_{1}} L^{\lambda_{1}}_{\bm n} \right\}}{\mathcal{Z}_{\lambda_{1}}(\st{L^{\lambda}_{\bm n}}_{\lambda\neq\lambda_{1}})} \, \prod_{\lambda \neq \lambda_{1}} P_{0}^{\lambda} (L^{\lambda}_{\bm n})
	=
	\underbrace{\frac{\exp\left\{ -\beta g_{\bm n} \right\}}{{Z}(\st{L^{\lambda}_{\bm n}})}}_{\textstyle = p_{\mathrm{eq}}(\bm n|\st{L^{\lambda}_{\bm n}})} \,
	\underbrace{\frac{Z(\st{L^{\lambda}_{\bm n}}) \, \exp\left\{ \beta f_{\lambda_{1}} L^{\lambda_{1}}_{\bm n} \right\}}{\mathcal{Z}_{\lambda_{1}}(\st{L^{\lambda}_{\bm n}}_{\lambda\neq\lambda_{1}})}}_{\textstyle =: P_{\mathrm{eq}}(L^{\lambda_{1}}_{\bm n}| \st{L^{\lambda}_{\bm n}}_{\lambda\neq\lambda_{1}})}
	\, \prod_{\lambda \neq \lambda_{1}} P_{0}^{\lambda} (L^{\lambda}_{\bm n}) \, .
	\label{eq:peqCG1}
\end{equation}
The first term is the equilibrium distribution of the closed CRN, while the second can be interpreted as the equilibrium distribution of the broken component, for given unbroken component.
In other words, the final equilibrium can be understood as a closed CRN equilibrium with an equilibrium probability distribution over the broken component.
Hence, the average amount of semigrand Gibbs free energy, $\Semigp_{\lambda_{1}}(\bm n) = G(\bm n) - f_{\lambda_{1}} L^{\lambda_{1}}_{\bm n}$, dissipated during the relaxation can be written as
\begin{equation}
	- \Delta \ave{\Semigp_{\lambda_{1}}} =
	\kt \, {\sum_{\bm n}} p_{\mathrm{eq}}(\bm n|\st{L^{\lambda}_{\bm n}}) \,
	{\textstyle\prod_{\lambda}} P^{\lambda}_{0}(L^{\lambda}_{\bm n}) \,
	\ln \frac{P^{\lambda_{1}}_{0}(L^{\lambda_{1}}_{\bm n})}{P_{\mathrm{eq}}(L^{\lambda_{1}}_{\bm n}| \st{L^{\lambda}_{\bm n}}_{\lambda\neq\lambda_{1}})} \, ,
	\label{}
\end{equation}
upon application of Eqs.~\eqref{eq:Gt-Gteq=D} with the distributions \eqref{eq:peq(n)} and \eqref{eq:peqCG1}.
When rewriting this expression as a sum over all values of the components and performing the summation over the states of $p_{\mathrm{eq}}(\bm n|\st{L^{\lambda}})$ we finally obtain
\begin{equation}
	\begin{split}
		- \Delta \ave{\Semigp_{\lambda_{1}}}
		& = {\sum_{\st{L_{\lambda}}_{\lambda \neq \lambda_{1}}}} \, P^{\lambda}_{0}(L_{\lambda}) \,
		\left[ {\sum_{L_{\lambda_{1}}}} \, P^{\lambda_{1}}_{0}(L_{\lambda_{1}}) \kt \ln \frac{P^{\lambda_{1}}_{0}(L_{\lambda_{1}})}{P_{\mathrm{eq}}(L_{\lambda_{1}}|\st{L_{\lambda}}_{\lambda\neq\lambda_{1}})} \right] \\
		& = {\sum_{\st{L_{\lambda}}_{\lambda \neq \lambda_{1}}}} \, P^{\lambda}_{0}(L_{\lambda}) \left[ - \Delta \ave{\Semigp_{\lambda_{1}}(\st{L_{\lambda}}_{\lambda\neq\lambda_{1}})} \right] \, .
	\end{split}
	\label{}
\end{equation}
\end{widetext}
In the first line we recognize the relative entropy between the initial probability of the broken component, $P^{\lambda_{1}}_{0}(L_{\lambda_{1}})$, and the equilibrium one, ${P_{\mathrm{eq}}(L_{\lambda_{1}}|\st{L_{\lambda}}_{\lambda\neq\lambda_{1}})}$.
It it is equal to the difference of semigrand Gibbs free energy at given component, as highlighted in the second line.
We thus see that the dissipation following the relaxation from one equilibrium to the other is completely characterized by the equilibration of the initially constrained degrees of freedom.

This procedure can of course be repeated when a further species is chemostatted and it breaks another conservation law.
The dissipation is quantified by a difference of semigrand Gibbs free energy, which accounts for the relaxation of the degree of freedom which has been released.
When the chemostatting breaks all conservation laws without generating fundamental forces, the CRN finally reaches the global minimum of available semigrand Gibbs free energy, Fig.~\ref{fig:relaxation}.
In this case, the potential becomes the grand potential used in Ref.~\cite{schmiedl07} and discussed in \S~\ref{sec:frCW}, \emph{cf.} Eqs.~\eqref{eq:GrandSeifert}, \eqref{eq:semigpL}, \eqref{eq:Semigp}, and \eqref{eq:mueq}.

\subsection{$W_{\mathrm{d}}$--$\Semigp$ Gauge}
\label{sec:dwGauge}

The driving work and the stochastic semigrand Gibbs potential are defined up to a gauge---distinct from that involving $G$ and $W_{\mathrm{c}}$---, which corresponds to the choice of the components.
Let us consider a basis change in the space of conservation laws
\begin{equation}
	\bm \ell_{\lambda}
	\rightarrow
	\bm \ell'_{\lambda}
	= {\textstyle\sum_{\lambda'}} \Omega_{\lambda\lambda'} \bm \ell_{\lambda'} \, ,
	\label{eq:basisChange}
\end{equation}
with $\Omega_{\lambda_{\mathrm{u}}\lambda_{\mathrm{b}}} = 0$ for all $\lambda_{\mathrm{u}}, \lambda_{\mathrm{b}}$, so that the unbroken ones preserve their properties.
Accordingly, the conjugated intensive variables transform as
\begin{equation}
	f_{\lambda}
	\rightarrow
	f'_{\lambda}
	= {\textstyle\sum_{\lambda'}} f_{\lambda'} \overline{\Omega}_{\lambda' \lambda} \, ,
\end{equation}
see Eq.~\eqref{eq:mueq}, where $\overline{\Omega}$ denotes the inverse of $\Omega$.
We now notice that when the sum involves only the broken conservation laws, such a bilinear form becomes
\begin{equation}
	{\textstyle\sum_{\lambda_{\mathrm{b}}}} f_{\lambda_{\mathrm{b}}} \bm \ell_{\lambda_{\mathrm{b}}}
	\rightarrow 
	{\textstyle\sum_{\lambda_{\mathrm{b}}}} f_{\lambda_{\mathrm{b}}} \bm \ell_{\lambda_{\mathrm{b}}}
	-
	{\textstyle\sum_{\lambda_{\mathrm{u}}}} \mathfrak{f}_{\lambda_{\mathrm{u}}} \bm \ell_{\lambda_{\mathrm{u}}} \, ,
	\label{}
\end{equation}
where 
\begin{equation}
	\mathfrak{f}_{\lambda_{\mathrm{u}}}
	:=
	\sum_{\lambda_{\mathrm{u}}'\lambda_{\mathrm{b}}'}
	f_{\lambda_{\mathrm{b}}'}
	\overline{\Omega}_{\lambda_{\mathrm{b}}'\lambda_{\mathrm{u}}'}
	\Omega_{\lambda_{\mathrm{u}}'\lambda_{\mathrm{u}}}
	\, .
	\label{eq:gaugeTerm}
\end{equation}
Therefore, the instantaneous driving work rate (the integrand of Eq.~\eqref{eq:dw} rewritten with Eq.~\eqref{eq:f}), and the semigrand potential, become
\begin{equation}
	\dot{W}_{\mathrm{d}}(\bm n) \rightarrow \dot{W}_{\mathrm{d}}(\bm n) + {\textstyle\sum_{\lambda_{\mathrm{u}}}} \partial_{t} \mathfrak{f}_{\lambda_{\mathrm{u}}} L^{\lambda_{\mathrm{u}}}_{\bm n} \, ,
	\label{eq:dwGauge}
\end{equation}
and
\begin{equation}
	\Semigp(\bm n) \rightarrow \Semigp(\bm n) + {\textstyle\sum_{\lambda_{\mathrm{u}}}} \mathfrak{f}_{\lambda_{\mathrm{u}}} L^{\lambda_{\mathrm{u}}}_{\bm n} \, ,
	\label{eq:SemigpGauge}
\end{equation}
respectively.
In contrast, the nonconservative forces---and thus the nonconservative work---is left invariant
\begin{equation}
	\bm{\mathcal{F}}_\Yf
	\rightarrow 
	\bm{\mathcal{F}}_\Yf
	+ 
	{\textstyle\sum_{\lambda_{\mathrm{u}}}} \mathfrak{f}_{\lambda_{\mathrm{u}}} \bm \ell_{\lambda_{\mathrm{u}}}^{\yfrc}
	=
	\bm{\mathcal{F}}_\Yf
	\, ,
	\label{}
\end{equation}
since $\bm \ell_{\lambda_{\mathrm{u}}}^{\yfrc} = \bm 0$.
Crucially, the gauge terms in $W_{\mathrm{d}}$ and $-\Delta \Semigp$ cancel and the EP is unaltered.
After all, the physical process is not modified.
Notice also that since the gauge term is nonfluctuating, it vanishes for cyclic protocols when integrated over a period.

We thus conclude that driving work and semigrand Gibbs potential are not univocally defined as they are affected by a gauge freedom.
The gauge affecting the potential--work connection in stochastic thermodynamics led to debates, see Ref.~\cite{campisi11} and references therein.
As observed in the latter reference, the problem is rooted in what can be experimentally measured as work, as different experimental set-ups entail different gauge choices.
In our chemical framework, different choices of the broken components, involve expressions of the work in which different species appear and whose abundances need to be measured to estimate the work.

\begin{example}
	To illustrate the potential--work gauge we use the CRN in Fig.~\ref{fig:openMM}.
	Let us consider the transformation of the set conservation laws, Eq.~\eqref{eq:clMM}, identified by the matrix
	\begin{equation}
		\Omega =
		\begin{pmatrix}
			1 & -1 \\
			0 & 1
		\end{pmatrix}
		\, ,
		\label{}
	\end{equation}
	according to which the conservation laws become
	\begin{subequations}
		\begin{align}
			\bm \ell'_{\ce{E}} =
			\bm \ell_{\ce{E}} & =
			\kbordermatrix{
				& \greyt{\ce{E}} & \greyt{\ce{E^{\ast}}} & \greyt{\ce{E^{\ast\ast}}} & \greyt{\ce{A}} & \greyt{\ce{B}} \\
				& 1 & 1 & 1 & 0 & 0
			} \, , \label{eq:clMMuGauge} \\
			\bm \ell'_{\mathrm{b}} =
			\bm \ell_{\mathrm{b}} - \bm \ell_{\ce{E}} & =
			\kbordermatrix{
				& \greyt{\ce{E}} & \greyt{\ce{E^{\ast}}} & \greyt{\ce{E^{\ast\ast}}} & \greyt{\ce{A}} & \greyt{\ce{B}} \\
				& -1 & 0 & 0 & 1 & 1
			} \, .  \label{eq:clMMbGauge}
		\end{align}
		\label{eq:clMMGauge}
	\end{subequations}
	Using Eqs.~\eqref{eq:f=muMM}, the gauge term reads
	\begin{equation}
		\mathfrak{f}_{\lambda_{\mathrm{u}}}(\pi_{t})
		= \mu_{\ce{A}}(\pi_{t})
		\label{eq:gaugeTermEx}
	\end{equation}
	from which we can easily derive the expression for the new driving work rate
	\begin{equation}
		\dot{W}_{\mathrm{d}}(\bm n)
		= (n_{\ce{E}} - n_{\ce{A}} - n_{\ce{B}}) \partial_{t} \mu_{\ce{A}} \, .
		\label{eq:MMdcwGauge}
	\end{equation}
	The semigrand Gibbs free energy easily follows.
	We can now highlight the difference between the two definitions of driving work, Eqs.~\eqref{eq:MMdcw} and \eqref{eq:MMdcwGauge}:
	while the first entails the measurement of the population of $\ce{A}$, $\ce{B}$, and of the activated complexes $\ce{E^{\ast}}$ and $\ce{E^{\ast\ast}}$, the latter entails that of $\ce{A}$, $\ce{B}$, and of the free enzyme $\ce{E}$.
	The values of the two expressions will differ except for cyclic protocols integrated over a period.
	\qed
\end{example}

\section{Fluctuation Theorems}
\label{sec:fr}

\begin{figure}[t]
	\centering
	\includegraphics[width=.48\textwidth]{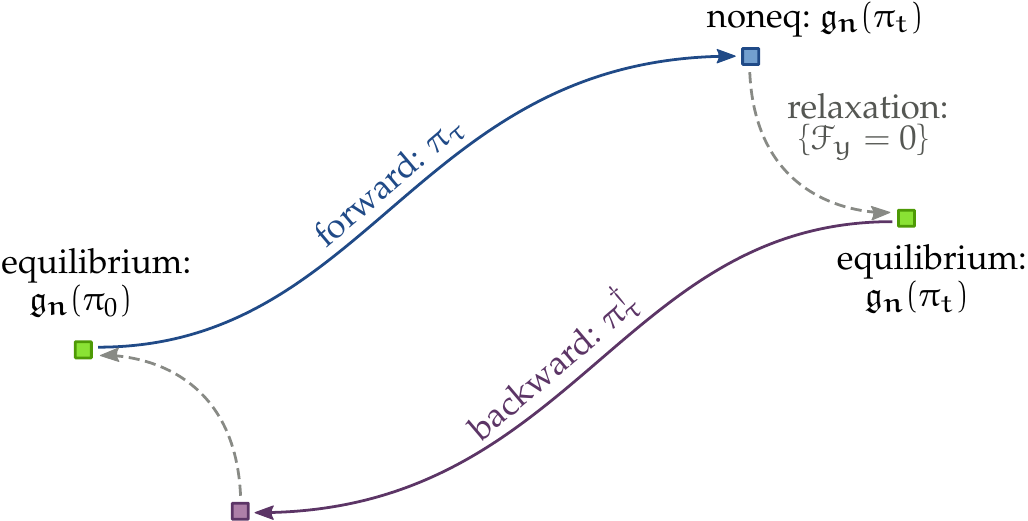}
	\caption{
		Schematic representation of the forward and backward processes.
		The relaxation to the equilibrium obtained by shutting down the driving and turning off the forces at time $t$ (resp. $0$) for the forward (resp. backward) process, merely relates the two processes but it is irrelevant for the FT.
	}
	\label{fig:fr}
\end{figure}

We now proceed to show that the driving work and the nonconservative chemical work satisfy a finite-time detailed FT.
The FT holds for any process, referred to as \emph{forward}, if the open CRN is initially prepared at equilibrium, Eq.~\eqref{eq:peq(n)Open}.
For the sake of simplicity, and without loss of generality, we assume that the initial distribution of unbroken components is $P(\st{L^{\lambda_{\mathrm{u}}}_{\bm n}}) = \prod_{\lambda_{\mathrm{u}}} \delta\big[L^{\lambda_{\mathrm{u}}}_{\bm n},L_{\lambda_{\mathrm{u}}}\big]$.
Let $\pi_{0}$ be the initial value of the protocol, which corresponds to equilibrium ruled by $\semigp(\pi_{0})$.
At time $0$, the driving is activated and the CRN evolves controlled by the protocol $\pi_{\tau}$, for $\tau \in [0,t]$.
The corresponding \emph{backward} process is again initially prepared at the equilibrium---where $\bm{\mathcal{F}}_{\Yf} = \bm 0$---, but the chemical potentials $\bm \mu_{\Yp}$ must have the same value they have at time $t$ in the forward process.
This guarantees that the equilibrium distribution is ruled by $\semigp_{\bm n}(\pi_{t})$.
The backward process is driven by the time-reversed protocol, $\pi^{\dagger}_{\tau} := \pi_{t-\tau}$, for $\tau \in [0,t]$ (Fig.~\ref{fig:fr}).

The \emph{finite-time detailed FT} establishes the relationship between the forward and backward process
\begin{equation}
	\hspace{-2.1em}
	\frac{\mathcal{P}_{t}(W_{\mathrm{d}}, \st{W^{\mathrm{nc}}_{\iyf}})}
	{\mathcal{P}_{t}^{\dagger}(-W_{\mathrm{d}}, \st{-W^{\mathrm{nc}}_{\iyf}})}
	= \exp\left\{ \beta \big( W_{\mathrm{d}} + {\textstyle\sum_{\iyf}} W^{\mathrm{nc}}_{\iyf} - \Delta \Semigp_{\mathrm{eq}} \big) \right\} \, ,
	\label{eq:dftAwesome}
\end{equation}
where $\mathcal{P}_{t}(W_{\mathrm{d}},\st{W^{\mathrm{nc}}_{\iyf}})$ is the probability of observing $W_{\mathrm{d}}$ driving work and $\st{W^{\mathrm{nc}}_{\iyf}}$ nonconservative contributions along the forward process, Eqs.~\eqref{eq:dw} and \eqref{eq:pCW}.
Instead, $\mathcal{P}^{\dagger}_{t}(-W_{\mathrm{d}},\st{-W^{\mathrm{nc}}_{\iyf}})$ is the probability of observing $-W_{\mathrm{d}}$ driving work and $\st{-W^{\mathrm{nc}}_{\iyf}}$ nonconservative contributions along the backward process.
Finally,
\begin{equation}
	\Delta \Semigp_{\mathrm{eq}} = - \kt \ln \frac{\mathcal{Z}(\pi_{t},\st{L_{\lambda_{\mathrm{u}}}})}{\mathcal{Z}(\pi_{0},\st{L_{\lambda_{\mathrm{u}}}})} \, ,
	\label{eq:deltatGeq}
\end{equation}
is the difference of equilibrium semigrand Gibbs potential between the backward and forward initial equilibrium states.
When integrating this expression over all possible values of $W_{\mathrm{d}}$ and $\st{W^{\mathrm{nc}}_{\iyf}}$ we recover a Jarzynski-like integral FT
\begin{equation}
	\hspace{-2.1em}
	\ave{\exp\left\{ -\beta \big( W_{\mathrm{d}} + {\textstyle\sum_{\iyf}} W^{\mathrm{nc}}_{\iyf} \big) \right\}}
	= \exp\left\{ - \beta \Delta \Semigp_{\mathrm{eq}} \right\}  \, .
	\label{eq:jarzyAwesome}
\end{equation}
We emphasize that in contrast to the FT for the chemical work discussed in the first part of \S~\ref{sec:frCW}, the driving and nonconservative work contributions require that the process starts from the equilibrium state ruled by $\Semigp$, which is that of open CRNs.
As a consequence, there is no break of conservation laws happening during the process, and $\Semigp_{\mathrm{eq}}$ is nonfluctuating.
The proof of the FT \eqref{eq:dftAwesome} is given in App.~\ref{sec:DFTproofs}, and it hinges on the generating function techniques presented in Ref.~\cite{rao18:shape}.

We now discuss some special yet interesting cases of the FT \eqref{eq:dftAwesome}.
In unconditionally detailed-balance CRNs, the nonconservative work vanishes and we obtain
\begin{equation}
	\frac{\mathcal{P}_{t}(W_{\mathrm{d}})}
	{\mathcal{P}_{t}^{\dagger}(-W_{\mathrm{d}})}
	= \exp\left\{ \beta \big( W_{\mathrm{d}} - \Delta \Semigp_{\mathrm{eq}} \big) \right\} \, .
	\label{eq:dftAwesomeUDB}
\end{equation}
This is the analogue of Crooks' FT for CRNs \cite{crooks98,crooks99}, since solely the work due to external manipulations is involved.
In contrast, for autonomous processes, the driving chemical work vanishes and the FT can be formulated as
\begin{equation}
	\frac{\mathcal{P}_{t}(\st{\mathcal{I}_{\iyf}})}{\mathcal{P}_{t}(\st{-\mathcal{I}_{\iyf}})}
	= \exp\left\{ \beta {\textstyle\sum_{\iyf}} \mathcal{F}_{\iyf} \mathcal{I}_{\iyf} \right\} \, ,
	\label{eq:dftAwesomeAuto}
\end{equation}
which evidences the symmetry that the fluctuations of the fundamental currents (see Eq.~\eqref{eq:extCrnt}) satisfy.

The FT in Eq.~\eqref{eq:dftAwesome} is inspired by an analogous result derived in Refs.~\cite{bulnescuetara14,rao18:shape} in the context of generic Markov jump processes.
It is a major result of this paper and its importance is manifold.
It holds for processes of finite duration $t$, and it is expressed in terms of measurable chemical quantities.
Its only constraint is the initial state, which must be equilibrium.
It reveals the most appropriate boundary conditions under which Jarzynski--Crooks-like FTs can be formulated for CRNs:
equilibrium distribution of open CRNs.
Most important, it evidences the merits of our stoichiometric approach based on the identification of conservation laws:
it allowed us to characterize the potential describing the equilibrium distribution of open CRNs, and to formulate the decomposition of the EP which supports our FTs, Eq.~\eqref{eq:epAwesome}.

\paragraph*{Remark}
A physical interpretation of the argument of the exponential in Eq.~\eqref{eq:dftAwesome}, follows from the following observation:
if, at time $t$, the driving is stopped and the fundamental forces \eqref{eq:fundForces} turned off---\emph{viz.} set to zero by an appropriate choice of $\bm \mu_\Yf$: $\bm \mu^{\ast}_\Yf := \bm \mu_\Yp \cdot {\textstyle\sum_{\lambda_{\mathrm{b}}}} \overline{\bm \ell}^\yrf_{\lambda_{\mathrm{b}}} \bm \ell^\yfrc_{\lambda_{\mathrm{b}}}$---the CRN relaxes to the initial condition of the backward process.
During the relaxation neither $W_{\mathrm{d}}$ nor $\st{W^{\mathrm{nc}}_{\iyf}}$ are performed and the related EP is $T \Sigma_{\mathrm{relax}} = \Semigp(\bm n,\pi_{t}) + \kt \ln \mathcal{Z}(\pi_{t},\st{L_{\lambda_{\mathrm{u}}}})$.
The argument of the exponential can thus be interpreted as the EP of the fictitious combined process ``forward process + relaxation to the final equilibrium''.

\paragraph*{Remark}
For autonomous CRNs and arbitrary initial conditions, the steady-state FT follows
\begin{equation}
	\frac{\mathcal{P}(\st{\dot{\mathcal{I}}_{\iyf}})}{\mathcal{P}(\st{-\dot{\mathcal{I}}_{\iyf}})}
	\overset{t\rightarrow \infty}{=}
	\exp\left\{ t \beta {\textstyle\sum_{\iyf}} \mathcal{F}_{\iyf} \dot{\mathcal{I}}_{\iyf} \right\} \, ,
	\label{eq:dftAwesomeSS}
\end{equation}
where $\mathcal{P}(\st{\dot{\mathcal{I}}_{\iyf}})$ is the probability of observing average rates of fundamental external currents $\big\{\frac{1}{t} \int_{0}^{t} \de \tau \, I_{\iyf}(\tau)\big\}$ equal to $\st{\dot{\mathcal{I}}_{\iyf}}$.
Eq.~\eqref{eq:dftAwesomeSS} can be proved using the large deviation technique used in Ref.~\cite{gaspard04} in combination with the local detailed balance \eqref{eq:ldbAwesome}.

\subsection*{FT along Stoichiometric Cycles}
An alternative yet equivalent formulation of the FT \eqref{eq:dftAwesome} is that given in terms of nonconservative contributions along emergent stoichiometric cycles, Eq.~\eqref{eq:gamma}:
\begin{equation}
	\hspace{-1.8em}
	\frac{\mathcal{P}_{t}(W_{\mathrm{d}},\st{\Gamma_{\ecy}})}
	{\mathcal{P}_{t}^{\dagger}(-W_{\mathrm{d}},\st{-\Gamma_{\ecy}})}
	= \exp\left\{ \beta \big( W_{\mathrm{d}} + {\textstyle\sum_{\ecy}} \Gamma_{\ecy} - \Delta \Semigp_{\mathrm{eq}} \big) \right\} \, ,
	\label{eq:dftSuperAwesome}
\end{equation}
where $\mathcal{P}_{t}(W_{\mathrm{d}},\st{\Gamma_{\ecy}})$ is the probability of observing $W_{\mathrm{d}}$ driving work and $\st{\Gamma_{\ecy}}$ nonconservative contributions along the forward process.
We discuss its proof App.~\ref{sec:DFTproofs}.

\paragraph*{Remark}
As for the fundamental currents, the local detailed balance \eqref{eq:ldbSuperAwesome} can be used to prove a steady-state FT for currents along emergent stoichiometric cycles
\begin{equation}
	\frac{\mathcal{P}(\st{\dot{\mathcal{J}}_{\eta}})}{\mathcal{P}(\st{-\dot{\mathcal{J}}_{\eta}})}
	\overset{t\rightarrow \infty}{=}
	\exp\left\{ t \beta {\textstyle\sum_{\eta}} \mathcal{A}_{\eta} \dot{\mathcal{J}}_{\eta} \right\} \, ,
	\label{eq:dftSuperAwesomeSS}
\end{equation}
which is valid for autonomous CRNs and arbitrary initial conditions.
$\mathcal{P}(\st{\dot{\mathcal{J}}_{\eta}})$ is the probability of observing average rates of emergent cycle currents $\big\{\frac{1}{t} \int_{0}^{t} \de \tau \, {\textstyle\sum_{\rho}} \zeta_{\ecy,\rho} J_{\rho}(\tau) \big\}$ equal to $\st{\dot{\mathcal{J}}_{\eta}}$.
In contrast to the analogous FT obtained in Ref.~\cite{andrieux04}, Eq.~\eqref{eq:dftSuperAwesomeSS} is achieved using a stoichiometric approach based on the identification of stoichiometric cycles.
For this reason, it accounts for the minimal set of nonzero macroscopic affinities.

\section{Ensemble Average Rates Description}
\label{sec:ead}

We now summarize our main results for ensemble average rates and discuss the relaxation to equilibrium of detailed-balanced CRNs.
We also highlight the difference between an approach that does and does not take into account the topology of the CRN.
We do so by recapitulating the procedure to decompose the EP into its fundamental contributions.
We end by formulating a nonequilibrium Landauer's principle.

\subsection{Traditional Description}

\paragraph*{Enthalpy Balance}
The enthalpy balance follows from the time derivative of the average enthalpy, Eq.~\eqref{eq:H},
\begin{equation}
	\dt {\textstyle\sum_{\bm n}} p_{\bm n} (\bm h \cdot \bm n)
	\equiv \dt \avef{H}
	= \avef{\dot{Q}} + \avef{\dot{W}_{\mathrm{c}}} \, .
	\label{eq:aveHbalance}
\end{equation}
It characterizes the average rate of change of enthalpy in the same way Eq.~\eqref{eq:firstLaw} characterizes the enthalpy change along stochastic trajectories.
The average heat flow rate is given by
\begin{equation}
	\avef{\dot{Q}}
	= \avef{\dot{Q}^{\mathrm{thr}}} + \avef{\dot{Q}^{\mathrm{chm}}} \, .
	\label{eq:aveQ}
\end{equation}
The first term quantifies the average rate of heat of reaction,
\begin{equation}
	\avef{\dot{Q}^{\mathrm{thr}}}
	= {\textstyle\sum_{\rho}} \big[ \bm h \cdot \stoich_{\rho} + \bm h_\Y \cdot \stoich^\Y_{\rho} \big] \avef{J_{\rho}} \, ,
	\label{}
\end{equation}
where $\avef{J_{\rho}} = \sum_{\bm n} \rate{\rho}{\bm n} p_{\bm n}$ is the average reaction current.
The second term is the average heat flow in the chemostats,
\begin{equation}
	\avef{\dot{Q}^{\mathrm{chm}}} = T \bm s_\Y \cdot \avef{\bm I^\Y} \, ,
	\label{}
\end{equation}
where $\avef{\bm I^\Y} = \sum_{\rho} (- \stoich^{\Y}_{\rho}) \avef{J_{\rho}}$ are the average external currents, Eq.~\eqref{eq:aveCurrentGivenState}.
Instead, the ensemble average chemical work rate,
\begin{equation}
	\avef{\dot{W}_{\mathrm{c}}} = \bm \mu_\Y \cdot \avef{\bm I^\Y} \, ,
	\label{eq:aveWc}
\end{equation}
quantifies the average rate of exchange of Gibbs free energy with the chemostats.

\paragraph*{Entropy Production Rate}
At the ensemble average level, the second law of thermodynamics manifests itself in the non-negative average EP rate
\begin{equation}
	\hspace{-1.0em}
	\begin{split}
		& \avef{\dot{\Sigma}}
		= \dt \avef{S} - \tfrac{1}{T} \avef{\dot{Q}} \\
		& = \kb \sum_{\bm n,\rho} \rate{\rho}{\bm n} p_{\bm n}
		\ln \frac{\rate{\rho}{\bm n} p_{\bm n}}{\rate{\rho}{\bm n + \stoich_{\rho}} \, p_{\bm n + \stoich_{\rho}}} \ge 0 \, .
	\end{split}
	\label{eq:epr}
\end{equation}
where $\avef{S} = \sum_{\bm n} p_{\bm n} S(\bm n)$, Eq.~\eqref{eq:S}.
Using the expression for the transition affinity, Eq.~\eqref{eq:affinity}, it can be recast into,
\begin{equation}
	T \avef{\dot{\Sigma}}
	= \avef{\dot{W}_{\mathrm{c}}} - \dt \avef{G} \, ,
	\label{eq:eprCW}
\end{equation}
where the chemical work rate and the average Gibbs potential are given in Eqs.~\eqref{eq:aveWc} and \eqref{eq:aveG}, respectively.
Equivalently, Eqs.~\eqref{eq:aveHbalance}, \eqref{eq:epr}, and \eqref{eq:eprCW} can be obtained by directly averaging Eqs.~\eqref{eq:firstLaw}, \eqref{eq:epAffinityTrue}, and \eqref{eq:ep=W-deltaG}, respectively, over all stochastic trajectories.

For closed CRNs, Eq.~\eqref{eq:eprCW} reduces to $\dt \avef{G} = - T \avef{\dot{\Sigma}} \leq 0$.
This relation, together with Eq.~\eqref{eq:delta<G>}, shows that:
\emph{(i)} $\avef{G}$ is a Lyapunov function, and hence that closed CRNs relax to equilibrium, Eq.~\eqref{eq:peq(n)};
\emph{(ii)} $\ave{G} - \avef{G_{\mathrm{eq}}}_{\mathrm{L}} = T \avef{\Sigma}$ is the average dissipation during the relaxation to equilibrium.

\subsection{CRN-specific Description}
\label{sec:aveAwesome}

\paragraph*{Entropy Production Rate}
We now summarize the procedure to recover the EP decomposition \eqref{eq:epAwesome} at the ensemble average level.
\textit{(i)} Identify the broken and unbroken conservation laws, $\st{\bm \ell_{\lambda_{\mathrm{u}}}, \bm \ell_{\lambda_{\mathrm{b}}}}$, \S~\ref{sec:cl}.
\textit{(ii)} Identify a set of $\nof{\lambda_{\mathrm{b}}}$ exchanged species, $\rf$, for which the matrix whose rows are $\st{\bm \ell^\yrf_{\lambda_{\mathrm{b}}}}$ is nonsingular.
The columns of its inverse are denoted by $\st{\overline{\bm \ell}^\yrf_{\lambda}}$.
Physically, each species $\rf$ breaks exactly one conservation law.
The remaining exchanged species form the set denoted by $\frc$.
\textit{(iii)} The \emph{nonequilibrium semigrand Gibbs potential} follows from the average of Eq.~\eqref{eq:Semigp},
\begin{equation}
	\avef{\Semigp} = 
	{\textstyle\sum_{\bm n}} p_{\bm n} \left[ \kt \ln p_{\bm n} + \semigp_{\bm n} \right] \, .
	\label{eq:nonEqSemigp}
\end{equation}
It depends on the vector $\avef{\bm M^\yrf}$ which describes the average population of the combination of moieties whose conservation is broken by the chemostats, \S~\ref{sec:cl} and Eq.~\eqref{eq:moietyV}.
\textit{(iv)} The change in time of $\avef{\semigp}$ due to the time-dependent driving describes the average driving work rate, Eq.~\eqref{eq:dw},
\begin{equation}
	\avef{\dot{W}_{\mathrm{d}}}
	= - \big[ \partial_{t} \bm \mu_\Yp \big] \cdot \avef{\bm M^\yrf} \, .
	\label{eq:aveDW}
\end{equation}
It quantifies the average amount of work spent to change the chemical potentials of the chemostats $\Rf$.
\textit{(v)} The second group of exchanged species, $\frc$, is used to identify the minimal set of fundamental nonconservative forces, $\bm{\mathcal{F}}_\Yf \equiv \st{\mathcal{F}_{\iyf}}$, Eq.~\eqref{eq:fundForces}.
The average nonconservative chemical work rate follows from the product of these forces and their corresponding instantaneous external currents, Eq.~\eqref{eq:instChemoCurr},
\begin{equation}
	\avef{\dot{W}^{\mathrm{nc}}_{\iyf}} := \mathcal{F}_{\iyf} \avef{I_{\iyf}} \, .
	\label{eq:aveFW}
\end{equation}
They quantify the average work per unit time spent to sustain a net current of species $\frc$ across the CRN.
\textit{(vi)} The average EP rate decomposed as in Eq.~\eqref{eq:epAwesome} finally follows from Eqs.~\eqref{eq:nonEqSemigp}--\eqref{eq:aveFW},
\begin{equation}
	T \avef{\dot{\Sigma}}
	= - \dt \avef{\Semigp} + \avef{\dot{W}_{\mathrm{d}}} + {\textstyle\sum_{\iyf}} \avef{\dot{W}^{\mathrm{nc}}_{\iyf}} \, .
	\label{eq:eprAwesome}
\end{equation}
Its three fundamental contributions appear:
a conservative force contribution, a time-dependent driving contribution, a minimal set of nonconservative terms.

For open autonomous detailed-balanced CRNs, $\bm{\mathcal{F}}_\Yf = \bm 0$, $\partial_{t} \bm \mu_{\Yp} = 0$, and hence Eq.~\eqref{eq:eprAwesome} reduces to $\dt \avef{\Semigp} = - T \avef{\dot{\Sigma}} \leq 0$.
Recalling Eq.~\eqref{eq:Gt-Gteq=D}, this relation shows that:
\emph{(i)} $\avef{\Semigp}$ is a Lyapunov function, and hence that these CRNs relax to equilibrium, Eq.~\eqref{eq:peq(n)Open};
\emph{(ii)} $\ave{\Semigp} - \avef{\Semigp_{\mathrm{eq}}}_{\mathrm{L_u}} = T \avef{\Sigma}$ is the average dissipation during the relaxation to equilibrium.

\paragraph*{Enthalpy Balance}
By averaging Eq.~\eqref{eq:Hawesome}, the CRN-specific average enthalpy balance also ensues
\begin{equation}
	\dt \avef{\mathcal{H}} = \avef{\dot{Q}} + \avef{\dot{W}_{\mathrm{d}}} + {\textstyle\sum_{\iyf}} \avef{\dot{W}^{\mathrm{nc}}_{\iyf}} \, ,
	\label{eq:aveHawesome}
\end{equation}
which strengthen the interpretation of $\avef{\dot{W}_{\mathrm{d}}}$ and $\st{\avef{\dot{W}^{\mathrm{nc}}_{\iyf}}}$ as average work rate contributions.

\subsection{Average EP along Stoichiometric Cycles}
The average EP decomposition expressed in terms of emergent cycles currents and affinities can be achieved through an analogous recipe.
\textit{(i)} Identify broken and unbroken conservation laws, $\st{\bm \ell_{\lambda_{\mathrm{u}}}, \bm \ell_{\lambda_{\mathrm{b}}}}$, as well as stoichiometric and emergent stoichiometric cycles, $\st{\bm c^{\cy}, \bm c^{\ecy}}$ \S\S~\ref{sec:cl} and \ref{sec:cycles}.
Steps \textit{(ii)}--\textit{(iv)} as above.
\textit{(v)} Identify the emergent stoichiometric cycles affinities, Eq.~\eqref{eq:emergentA}, as well as their corresponding average currents ${\textstyle\sum_{\rho}} \zeta_{\ecy,\rho} \avef{J_{\rho}}$, Eq.~\eqref{eq:zeta}.
(\textit{vi}) The average EP rate follows from Eqs.~\eqref{eq:nonEqSemigp}, \eqref{eq:aveDW}, and the emergent stoichiometric cycles currents and affinities,
\begin{equation}
	T \avef{\dot{\Sigma}}
	= - \dt \avef{\Semigp} + \avef{\dot{W}_{\mathrm{d}}} + {\textstyle\sum_{\ecy}} \avef{\dot{\Gamma}_{\ecy}} \, ,
	\label{eq:eprSuperAwesome}
\end{equation}
where,
\begin{equation}
	\avef{\dot{\Gamma}_{\ecy}} = \mathcal{A}_{\ecy} {\textstyle\sum_{\rho}} \zeta_{\ecy,\rho} \avef{J_{\rho}} \, ,
	\label{eq:aveGamma}
\end{equation}
as in Eqs.~\eqref{eq:epSuperAwesome} and \eqref{eq:gamma}.

\subsection{Nonequilibrium Landauer's Principle}
\label{sec:lpAwesome}

\begin{figure}[t]
	\centering
	\includegraphics[width=.45\textwidth]{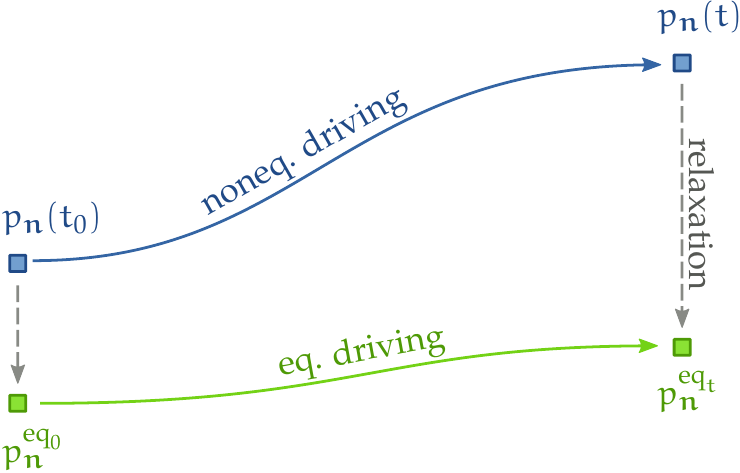}
	\caption{
		Pictorial representation of the transformation between two nonequilibrium probability distributions.
		The nonequilibrium transformation (blue line) is compared with the equilibrium one (green line).
		The latter is obtained by shutting down the driving and turning off the forces at each time (dashed gray lines).
	}
	\label{fig:landauer}
\end{figure}

We can now formulate the nonequilibrium Landauer's principle for the driving and nonconservative work.
We have already seen that when the driving is stopped and all forces are turned off, the CRN relaxes to equilibrium by minimizing the nonequilibrium semigrand Gibbs potential.
Equation~\eqref{eq:Gt-Gteq=D} can be thus combined with Eq.~\eqref{eq:eprAwesome}, and by integrating over time, we obtain
\begin{multline}
	\avef{W_{\mathrm{d}}} + {\textstyle\sum_{\iyf}} \avef{W^{\mathrm{nc}}_{\iyf}} \\
	= \Delta \avef{\mathcal{G}_{\mathrm{eq}}}_{\mathrm{L_u}} + \kt \, \Delta \mathcal{D} ( p \| p_{\mathrm{eq}} ) + T \underbrace{\avef{\Sigma}}_{\ge 0} \, .
	\label{eq:minWorkAwesome}
\end{multline}
This fundamental result shows that the minimal cost for transforming a CRN from an arbitrary nonequilibrium state to another is bounded by a relative entropy difference, as depicted in Fig.~\ref{fig:landauer}.
This entropy is an information-theoretical measure of the dissimilarity between two probability distributions: the actual nonequilibrium one and its corresponding equilibrium, which is used as a reference.
For processes starting at equilibrium, $\kt \Delta \mathcal{D} = \kt \mathcal{D}( p_{\mathrm{f}} \| p_{\mathrm{eq}_{\mathrm{f}}} ) \ge 0$ quantifies the minimal cost of producing the final nonequilibrium state.
In contrast, for processes relaxing to equilibrium, $\kt \Delta \mathcal{D} = - \kt \mathcal{D}( p_{\mathrm{i}} \| p_{\mathrm{eq}_{\mathrm{i}}} ) \le 0$ quantifies the maximum amount of work that can be extracted from the initial nonequilibrium state.
For transformations in absence of nonconservative forces ($\bm{\mathcal{F}}_{\Yf} = \bm{0}$), we obtain the chemical version of the result of Ref.~\cite{esposito11}.
The original Landauer's principle \cite{landauer00} is recovered when considering erasure in a two state system (\textsf{0} and \textsf{1}) with identical energies.
In this process, the initial equilibrium state (system equally likely to be found in \textsf{0} or \textsf{1}) is transformed into a metastable nonequilibrium one (system found with probability one in \textsf{0}) via a cyclic protocol.
The difference of relative entropy is $\Delta \mathcal{D}=\ln 2$, and thus $\avef{W_{\mathrm{d}}} \geq \kt \ln 2$.
Finally, Kelvin's formulation of the second law is recovered for transformation between equilibrium states in absence of nonconservative forces, $\avef{W_{\mathrm{d}}} \ge \Delta \avef{\mathcal{G}_{\mathrm{eq}}}_{\mathrm{L_u}}$.

\paragraph*{Remark}
To obtain the Landauer's principle for $\avef{\dot{W}_{\mathrm{d}}}$ and $\st{\avef{\dot{W}^{\mathrm{nc}}_{\iyf}}}$, the equilibrium states of the open CRN have been used as reference states, see Fig.~\ref{fig:landauer}.
Alternatively, one could use the equilibrium states of the closed CRN, which are obtained by shutting down all exchange reactions.
If one does so and uses Eq.~\eqref{eq:eprCW}, an analogous Landauer's principle for the chemical work can be derived,
\begin{equation}
	\avef{W_{\mathrm{c}}} = \Delta \avef{G_{\mathrm{eq}}}_{\mathrm{L}}	+ \kt \, \Delta \mathcal{D} ( p \| p_{\mathrm{eq}} ) + T \ave{\Sigma} \, .
\end{equation}
The traditional thermodynamic work relation $\avef{W_{\mathrm{c}}} \ge \Delta \avef{G_{\mathrm{eq}}}_{\mathrm{L}}$ is recovered for processes whose initial and final states are equilibrium ones.

\subsection{Connection with Deterministic Descriptions}
\label{sec:connection}

For CRNs with very abundant populations of species, a deterministic dynamical description in terms of nonlinear rate equations is justified.
The corresponding nonequilibrium thermodynamics was analyzed in Ref.~\cite{rao16:crnThermo}, where the counterparts of Eqs.~\eqref{eq:aveHbalance}, \eqref{eq:eprCW}, and \eqref{eq:epAwesomeUDB}, can be found.
Following a procedure similar to that described in this paper, one can also formulate the deterministic analog of the EP decomposition \eqref{eq:eprAwesome}.

One can also recover the deterministic thermodynamic description from the ensemble average one by performing the thermodynamic limit---$\bm n \gg \bm 1$, $V \gg 1$, with $\bm n / V =: [\mathbf{z}]$ finite, see App.~\ref{sec:potentials}---and assuming that $p_{\bm n} \simeq \delta_{\bm n,[\mathbf{z}]V}$, \emph{i.e.} the distribution is very peaked around the population that is solution of the rate equations, $[\mathbf{z}]V$.

We conclude with two remarks.

\paragraph*{Remark}
Not all results valid for stochastic CRNs hold for the deterministic ones.
An example is provided by the \emph{adiabatic--nonadiabatic EP decomposition} introduced in Ref.~\cite{esposito07} for generic stochastic processes:
it is valid for deterministic CRNs only for \emph{complex-balanced} CRNs, see Refs.~\cite{ge16,rao16:crnThermo}.

\paragraph*{Remark}
As briefly mentioned in \S~\ref{sec:ocn}, there is an alternative way of modeling open CRNs in which the exchanged species $\exchanged$ are treated as particle \emph{reservoir} with very large population.
All main results of our paper---\emph{i.e.} the EP decomposition \eqref{eq:epAwesome}, the finite-time detailed FT \eqref{eq:dftAwesome}, and the Landauer's principle \eqref{eq:minWorkAwesome}---still hold.
The only difference lies in the fact that the different definitions of stoichiometric matrices, Eq.~\eqref{eq:SMdirect}, also entail slightly different definitions of broken conservation law.
Besides that, the procedure described in \S~\ref{sec:aveAwesome} can be followed in the same way.

\section{Application}
\label{sec:examples}

\begin{figure}[t]
	\centering
	\includegraphics[width=.40\textwidth]{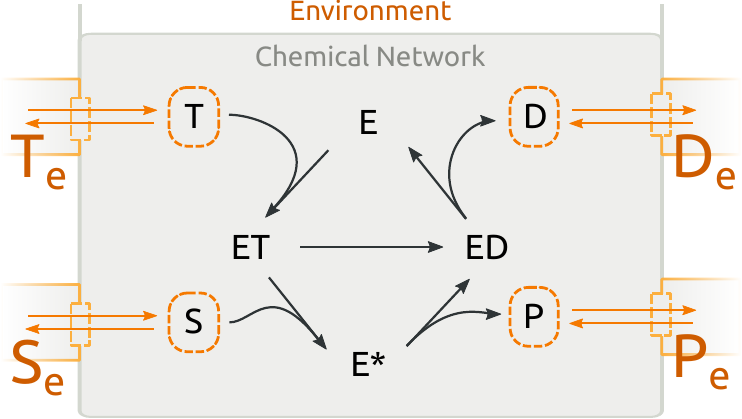}
	\caption{
		Pictorial illustration of the open CRN in Eqs.~\eqref{ce:transduction} and \eqref{ce:transductionExchange}, from which one can see the more clearly the active catalytic mechanism.
	}
	\label{fig:transducer}
\end{figure}

We now illustrate our EP decompositions \eqref{eq:epAwesome} and \eqref{eq:epSuperAwesome} on a CRN displaying more than one fundamental force, which allows us to introduce the phenomenology of free energy transduction.
We consider the following active catalytic mechanism
\begin{equation}
	\begin{gathered}
		\ce{T + E <=>[k_{+1}] ET <=>[k_{+5}] ED <=>[k_{+4}] E + D} \\
		\ce{ET + S <=>[k_{+2}] E^{\ast} <=>[k_{+3}] ED + P} \, .
	\end{gathered}
	\label{ce:transduction}
\end{equation}
It describes the $\ce{T}$-driven catalysis of $\ce{S}$ into $\ce{P}$, having $\ce{D}$ as a byproduct, see Fig.~\ref{fig:transducer}.
All substrates and products are regarded as exchanged species,
\begin{equation}
	\hspace{-1.0em}
	\ce{S <=>[][k_{+\mathrm{s}}] S_{\mathrm{e}}} \, , \; \ce{P <=>[][k_{+\mathrm{p}}] \ce{P}_{\mathrm{e}} } \, , \; \ce{T <=>[][k_{+\mathrm{t}}] \ce{T}_{\mathrm{e}} } \, , \; \ce{D <=>[][k_{+\mathrm{d}}] \ce{D}_{\mathrm{e}} } \, .
	\label{ce:transductionExchange}
\end{equation}
The stoichiometric matrices $\stoichM$ and $\stoichM^{\Y}$ read
\begin{equation}
	\hspace{-2em}
	\kbordermatrix{
		& \greyt{+1} & \greyt{+2} & \greyt{+3} & \greyt{+4} & \greyt{+5} & & \greyt{+\mathrm{s}} & \greyt{+\mathrm{p}} & \greyt{+\mathrm{t}} & \greyt{+\mathrm{d}} \\
		\greyt{\ce{E}} 			& -1 & 0 & 0 & 1 & 0 & 	\omit\vrule & 0 & 0 & 0 & 0 \\
		\greyt{\ce{ET}}			& 1 & -1 & 0 & 0 & -1 & \omit\vrule & 0 & 0 & 0 & 0 \\
		\greyt{\ce{E^{\ast}}} 	& 0 & 1 & -1 & 0 & 0 & 	\omit\vrule & 0 & 0 & 0 & 0 \\
		\greyt{\ce{ED}}			& 0 & 0 & 1 & -1 & 1 & 	\omit\vrule & 0 & 0 & 0 & 0 \\
		\greyt{\ce{S}}			& 0 & -1 & 0 & 0 & 0 & 	\omit\vrule & 1 & 0 & 0 & 0 \\
		\greyt{\ce{P}}			& 0 & 0 & 1 & 0 & 0 & 	\omit\vrule & 0 & 1 & 0 & 0 \\
		\greyt{\ce{T}}			& -1 & 0 & 0 & 0 & 0 & 	\omit\vrule & 0 & 0 & 1 & 0 \\
		\greyt{\ce{D}}			& 0 & 0 & 0 & 1 & 0 & 	\omit\vrule & 0 & 0 & 0 & 1
	} \, ,
	\label{eq:smActiveOpen}
\end{equation}
---in which the stoichiometric matrix of the closed CRN is highlighted---and
\begin{equation}
	\hspace{-2em}
	\kbordermatrix{
		& \greyt{+1} & \greyt{+2} & \greyt{+3} & \greyt{+4} & \greyt{+5} & & \greyt{+\mathrm{s}} & \greyt{+\mathrm{p}} & \greyt{+\mathrm{t}} & \greyt{+\mathrm{d}} \\
		\greyt{\ce{S_{\mathrm{e}}}}	& 0 & 0 & 0 & 0 & 0 & \omit\vrule	& -1 & 0 & 0 & 0 \\
		\greyt{\ce{P_{\mathrm{e}}}}	& 0 & 0 & 0 & 0 & 0 & \omit\vrule	& 0 & -1 & 0 & 0 \\
		\greyt{\ce{T_{\mathrm{e}}}}	& 0 & 0 & 0 & 0 & 0 & \omit\vrule	& 0 & 0 & -1 & 0 \\
		\greyt{\ce{D_{\mathrm{e}}}}	& 0 & 0 & 0 & 0 & 0 & \omit\vrule	& 0 & 0 & 0 & -1
	} \, ,
	\label{eq:smActiveOpenY}
\end{equation}
respectively.

We now follow the procedure described in \S~\ref{sec:ead}, and characterize all terms of Eq.~\eqref{eq:epAwesome}.
\emph{(i)} The closed CRN has three independent conservation laws:
\begin{subequations}
	\begin{align}
		&\hspace{-1em} \bm \ell_{\ce{E}} = 
		\kbordermatrix{
			& \greyt{\ce{E}} & \greyt{\ce{ET}} & \greyt{\ce{E^{\ast}}} & \greyt{\ce{ED}} & \greyt{\ce{S}} & \greyt{\ce{P}} & \greyt{\ce{D}} & \greyt{\ce{T}} & \greyt{\ce{S_{\mathrm{e}}}}	& \greyt{\ce{P_{\mathrm{e}}}}	& \greyt{\ce{T_{\mathrm{e}}}}	& \greyt{\ce{D_{\mathrm{e}}}} \\
			& 1 & 1 & 1 & 1 & 0 & 0 & 0 & 0 & 0 & 0 & 0 & 0
		} \, , \label{eq:clActiveE} \\
		&\hspace{-1em} \bm \ell_{\ce{S}} =
		\kbordermatrix{
			& \greyt{\ce{E}} & \greyt{\ce{ET}} & \greyt{\ce{E^{\ast}}} & \greyt{\ce{ED}} & \greyt{\ce{S}} & \greyt{\ce{P}} & \greyt{\ce{D}} & \greyt{\ce{T}} & \greyt{\ce{S_{\mathrm{e}}}}	& \greyt{\ce{P_{\mathrm{e}}}}	& \greyt{\ce{T_{\mathrm{e}}}}	& \greyt{\ce{D_{\mathrm{e}}}} \\
			& 0 & 0 & 1 & 0 & 1 & 1 & 0 & 0 & 1 & 1 & 0 & 0
		} \, , \label{eq:clActiveS} \\
		&\hspace{-1em} \bm \ell_{\mathrm{T}} =
		\kbordermatrix{
			& \greyt{\ce{E}} & \greyt{\ce{ET}} & \greyt{\ce{E^{\ast}}} & \greyt{\ce{ED}} & \greyt{\ce{S}} & \greyt{\ce{P}} & \greyt{\ce{D}} & \greyt{\ce{T}} & \greyt{\ce{S_{\mathrm{e}}}}	& \greyt{\ce{P_{\mathrm{e}}}}	& \greyt{\ce{T_{\mathrm{e}}}}	& \greyt{\ce{D_{\mathrm{e}}}} \\
			& 0 & 1 & 1 & 1 & 0 & 0 & 1 & 1 & 0 & 0 & 1 & 1
		} \, . \label{eq:clActiveT}
	\end{align}
	\label{eq:clActive}
\end{subequations}
The first corresponds to the enzyme moiety and it is unbroken in the open CRN.
In contrast, the last two correspond to the moieties $\ce{S}$--$\ce{P}$ and $\ce{T}$--$\ce{D}$, which are broken in the open CRN.
\emph{(ii)} We choose $\ce{S_{\mathrm{e}}}$ and $\ce{T_{\mathrm{e}}}$ as chemostatted species $\Rf$, since the entries of $\bm \ell_{\mathrm{S}}$ and $\bm \ell_{\mathrm{T}}$ corresponding to these species identify a nonsingular matrix---it is an identity matrix.
\emph{(iii)} The moiety population vector reads
\vspace{-2.6ex}
\begin{equation}
	\bm M^\yrf_{\bm n} = 
	\kbordermatrix{
	& \\
	\greyt{\ce{S}} & n_{\ce{E^{\ast}}} + n_{\ce{S}} + n_{\ce{P}} \\ 
	\greyt{\ce{T}} & n_{\ce{ET}} + n_{\ce{E^{\ast}}} + n_{\ce{ED}} + n_{\ce{T}} + n_{\ce{D}}
	} \, ,
	\label{eq:MActiveT}
\end{equation}
from which the semigrand Gibbs potential $\Semigp$ follows, Eqs.~\eqref{eq:Semigp} and \eqref{eq:nonEqSemigp}.
\emph{(iv)} The driving work rate follows from the scalar product of the vector above and
\vspace{-2.6ex}
\begin{equation}
	- \partial_{t} \bm \mu_\Yp = 
	\kbordermatrix{
	& \\
	\greyt{\ce{S_{\mathrm{e}}}} & - \partial_{t} \mu_{\ce{S_{\mathrm{e}}}} \\ 
	\greyt{\ce{T_{\mathrm{e}}}} & - \partial_{t} \mu_{\ce{T_{\mathrm{e}}}}
	} \, ,
	\label{}
\end{equation}
Eqs.~\eqref{eq:dw} and \eqref{eq:aveDW}.
\emph{(v)} The chemostatted species $\ce{P_{\mathrm{e}}}$ and $\ce{D_{\mathrm{e}}}$ form the set $\Frc$ and determine the fundamental forces,
\vspace{-2.6ex}
\begin{equation}
	\bm{\mathcal{F}}_\Yf =
	\begin{pmatrix}
		\mathcal{F}_{\ce{P_{\mathrm{e}}}} \\
		\mathcal{F}_{\ce{D_{\mathrm{e}}}}
	\end{pmatrix} =
	\kbordermatrix{
	& \\
	\greyt{\ce{P_{\mathrm{e}}}} & \mu_{\ce{P_{\mathrm{e}}}} - \mu_{\ce{S_{\mathrm{e}}}} \\ 
	\greyt{\ce{D_{\mathrm{e}}}} & \mu_{\ce{D_{\mathrm{e}}}} - \mu_{\ce{T_{\mathrm{e}}}}
	} \, ,
	\label{}
\end{equation}
Eq.~\eqref{eq:fundForces}.
Together with the instantaneous external currents
\vspace{-2.6ex}
\begin{equation}
	\bm I^\Yf =
	\begin{pmatrix}
		{I}_{\ce{P}_{\mathrm{e}}} \\
		{I}_{\ce{D_{\mathrm{e}}}}
	\end{pmatrix} =
	\kbordermatrix{
	& \\
	\greyt{\ce{P_{\mathrm{e}}}} & J_{+\mathrm{p}} - J_{-\mathrm{p}} \\ 
	\greyt{\ce{D_{\mathrm{e}}}} & J_{+\mathrm{d}} - J_{-\mathrm{d}}
	} \, ,
	\label{}
\end{equation}
they identify the nonconservative contributions, Eq.~\eqref{eq:pCW}.
The first one, $\mathcal{F}_{\ce{P}_{\mathrm{e}}} I_{\ce{P}_{\mathrm{e}}}$, characterizes the work spent to convert $\ce{S}$ into $\ce{P}$, while the second, $\mathcal{F}_{\ce{D_{\mathrm{e}}}} I_{\ce{D_{\mathrm{e}}}}$, that due to the consumption of $\ce{T}$.
The sum of these terms and the driving work integrated over time contribute to the EP as in Eq.~\eqref{eq:epAwesome}.

The similar EP decomposition written in terms of nonconservative contributions along stoichiometric cycles follows when these latter are identified.
The kernel of stoichiometric matrix of the closed CRN is empty, while that of the open is spanned by
\begin{subequations}
	\begin{align}
		&\hspace{-1.2em} \bm c_{1} =
		\kbordermatrix{
			& \greyt{+1} & \greyt{+2} & \greyt{+3} & \greyt{+4} & \greyt{+5} & \greyt{+\mathrm{s}} & \greyt{+\mathrm{p}} & \greyt{+\mathrm{t}} & \greyt{+\mathrm{d}} \\
			& 1 & 0 & 0 & 1 & 1 & 0 & 0 & 1 & -1
		} \, , \label{eq:cycleActiveFutile} \\
		&\hspace{-1.2em} \bm c_{2} =
		\kbordermatrix{
			& \greyt{+1} & \greyt{+2} & \greyt{+3} & \greyt{+4} & \greyt{+5} & \greyt{+\mathrm{s}} & \greyt{+\mathrm{p}} & \greyt{+\mathrm{t}} & \greyt{+\mathrm{d}} \\
			& 1 & 1 & 1 & 1 & 0 & 1 & -1 & 1 & -1
		} \, , \label{eq:cycleActiveGood}
		\end{align}
	\label{eq:cycleActive}
\end{subequations}
which are regarded as emergent stoichiometric cycles.
Along the first, the enzyme converts one molecule of $\ce{T}$ into one of $\ce{D}$, while for the second it processes $\ce{T}$ and $\ce{S}$ and produces $\ce{D}$ and $\ce{P}$,
\begin{subequations}
	\begin{align}
		\bm C^\Y_{1} & =
		\kbordermatrix{
			& \greyt{\ce{S_{\mathrm{e}}}} & \greyt{\ce{P_{\mathrm{e}}}} & \greyt{\ce{T_{\mathrm{e}}}} & \greyt{\ce{D_{\mathrm{e}}}} \\
			& 0 & 0 & 1 & -1
		} \, , \label{eq:CActiveFutile} \\
		\bm C^\Y_{2} & =
		\kbordermatrix{
			& \greyt{\ce{S_{\mathrm{e}}}} & \greyt{\ce{P_{\mathrm{e}}}} & \greyt{\ce{T_{\mathrm{e}}}} & \greyt{\ce{D_{\mathrm{e}}}} \\
			& 1 & -1 & 1 & -1
		} \, . \label{eq:CActiveGood}
		\end{align}
	\label{eq:CActive}
\end{subequations}
At this point we can proceed from step \emph{(v)} and determine the affinities,
\begin{subequations}
	\begin{align}
		\mathcal{A}_{1} & = \mu_{\ce{T_{\mathrm{e}}}} - \mu_{\ce{D_{\mathrm{e}}}} \\
		\mathcal{A}_{2} & = \mu_{\ce{T_{\mathrm{e}}}} + \mu_{\ce{S_{\mathrm{e}}}} - \mu_{\ce{D_{\mathrm{e}}}} - \mu_{\ce{P_{\mathrm{e}}}} \, ,
	\end{align}
	\label{eq:affinitiesActive}
\end{subequations}
as well as the related instantaneous currents,
\begin{subequations}
	\begin{align}
		\mathcal{J}_{1} & = J_{+\mathrm{p}} - J_{-\mathrm{p}} - J_{+\mathrm{d}} - J_{-\mathrm{d}} \\
		\mathcal{J}_{2} & = J_{-\mathrm{p}} - J_{+\mathrm{p}} \, .
	\end{align}
	\label{eq:currentsActive}
\end{subequations}
The nonconservative work follows from the products $\mathcal{A}_{1} \mathcal{J}_{1}$ and $\mathcal{A}_{2} \mathcal{J}_{2}$, and the decomposition in Eq.~\eqref{eq:epSuperAwesome} can be thus expressed.
The former characterizes the dissipation due to the futile consumption of $\ce{T}$, since $\ce{S}$ is not converted into $\ce{P}$.
The latter, instead, is the work spent to convert $\ce{T}$ and $\ce{S}$ into $\ce{D}$ and $\ce{P}$.

\begin{figure*}[t]
	\centering
	\subfloat[][Average External Currents]
	{\includegraphics[width=.45\textwidth]{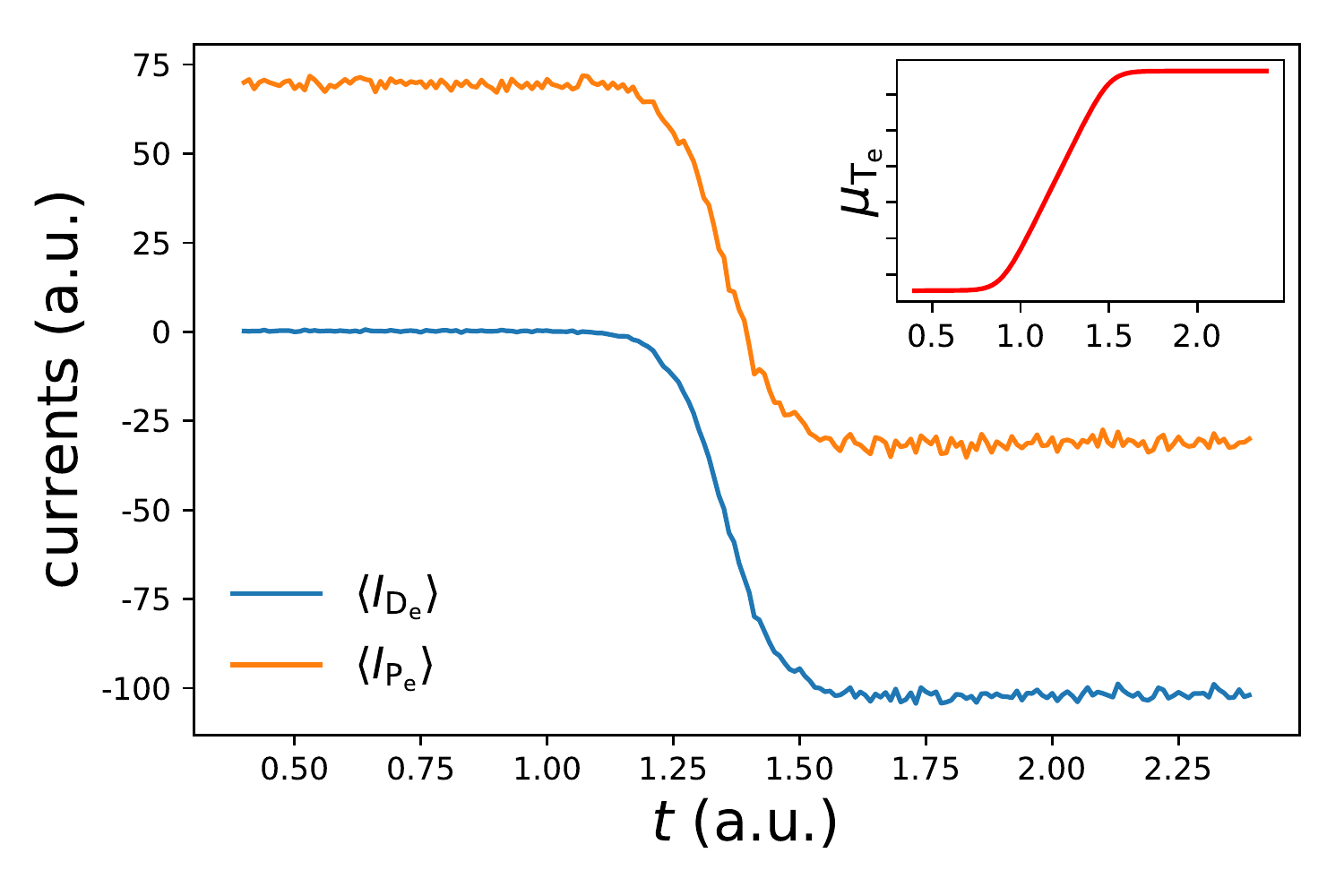} \label{fig:transdCrnts} } \quad
	\subfloat[][Average Work Contributions]
	{\includegraphics[width=.45\textwidth]{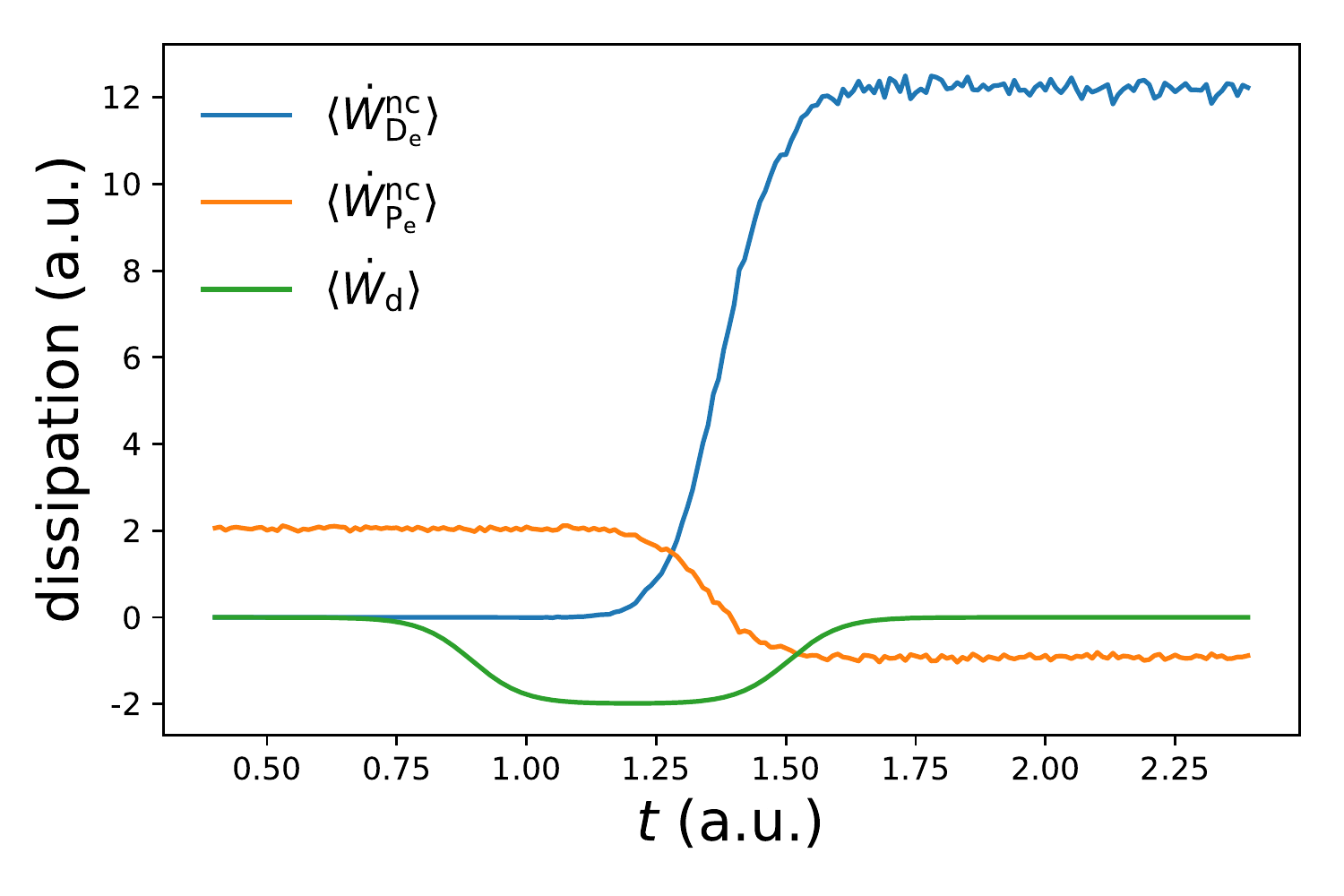} \label{fig:transdWork} }
	\caption{
		(a) average external currents and (b) average work rates \emph{vs.} time, for the CRN in Fig.~\ref{fig:transducer}.
		The plots are obtained using $10^{4}$ trajectories generated via the stochastic simulation algorithm.
		To simplify the illustration, all substrate and products are treated as chemostatted species.
		The concentrations of $\ce{S_{\mathrm{e}}}$, $\ce{P_{\mathrm{e}}}$, and $\ce{D_{\mathrm{e}}}$ are kept constant [$[\ce{S_{\mathrm{e}}}] = 10$, $[\ce{P_{\mathrm{e}}}] = 70$, and $[\ce{D_{\mathrm{e}}}] = 10$] whereas that of $\ce{T_{\mathrm{e}}}$ increases according to a logistic function: $[\ce{T_{\mathrm{e}}}] = [\ce{T_{\mathrm{e}}}]_{\mathrm{max}}/(1+\exp\left\{ - \kappa (t-t_{0}) \right\})$ [$[\ce{T_{\mathrm{e}}}]_{\mathrm{max}} = 200$, $\kappa = 20$, $t_{0} = 1.5$].
		This mimics the process in which the force that sustain the active catalysis, $\mathcal{F}_{\ce{D_{\mathrm{e}}}}$, is switched on from $0$ to a finite value after $t_{0}$.
		The change of chemical potential $\mu_{\ce{T_{\mathrm{e}}}}$ is plotted in red in the inset.
		The choice of the rate constants is as follows:
		$k_{+1} = 10^{3}$;
		$k_{+2} = 10^{3}$;
		$k_{+3} = 10^{3}$;
		$k_{+4} = 10^{3}$;
		$k_{+5} = 10^{2}$;
		whereas the backward rates are obtained by means of Eq.~\eqref{eq:lnk} using the following values for the standard-state chemical potentials:
		$\mu^{\circ}_{\ce{E}} = 1$;
		$\mu^{\circ}_{\ce{ET}} = 3$;
		$\mu^{\circ}_{\ce{E^{\ast}}} = 4$;
		$\mu^{\circ}_{\ce{ED}} = 2$;
		$\mu^{\circ}_{\ce{S_{\mathrm{e}}}} = 1$;
		$\mu^{\circ}_{\ce{P_{\mathrm{e}}}} = 2$;
		$\mu^{\circ}_{\ce{T_{\mathrm{e}}}} = 10$;
		$\mu^{\circ}_{\ce{D_{\mathrm{e}}}} = 1$.
		Since reactions are unimolecular the constant term $-\kt \bm 1 \ln [\ce{s}]$ is ignored.
		Finally, $\kt = 1$ and the value of the enzyme moiety is $L_{\mathrm{E}} = 10$.
	}
	\label{fig:transduction}
\end{figure*}

This system can be used to illustrate free energy transduction when one considers the autonomous regime where $\mathcal{F}_{\ce{D_{\mathrm{e}}}} < 0$, $\mathcal{F}_{\ce{P_{\mathrm{e}}}} > 0$, but $\avef{\dot{W}^{\mathrm{nc}}_{\ce{D_{\mathrm{e}}}}} > - \avef{\dot{W}^{\mathrm{nc}}_{\ce{P_{\mathrm{e}}}}} > 0$.
Namely, the external current of $\ce{P_{\mathrm{e}}}$ flows towards the chemostat, $\avef{I_{\ce{P_{\mathrm{e}}}}} < 0$ ($\ce{P_{\mathrm{e}}}$ produced), despite the fact that its force is positive, $\mathcal{F}_{\ce{P_{\mathrm{e}}}} > 0$.
This can happen thanks to the free energy provided by the conversion of $\ce{T_{\mathrm{e}}}$ into $\ce{D_{\mathrm{e}}}$, $\avef{\dot{W}^{\mathrm{nc}}_{\ce{D_{\mathrm{e}}}}} > 0$.
In Fig.~\ref{fig:transduction} we illustrate the behavior of the average external currents and work contributions as function of time when the transducer in Fig.~\ref{fig:transducer} is smoothly switched from a nontransducing regime to a transduction one.
At early times, $\mathcal{F}_{\ce{D_{\mathrm{e}}}} = 0$, $\mathcal{F}_{\ce{P_{\mathrm{e}}}} > 0$, and one observes only a consumption of $\ce{P_{\mathrm{e}}}$:
$\avef{I_{\ce{P_{\mathrm{e}}}}} > 0$ and $\avef{I_{\ce{D_{\mathrm{e}}}}} \simeq 0$ (respectively, orange and blue curves in Fig.~\ref{fig:transdCrnts}).
Consequently, the nonconservative work contributions are $\avef{\dot{W}^{\mathrm{nc}}_{\ce{P_{\mathrm{e}}}}} > 0$ and $\avef{\dot{W}^{\mathrm{nc}}_{\ce{D_{\mathrm{e}}}}} = 0$ (respectively, orange and blue curves in Fig.~\ref{fig:transdWork}).
In contrast, when the \emph{motive} force $\mathcal{F}_{\ce{D_{\mathrm{e}}}}$ is switched on (at large times), the current $\avef{I_{\ce{P_{\mathrm{e}}}}}$ turns negative whereas the \emph{motive} current $\avef{I_{\ce{P_{\mathrm{e}}}}}$ aligns itself with its corresponding force.
We thus observe $\avef{\dot{W}^{\mathrm{nc}}_{\ce{D_{\mathrm{e}}}}} > - \avef{\dot{W}^{\mathrm{nc}}_{\ce{P_{\mathrm{e}}}}} > 0$.
At intermediate times, driving work is extracted following the smooth increase of the \emph{motive} force (green curve in Fig.~\ref{fig:transdWork}).

\section{Conclusions and Perspectives}

In this paper we presented a thorough description of nonequilibrium thermodynamics of stochastic CRNs.
The fundamental results of traditional irreversible chemical thermodynamics (\emph{viz.} enthalpy and entropy balance) are formulated at the level of single trajectories, Eqs.~\eqref{eq:H} and \eqref{eq:epDef}.
By making use of the CRN topology and by identifying conservation laws we decompose the EP into two fundamental work contributions and a semigrand potential difference, Eqs.~\eqref{eq:epAwesome} and \eqref{eq:eprAwesome}.
The driving work describes the thermodynamic cost of manipulating the CRN by changing the chemical potentials of its chemostats.
Instead, the nonconservative work quantifies the cost of sustaining chemical currents through the CRN.
These currents prevent the CRN from reaching equilibrium, but when the related fundamental forces vanish (and the chemical potentials of the reservoirs are kept constant in time), the CRN relaxes to equilibrium by minimizing the semigrand Gibbs potential.
We elucidate the relationship between this thermodynamic potential and the dynamical potentials used in chemical reaction network theory.
Our EP decomposition written in terms of stoichiometric cycles affinities generalizes previous decompositions formulated for linear CRNs or steady-state dynamics.

Two detailed FTs follow from our EP decompositions, Eqs.~\eqref{eq:dftAwesome} and \eqref{eq:dftSuperAwesome}.
They are valid at any time and entirely expressed in terms of physical quantities.
Hence, they offer the possibility of validating experimentally our findings, and, from a wider perspective, of validating the foundations of stochastic thermodynamics beyond electronic devices or colloidal particles \cite{proesmans16,ciliberto17}.
Finally, we derive a nonequilibrium Landauer's principle for the work contributions, Eq.~\eqref{eq:minWorkAwesome}, which quantifies the minimum thermodynamic cost involved in transformations between arbitrary nonequilibrium states. 
In contrast to early formulations of the latter principle, we consider not only the cost of external manipulations, but also that related to sustained currents across the system.

Our EP decomposition identifies the fundamental dissipative contributions in CRNs of arbitrary complexity, and it can be thus used to analyze free energy conversion in CRNs beyond single biocatalysts, molecular motors, or sensory systems, which are usually described by linear CRNs \cite{seifert11,rao15:proofreading,bo15,altaner15}.
The nonconservative work contributions capture Hill's idea of free energy transduction and extend it to nonlinear CRNs with an arbitrary number of chemical forces.
[As illustrated in \S~\ref{sec:examples}, transduction occurs whenever one contribution becomes negative, thus requiring the other ones to be positive and larger than the former in absolute value by virtue of the second law of thermodynamics.]
In turn, the driving work contribution allows to generalize transduction to CRNs with reservoirs externally controlled in time.
Hence, our framework can be used to analyze pumping in CRNs \cite{astumian11,esposito15:pumps}, namely mechanisms whose periodic external control sustains a chemical current against its spontaneous direction.

In biochemical information-handling systems \cite{andrieux08:copolymerization,horowitz14,bo15,ouldridge17} and chemical computing \cite{soloveichik08}, information is stored and processed at the molecular level.
Conservation laws play a crucial role since they enable to store information in the form of nontrivial probability distributions \cite{poole17} (see \emph{e.g.} Eq.~\eqref{eq:peq(n)Open}).
Early applications of the nonequilibrium Landauer's principle proved successful for characterizing the thermodynamic cost of information processing in simple mechanisms \cite{sartori15,ouldridge17:persistent}.
Our generalization of this principle could be thus used to analyze biochemical information-handling systems of far greater complexity.
This endeavor is important in the light of the current understanding that biological systems evolved by optimizing the gathering and representation of information \cite{bialek12,tkacik16}.

Noise is known to play an important role in many biochemical processes.
Since a complete stochastic description remains both analytically and computationally demanding, developing hybrid stochastic--deterministic descriptions would be of great importance \cite{bressloff14,rao16:crnThermo,winkelmann17}.
Also, many of these processes are regulated by enzymes, thus extending the present theory beyond mass-action kinetics, as already done for deterministic CRNs \cite{wachtel18}, is also necessary.

\begin{acknowledgments}
	R.R. warmly thanks A.~Wachtel, A.~Lazarescu and M.~Polettini for valuable discussions.
	This work was funded by the National Research Fund of Luxembourg (AFR PhD Grant 2014-2, No.~9114110) and the European Research Council project NanoThermo (ERC-2015-CoG Agreement No.~681456). 
\end{acknowledgments}

\appendix

\begin{table}[tbh]
	\begin{tabular}{lcr}
		\toprule
		\textbf{symbol} & \textbf{physical quantity} & \textbf{equation} \\
		\midrule
		$\stoich_{\rho}$ 					& stoichiometric vectors 						& \eqref{eq:stoichiometricVector} \\
		$\stoichM$ 							& stoichiometric matrix 						& \eqref{eq:SMopen}--\eqref{eq:SMopenY} \\
		$\rate{\rho}{\bm n}$ 				& stochastic reaction rates 					& \eqref{eq:srr} \\
		$J_{\rho}(\tau)$ 					& instantaneous reaction fluxes					& \eqref{eq:instTrnsCrnt} \\
		$\bm \ell$ 							& conservation laws								& \eqref{eq:cl} \\
		$L_{\bm{n}}$						& component										& \eqref{eq:component} \\
		$\bm c$ 							& stoichiometric cycles							& \eqref{eq:cycle} \\
		$\bm{\mu}_{\Y}$						& chemostats chemical potential					& \eqref{eq:muYe} \\
		$g_{\bm{n}}$						& Gibbs free energy of $\bm{n}$					& \eqref{eq:g} \\
		$G(\bm{n})$							& stochastic Gibbs potential					& \eqref{eq:G} \\
		$\avef{G(\bm{n})}$					& nonequilibrium Gibbs potential				& \eqref{eq:aveG} \\
		$Z$									& closed-CRN partition function					& \eqref{eq:Zclosed} \\
		$A_{\rho}(\bm{n})$					& reaction affinity								& \eqref{eq:affinity} \\
		$s_{\bm{n}}$						& entropy of $\bm{n}$							& \eqref{eq:s} \\
		$S(\bm{n})$							& stochastic entropy							& \eqref{eq:S} \\
		$\avef{S(\bm{n})}$					& Gibbs--Shannon entropy						& \eqref{eq:gibbsShannon} \\
		$H(\bm{n})$							& enthalpy										& \eqref{eq:H} \eqref{eq:aveHbalance} \\
		$Q$	($\avef{\dot{Q}}$)				& heat flow (rate)								& \eqref{eq:Q} \eqref{eq:aveQ} \\
		$W_{\mathrm{c}}$ ($\avef{\dot{W}_{\mathrm{c}}}$)	& chemical work	(rate)							& \eqref{eq:cw} \eqref{eq:aveWc} \\
		$\bm{I}^{\Y}$						& instantaneous external currents 				& \eqref{eq:instChemoCurr} \\
		$\Sigma$ ($\avef{\dot{\Sigma}}$)	& entropy production (rate)						& \eqref{eq:ep} \eqref{eq:epr} \\
		$\bm M^\yrf_{\bm n}$				& moiety population vector						& \eqref{eq:moietyV} \\
		$\bm{\mathcal{F}}_\Yf$				& fundamental forces							& \eqref{eq:fundForces} \\
		$\bm{\mathcal{I}}^\Yf$				& fundamental external currents 				& \eqref{eq:extCrnt} \\
		$\semigp_{\bm{n}}$					& semigrand Gibbs free energy of $\bm{n}$		& \eqref{eq:semigp} \\
		$\Semigp(\bm{n})$					& stoch. semigrand Gibbs pot.					& \eqref{eq:Semigp} \\
		$\avef{\Semigp(\bm{n})}$			& noneq. semigrand Gibbs pot.					& \eqref{eq:nonEqSemigp} \\
		$\mathcal{Z}$						& open-CRN partition function					& \eqref{eq:Zopen} \\
		$W_{\mathrm{d}}$ ($\avef{\dot{W}_{\mathrm{d}}}$)	& driving chem. work	(rate)					& \eqref{eq:dw} \eqref{eq:aveDW} \\
		$W^{\mathrm{nc}}_{\iyf}$ ($\avef{\dot{W}^{\mathrm{nc}}_{\iyf}}$)	& nonconservative chem. work (rate)			& \eqref{eq:pCW} \eqref{eq:aveFW} \\
		$\mathcal{H}(\bm{n})$				& semigrand enthalpy							& \eqref{eq:calH} \eqref{eq:aveHawesome} \\
		$\mathcal{A}_{\ecy}$				& stoichiometric cycle affinity					& \eqref{eq:emergentA} \\
		$\mathcal{J}_{\ecy}$				& stoichiometric cycle current					& \eqref{eq:cycleCurrents} \\
		$\Gamma_{\ecy}$						& nonconservative cycles chem. work			& \eqref{eq:gamma} \eqref{eq:aveGamma} \\
		\bottomrule
	\end{tabular}
	\caption{
		\label{tab:glossary}
		List of symbols used throughout the text.
		The physical quantity that they denote and the equation number in which they are defined are also reported.
	}
\end{table}

\section{Thermodynamic Potentials}
\label{sec:potentials}

Using equilibrium statistical mechanics, we derive the equilibrium Gibbs free energy of a CRN in a given state $\bm n$.
Our derivation is similar to that found in Ref.~\cite[\S~3.2]{dirks07}, whereas for different approaches we refer the Reader to Refs.~\cite{mcquarrie76,diu89,beard08,ouldridge17:review}.

We regard the reacting species, labeled by $\sigma = 1,\dots,\nof{\mathrm{z}}$, as solutes of an ideal dilute solution in a closed vessel.
Since the solvent, $\mathrm{s}$, is much more abundant than the solutes, $n_{\mathrm{s}} \gg \sum_{\sigma} n_{\sigma}$.
As in ideal solutions, interactions among solutes are negligible, and the partition function of the whole solution $\mathcal{Q}(T,\bm n,n_{\mathrm{s}})$ can be written as the product of single species partition functions, $\bm q \equiv \st{q_{\sigma}(T)}$ and $q_{\mathrm{s}}$.
By idealizing the solution as a lattice gas, in which each site is occupied by one molecule, we obtain
\begin{equation}
	\mathcal{Q}(T,\bm n,n_{\mathrm{s}}) = \frac{\left( n_{\mathrm{s}} + {\textstyle\sum_{\sigma}} n_{\sigma} \right)!}{n_{\mathrm{s}}! {\textstyle \prod_{\sigma}} n_{\sigma}!} \, q_{\mathrm{s}}(n_{\mathrm{s}}) \prod_{\sigma} q_{\sigma}^{n_{\sigma}} \, .
	\label{eq:QQ}
\end{equation}
The combinatoric term accounts for all possible permutations of molecules, in which the overcounting due to the indistinguishability of molecules of the same species is removed.
We note that the fact that different molecules might occupy different volumes is neglected.

Since we deal with dilute solutions, $\bm q \equiv \st{q_{\sigma}(T)}$ depends mainly on the temperature and the solutes--solvent interactions, whereas $q_{\mathrm{s}}$ depends also on the abundance of solvent, as well as the external pressure (which we omit for brevity).
Using Stirling's formula and the high relative abundance of the solvent, the combinatoric term can be approximated as
\begin{equation}
	\frac{\left( n_{\mathrm{s}} + {\textstyle\sum_{\sigma}} n_{\sigma} \right)!}{n_{\mathrm{s}}! {\textstyle \prod_{\sigma}} n_{\sigma}!} \simeq 
	{\textstyle\prod_{\sigma}} \frac{n_{\mathrm{s}}^{n_{\sigma}}}{n_{\sigma}!}
	\equiv \frac{n_{\mathrm{s}}^{\cdot \bm {n}}}{ \bm {n}! } \, .
	\label{}
\end{equation}
Using Eq.~\eqref{eq:QQ}, the \emph{Gibbs free energy} of a given state $\bm n$ is thus given by
\begin{equation}
	\begin{split}
		g_{\bm n}
		& = -\kt \ln \mathcal{Q}(T,\bm {n},n_{\mathrm{s}}) \\
		& = \left( \bm \mu^\circ - \bm 1 \kt \ln n_{\mathrm{s}} \right) \cdot \bm {n} + \kt \ln \bm {n}! + g_{\mathrm{s}} \, ,
	\end{split}
	\label{eq:giggio}
\end{equation}
where
\begin{equation}
	\bm \mu^\circ := - \kt \ln \bm {q}
	\label{}
\end{equation}
can be identified as \emph{standard chemical potentials}.
Since the contribution deriving from the solvent, $g_{\mathrm{s}} := - \kt \, \ln q_{\mathrm{s}}(n_{\mathrm{s}})$, is constant, it can be set to zero without loss of generality.
We emphasize that despite the idealizations that we introduced, Eq.~\eqref{eq:giggio} is consistent with a rigorous approach based on mean-force potentials, \emph{cf.}~\cite[Eq.~F.44.a]{diu89}.

The Gibbs free energy changes along internal reactions read
\begin{equation}
	\hspace{-0.6em}
	\begin{split}
		& \Delta_{\ir} g = g_{\bm n + \stoich_{\ir}} - g_{\bm n} \\
		& = \left( \bm \mu^\circ - \bm 1 \kt \ln n_{\mathrm{s}} \right) \cdot \stoich_{\ir}
		+ \kt \ln \frac{(\bm n + \stoich_{\ir})!}{\bm n!} \, .
	\end{split}
	\label{eq:deltag}
\end{equation}

\paragraph*{Thermodynamic Limit}
For $V\gg 1$, $\bm n \gg \bm 1$, and finite $[\mathbf{z}] = \bm n / V$, the Gibbs potential \eqref{eq:giggio} becomes
\begin{equation}
	g_{\bm n} / V \simeq \bm \mu \cdot [\mathbf{z}] - \kt [\mathbf{z}] \cdot \bm 1 \, ,
	\label{}
\end{equation}
where
\begin{equation}
	\bm \mu = \bm \mu^\circ + \kt \ln \left\{ [\mathbf{z}]/[\mathrm{s}] \right\}
	\label{eq:chemPot}
\end{equation}
are the \emph{chemical potentials} of solutes in an ideal dilute solution, and $[\mathrm{s}] = n_{\mathrm{s}}/V$ is the concentration of solvent.
We thus recover the Gibbs free energy density of ideal dilute solutions, see \emph{e.g.} \cite{fermi56,peliti11}.

When applying the same limit to the Gibbs free energy differences, Eq.~\eqref{eq:deltag}, we recover the \emph{Gibbs free energies of reaction},
\begin{equation}
	\Delta_{\ir} g \simeq \bm \mu \cdot \stoich_{\ir} \, .
	\label{eq:deltaCP}
\end{equation}
This result also justifies the form of the second term in the local detailed balance of exchange reactions, Eq.~\eqref{eq:ldb}.

\paragraph*{Remark}
The chemical potentials of ideal dilute solutions obtained in Eq.~\eqref{eq:chemPot} are expressed in terms of the concentration of the solvent.
By including this term in $\bm{\mu}^\circ$ and introducing a reference concentration for each species $[\mathbf{z}_{0}]$, we recover the common expression for the potential of ideal dilute solutions $\bm \mu = \hat{\bm{\mu}}^\circ + \kt \ln \left\{ [\mathbf{z}]/[\mathbf{z}_{0}] \right\}$, where the standard-state chemical potential $\hat{\bm{\mu}}^\circ := \bm \mu^\circ + \kt \ln \left\{ [\mathbf{z}_{0}]/[\mathrm{s}] \right\}$ is that measured at the reference concentration.

\bigskip

Summarizing, $g_{\bm n}$ given in Eq.~\eqref{eq:giggio} characterizes the free energy of each CRN state.
In the thermodynamic limit, the traditional potentials of ideal dilute solutions are recovered.

\begin{widetext}
\section{Proofs of Detailed Fluctuation Theorems}
\label{sec:DFTproofs}

To prove the finite time detailed FTs \eqref{eq:dftAwesome} we use a moment generating functions and change the notation in favor of one using brackets and operators.

Let $P_{t}(\bm n,W_{\mathrm{d}},\st{W^{\mathrm{nc}}_{\iyf}})$ be the joint probability of observing a trajectory ending in the state $\bm n$ along which the driving work is $W_{\mathrm{d}}$ while the nonconservative contributions are $\st{W^{\mathrm{nc}}_{\iyf}}$.
These probabilities, one for each $\bm n$, are stacked in the ket $\ket{P_{t}(W_{\mathrm{d}}, \st{W^{\mathrm{nc}}_{\iyf}})}$.
The time evolution of their moment generating function,
\begin{equation}
	\ket{\Lambda_{t} (\xi_{\mathrm{d}},\st{\xi_{\iyf}})} := \int \de W_{\mathrm{d}} {\textstyle \prod_{\iyf}} \de W^{\mathrm{nc}}_{\iyf}
	\exp\left\{ -\xi_{\mathrm{d}} W_{\mathrm{d}} - {\textstyle\sum_{\iyf}} \xi_{\iyf} W^{\mathrm{nc}}_{\iyf} \right\} \ket{P_{t}(W_{\mathrm{d}}, \st{W^{\mathrm{nc}}_{\iyf}})} \, ,
	\label{eq:mgfnApp}
\end{equation}
is ruled by the biased stochastic dynamics
\begin{equation}
	\dt \ket{\Lambda_{t}(\xi_{\mathrm{d}},\st{\xi_{\iyf}})} =
	{\mathcal{W}}_{t}(\xi_{\mathrm{d}},\st{\xi_{\iyf}}) \ket{\Lambda_{t}(\xi_{\mathrm{d}},\st{\xi_{\iyf}})} \, ,
	\label{eq:biasedME}
\end{equation}
where the entries of the biased generator are given by
\begin{equation}
	\mathcal{W}_{\bm m\bm n,t}(\xi_{\mathrm{d}},\set{\xi_{\iyf}}) =
	{\textstyle\sum_{\rho}} \rate{\rho}{\bm n}
	\left\{ \exp\left\{ - {\textstyle\sum_{\iyf}} \xi_{\iyf} \mathcal{F}_{\iyf} \big(-S^{\iyf}_{\rho} \big) \right\} \delta_{\bm m,\bm n + \stoich_{\rho}} - \delta_{\bm m,\bm n} \right\}
	- \xi_{\mathrm{d}} \partial_{t} \semigp_{\bm m} \delta_{\bm n,\bm m} \, .
	\label{eq:biasedGenerator}
\end{equation}
We denoted the entries of $\stoich^{\Yf}_{\rho}$ as $\st{S^{\iyf}_{\rho}}$.
As a consequence of the local detailed balance \eqref{eq:ldbAwesome}, the stochastic generator satisfies the following symmetry
\begin{equation}
	\mathcal{W}_{t}\transpose(\xi_{\mathrm{d}},\st{\xi_{\iyf}}) = \mathcal{B}_{t}^{-1} \, \mathcal{W}_{t}(\xi_{\mathrm{d}},\st{ 1 - \xi_{\iyf} }) \, \mathcal{B}_{t} \, ,
	\label{eq:symmetryOper}
\end{equation}
where the entries of $\mathcal{B}_{t}$ are given by
\begin{equation}
	\mathcal{B}_{\bm n\bm m,t} := \exp\left\{- \beta \semigp_{\bm m}(t) \right\} \delta_{\bm n,\bm m} \, .
	\label{eq:B}
\end{equation}
Introducing the partition function for the generic equilibrium state identified by the protocol at time $\tau$, $\mathcal{Z}_{\tau} \equiv \mathcal{Z}(\pi_{\tau},\st{L_{\lambda_{\mathrm{u}}}}) = \exp\{ - \beta \mathcal{G}_{\mathrm{eq}_{\tau}} \}$, the initial condition can be written as
\begin{equation}
	\ket{\Lambda_{0}(\xi_{\mathrm{d}},\st{\xi_{\iyf}})} = \ket{p_{\mathrm{eq}_{0}}} = \mathcal{B}_{0}/\mathcal{Z}_{0} \ket{1} \, .
\end{equation}
The ket $\ket{1}$ refers to the vector in the state space whose entries are all equal to one.

In order to proceed further, it is convenient to first prove a preliminary result.
Let us consider the generic biased dynamics, \emph{e.g.} Eq.~\eqref{eq:biasedME},
\begin{equation}
	\dt \ket{\Lambda_{t}(\xi)} = \mathcal{W}_{t}(\xi) \ket{\Lambda_{t}(\xi)} \, ,
\label{eq:generation}
\end{equation}
whose initial condition is $\ket{\Lambda_{0}(\xi)} = \ket{p(0)}$.
A formal solution of Eq.~\eqref{eq:generation} is $\ket{\Lambda_{t}(\xi)} = {\mathcal{U}}_{t}(\xi) \, \ket{p(0)}$, where the time-evolution operator reads $\mathcal{U}_{t}(\xi) = \mathcal{T}_{+} \exp\left\{ \int_{0}^{t} \de \tau \, {\mathcal{W}}_{\tau}(\xi) \right\}$, $\mathcal{T}_{+}$ being the time-ordering operator.
We clearly have $\dt \mathcal{U}_{t}(\xi) = {\mathcal{W}}_{t}(\xi) \mathcal{U}_{t}(\xi)$.
Let us now consider the following transformed evolution operator
\begin{equation}
	\tilde{\mathcal{U}}_{t}(\xi) := \mathcal{X}^{-1}_{t} \mathcal{U}_{t}(\xi) {\mathcal{X}}_{0} \, ,
	\label{eq:transformedUdef}
\end{equation}
${\mathcal{X}}_{t}$ being a generic invertible operator.
Its dynamics is ruled by the following biased stochastic dynamics
\begin{equation}
	\dt \tilde{\mathcal{U}}_{t}(\xi) =
	\dt \mathcal{X}^{-1}_{t} \mathcal{U}_{t}(\xi) {\mathcal{X}}_{0} + \mathcal{X}^{-1}_{t} \dt \mathcal{U}_{t}(\xi) {\mathcal{X}}_{0}
	= \left\{ \dt \mathcal{X}^{-1}_{t} {\mathcal{X}}_{t} + \mathcal{X}^{-1}_{t} {\mathcal{W}}_{t}(\xi) {\mathcal{X}}_{t} \right\} \tilde{\mathcal{U}}_{t}(\xi)
	\equiv \tilde{\mathcal{W}}_{t}(\xi) \, \tilde{\mathcal{U}}_{t}(\xi) \, ,
	\label{eq:transformedUdynamics}
\end{equation}
which allows us to conclude that the transformed time-evolution operator is given by
\begin{equation}
	\tilde{\mathcal{U}}(\xi) = \mathcal{T}_{+} \exp\left\{ \int_{0}^{t} \de \tau \, \tilde{\mathcal{W}}_{\tau}(\xi) \right\} \, .
	\label{eq:transformedU}
\end{equation}

From Eqs.~\eqref{eq:transformedUdef}, \eqref{eq:transformedUdynamics} and \eqref{eq:transformedU} we deduce that
\begin{equation}
	\mathcal{X}^{-1}_{t} \mathcal{U}_{t}(\xi) {\mathcal{X}}_{0} = \mathcal{T}_{+} \exp\left\{ \int_{0}^{t} \de \tau \, \left[ \de_{\tau} \mathcal{X}^{-1}_{\tau} {\mathcal{X}}_{\tau} + \mathcal{X}^{-1}_{\tau} {\mathcal{W}}_{\tau}(\xi) {\mathcal{X}}_{\tau} \right] \right\} \, .
	\label{eq:magic}
\end{equation}

We can now come back to our specific biased stochastic dynamics \eqref{eq:biasedME}.
The moment generating function of $P_{t}(W_{\mathrm{d}}, \st{W^{\mathrm{nc}}_{\iyf}})$ is given by
\begin{equation}
	\Lambda_{t}(\xi_{\mathrm{d}},\st{\xi_{\iyf}}) 
	=  \braket{1|\Lambda_{t}(\xi_{\mathrm{d}},\st{\xi_{\iyf}})} 
	= \braket{1| \mathcal{U}_{t}(\xi_{\mathrm{d}},\st{\xi_{\iyf}}) \mathcal{B}_{0} / \mathcal{Z}_{0} | 1}
	= \braket{1|\frac{\mathcal{B}_{t}}{\mathcal{Z}_{t}} \mathcal{B}_{t}^{-1} \, \mathcal{U}_{t}(\xi_{\mathrm{d}},\st{\xi_{\iyf}}) \, \mathcal{B}_{0} | 1} \frac{\mathcal{Z}_{t}}{\mathcal{Z}_{0}} \, ,
	\label{eq:proofFirst}
\end{equation}
where $\mathcal{U}_{t}(\xi_{\mathrm{d}},\st{\xi_{\iyf}})$ is the time-evolution operator of the biased stochastic dynamics \eqref{eq:biasedME}.
Note that $\bra{1}{\mathcal{B}_{t}}/{\mathcal{Z}_{t}}$ is the equilibrium initial distribution of the backward process $\bra{p_{\mathrm{eq}_{t}}}$.
Using the relation in Eq.~\eqref{eq:magic}, the last term can be rewritten as
\begin{equation}
	= \braket{p_{\mathrm{eq}_{t}} | \mathcal{T}_{+} \exp\left\{ \int_{0}^{t} \de \tau \, \left[ \partial_{\tau} \mathcal{B}_{\tau}^{-1} \mathcal{B}_{\tau} + \mathcal{B}_{\tau}^{-1} \, {\mathcal{W}}_{\tau}(\xi_{\mathrm{d}},\st{\xi_{\iyf}}) \, \mathcal{B}_{\tau} \right] \right\} |1}
	\exp\left\{ - \beta \Delta \Semigp_{\mathrm{eq}} \right\} \, ,
\end{equation}
where $\Delta \Semigp_{\mathrm{eq}}$ is defined in Eq.~\eqref{eq:deltatGeq}.
Since $\partial_{\tau} \mathcal{B}_{\tau}^{-1} \mathcal{B}_{\tau} = \diag\left\{ \partial_{\tau} \semigp_{\bm n} \right\}$ the first term in square bracket can be added to the diagonal entries of the second term, thus giving
\begin{equation}
	= \braket{p_{\mathrm{eq}_{t}} | \mathcal{T}_{+} \exp\left\{ \int_{0}^{t} \de \tau \, \left[ \mathcal{B}_{\tau}^{-1} \, {\mathcal{W}}_{\tau}(\xi_{\mathrm{d}} - 1,\st{\xi_{\iyf}}) \, \mathcal{B}_{\tau} \right] \right\} |1}
	\exp\left\{ - \beta \Delta \Semigp_{\mathrm{eq}} \right\} \, .
\end{equation}
The symmetry \eqref{eq:symmetryOper} allow us to recast the latter into
\begin{equation}
	= \braket{p_{\mathrm{eq}_{t}} | \mathcal{T}_{+}
	\exp\left\{ \int_{0}^{t} \de \tau \, \mathcal{W}\transpose_{\tau}\left( \xi_{\mathrm{d}} - 1, \st{ 1 - \xi_{\iyf} } \right) \right\} |1}
	\exp\left\{ - \beta \Delta \Semigp_{\mathrm{eq}} \right\} \, .
	\label{eq:intermediario}
\end{equation}
The crucial step comes as we transform the integration variable from $\tau$ to $\tau^{\dagger} = t - \tau$.
Accordingly, the time-ordering operator, $\mathcal{T}_{+} $, becomes an anti-time-ordering one $\mathcal{T}_{-}$, while the diagonal entries of the biased generator become
\begin{equation}
	\mathcal{W}_{\bm m\bm m,t-\tau^{\dagger}}(\xi_{\mathrm{d}},\st{\xi_{\iyf}})
	= {\textstyle\sum_{\rho}} \rate{\rho}{\bm m, t-\tau^{\dagger}} + \xi_{\mathrm{d}} \, \partial_{\tau^{\dagger}} \semigp_{\bm m}(t-\tau^{\dagger}) \\
\end{equation}
from which we conclude that
\begin{equation}
	\mathcal{W}_{\bm n\bm m,t-\tau^{\dagger}}(\xi_{\mathrm{d}},\st{\xi_{\iyf}})
	= {\mathcal{W}}_{\bm n\bm m,t-\tau^{\dagger}}(- \xi_{\mathrm{d}},\st{\xi_{\iyf}})
=: {\mathcal{W}}^{\dagger}_{\bm n\bm m,\tau^{\dagger}}(-\xi_{\mathrm{d}},\st{\xi_{\iyf}}) \, .
\end{equation}
$\mathcal{W}^{\dagger}_{\tau^{\dagger}}(\xi_{\mathrm{d}},\st{\xi_{\iyf}})$ is the biased generator of the dynamics subject to the time-reversed protocol, $\pi^{\dagger}$, \emph{i.e.} the dynamics of the backward process.
Equation~\eqref{eq:intermediario} thus becomes
\begin{equation}
	= \braket{p_{\mathrm{eq}_{t}} | \mathcal{T}_{-} \exp\left\{ \int_{0}^{t} \de \tau^{\dagger} \, {\mathcal{W}_{\tau^{\dagger}}^{\dagger}}\transpose \left( 1 - \xi_{\mathrm{d}},\st{ 1 - \xi_{\iyf} } \right) \right\} |1}
	\exp\left\{ - \beta \Delta \Semigp_{\mathrm{eq}} \right\} \, .
\end{equation}
Upon a global transposition, we can write
\begin{equation}
	= \braket{1 | \mathcal{T}_{+} \exp\left\{ \int_{0}^{t} \de \tau^{\dagger} \, {\mathcal{W}_{\tau^{\dagger}}^{\dagger}} \left( 1 - \xi_{\mathrm{d}},\st{ 1 - \xi_{\iyf} } \right) \right\} |p_{\mathrm{eq}_{t}}}
	\exp\left\{ - \beta \Delta \Semigp_{\mathrm{eq}} \right\} \, ,
\end{equation}
where we also used the relationship between transposition and time-ordering
\begin{equation}
	\mathcal{T}_{+} \left( {\textstyle\prod_{i}} A\transpose_{t_{\l}} \right) = \left( \mathcal{T}_{-} {\textstyle\prod_{i}} A_{t_{\l}} \right)\transpose \, ,
\end{equation}
in which $A_{t}$ is a generic operator.
From the last expression, we readily obtain
\begin{equation}
	\begin{split}
		& = \braket{1 | {\mathcal{U}^{\dagger}_{t}} \left( 1 - \xi_{\mathrm{d}},\st{ 1 - \xi_{\iyf} } \right) |p_{\mathrm{eq}_{t}}}
	\exp\left\{ - \beta \Delta \Semigp_{\mathrm{eq}} \right\} \\
		& = \Lambda^{\dagger}_{t}\left( 1 - \xi_{\mathrm{d}},\st{ 1 - \xi_{\iyf} } \right)
	\exp\left\{ - \beta \Delta \Semigp_{\mathrm{eq}} \right\} \, ,
	\end{split}
\end{equation}
where $\Lambda^{\dagger}_{t}\left( \xi_{\mathrm{d}},\st{ \xi_{\iyf} } \right)$ is the moment generating function of $P^{\dagger}_{t}(W_{\mathrm{d}}, \st{W^{\mathrm{nc}}_{\iyf}})$.
Summarizing, we have the following symmetry
\begin{equation}
	\Lambda_{t} (\xi_{\mathrm{d}},\st{\xi_{\iyf}}) = \Lambda^{\dagger}_{t}\left( 1 - \xi_{\mathrm{d}},\st{ 1 - \xi_{\iyf} } \right)
	\exp\left\{ - \beta \Delta \Semigp_{\mathrm{eq}} \right\} \, ,
	\label{eq:proofLast}
\end{equation}
whose inverse Laplace transform gives the FT in Eq.~\eqref{eq:dftAwesome}.

\subsection*{Fluctuation Theorem for Emergent Stoichiometric Cycles Currents}
\label{sec:proofCycles}

The finite-time detailed FT for nonconservative contributions along fundamental cycles, Eq.~\eqref{eq:dftSuperAwesome}, follows the same logic and mathematical steps described above.
The moment generating function which now must be taken into account is
\begin{equation}
	\ket{\Lambda_{t} (\xi_{\mathrm{d}},\st{\xi_{\eta}})} := \int \de W_{\mathrm{d}} {\textstyle \prod_{\eta}} \de \Gamma_{\eta}
	\exp\left\{ -\xi_{\mathrm{d}} W_{\mathrm{d}} - {\textstyle\sum_{\eta}} \xi_{\eta} \Gamma_{\eta} \right\} \ket{P_{t}(W_{\mathrm{d}}, \st{\Gamma_{\eta}})} \, ,
	\label{eq:mgfnCycles}
\end{equation}
which is ruled by the biased generator whose entries are
\begin{equation}
	\mathcal{W}_{\bm m\bm n,t}(\xi_{\mathrm{d}},\set{\xi_{\eta}}) =
	{\textstyle\sum_{\rho}} \rate{\rho}{\bm n}
	\left\{ \exp\left\{ - {\textstyle\sum_{\eta}} \xi_{\eta} \mathcal{A}_{\eta} \zeta_{\eta,\rho} \right\} \delta_{\bm m,\bm n + \stoich_{\rho}} - \delta_{\bm m,\bm n} \right\}
	- \xi_{\mathrm{d}} \partial_{t} \semigp_{\bm m} \delta_{\bm n,\bm m} \, .
	\label{eq:biasedGeneratorAff}
\end{equation}
The symmetry of the latter generator---on top of which the proof is constructed---is based on the expression of the local detailed balance given in Eq.~\eqref{eq:ldbAwesome},
\begin{equation}
	{\mathcal{W}}_{t}\transpose(\xi_{\mathrm{d}},\st{\xi_{\eta}}) = \mathcal{B}_{t}^{-1} \, \mathcal{W}_{t}(\xi_{\mathrm{d}},\st{ 1 - \xi_{\eta} }) \, \mathcal{B}_{t} \, ,
	\label{eq:symmetryOperAff}
\end{equation}
where the entries of $\mathcal{B}_{t}$ are given in Eq.~\eqref{eq:B}.
Following the steps from Eq.~\eqref{eq:proofFirst} to Eq.~\eqref{eq:proofLast}, with the definitions and equations in Eqs.~\eqref{eq:mgfnCycles}--\eqref{eq:symmetryOperAff}, proves the FT in Eq.~\eqref{eq:dftSuperAwesome}.
\end{widetext}

\bibliography{indiceLocale}

\end{document}